\documentclass{article}

\usepackage{arxiv}
\usepackage[title]{appendix}
\usepackage[utf8]{inputenc} 
\usepackage[T1]{fontenc}    
\usepackage{hyperref}       
\usepackage{url}            
\usepackage{booktabs}       
\usepackage{amsfonts}       
\usepackage{nicefrac}       
\usepackage{microtype}      
\usepackage{lipsum}

\usepackage{tikz}
\usepackage{amsmath}
\usepackage{amssymb,amsmath,epsfig,amsthm}
\usepackage{graphicx,latexsym,subfigure}
\usepackage{color,soul}
\usepackage{cite}
\usepackage{soul}
\usepackage{diagbox}

\usepackage{xifthen}
\usepackage{versions}
\usepackage{caption}

\usepackage{titlesec}
\usepackage{multirow, array}
 
\titleformat{\paragraph}{\bf\normalsize}{\theparagraph}{1em}{}\titlespacing*{\paragraph}{0pt}{3.25ex plus 1ex minus .2ex}{1.5ex plus .2ex}

\newcommand{\heading}[1]{{\noindent\bf{#1}}~}

\newcommand\eat[1]{}
\newcommand{\obssum}{{\noindent\bf{ Summary:}}}

\title{A Large Scale Study and Classification of VirusTotal Reports on Phishing and Malware URLs}

\author{
  Euijin Choo \\
  University of Alberta\\
  \texttt{euijin@ualberta.ca} \\
   \And
     Mohamed Nabeel  \\
  Qatar Computing Research Institute\\
  \texttt{mnabeel@hbku.edu.qa} \\
         \And
     Ravindu De Silva\\
     SCoRe Lab
  \\
  \texttt{ravindud@scorelab.org} \\
       \And
     Ting Yu \\
  Qatar Computing Research Institute\\
  \texttt{tyu@hbku.edu.qa} \\
     \And
     Issa Khalil \\
  Qatar Computing Research Institute\\
  \texttt{ikhalil@hbku.edu.qa} \\
}

\begin{document}
\maketitle

\begin{abstract}
VirusTotal (VT) provides aggregated threat intelligence on various entities including URLs, IP addresses, and binaries. It is widely used by researchers and practitioners to collect ground truth and evaluate the maliciousness of entities. In this work, we provide a comprehensive analysis of VT URL scanning reports containing the results of 95 scanners for 1.577 Billion URLs over two years. Individual VT scanners are known to be noisy in terms of their detection and attack type classification. To obtain high quality ground truth of URLs and actively take proper actions to mitigate different types of attacks, there are two challenges: (1) how to decide whether a given URL is malicious given noisy reports and (2) how to determine attack types (e.g., phishing or malware hosting) that the URL is involved in, given conflicting attack labels from different scanners. In this work, we provide a systematic comparative study on the behavior of VT scanners for different attack types of URLs. A common practice to decide the maliciousness is to use a cut-off threshold of scanners that report the URL as malicious. However, in this work, we show that using a fixed threshold is suboptimal, due to several reasons:  (1) correlations between scanners; (2) lead/lag behavior; (3) the specialty of scanners; (4) the quality and reliability of scanners. A common practice to determine an attack type is to use majority voting. However, we show that majority voting could not accurately classify the attack type of a URL due to the bias from correlated scanners. Instead, we propose a machine learning-based approach to assign an attack type to URLs given the VT reports. 

\end{abstract}

\section{Introduction}\label{sec:intro}





Machine learning models that detect or predict malicious URLs greatly rely on ground truth datasets for training and evaluation. It is thus paramount not only to have high quality ground truth URLs, but also to compile such ground truth in an efficient and scalable way to address the need for high volume ground truth (e.g., for deep learning-based models) and the need for relatively frequent model retraining. Many research works utilize VirusTotal (VT) to build such ground truth datasets ~\cite{li2019tealeave,sharif2018impending,vt1:phishing:2018,vt1:www:2018,vt3:oprea2018made,vt1:zuo2017smartgen}. VT is an attractive source due to its large coverage of URLs and aggregated threat intelligence from multiple scanners. Each scanner not only assesses whether a given URL is malicious but also labels what type of attacks (e.g., phishing, malware) the URL is involved in. However, most VT scanners are a black box to users. How they decide on the maliciousness and the type of attacks is rarely known to the public. To build a high quality ground truth of malicious URLs, it is necessary to derive a strategy to 
aggregate the reports for a URL from multiple scanners and make a final decision on the maliciousness and the type of attacks for the URL. 

To this end, we analyze all the VT URL reports generated from July 2019 to Jan. 2022. We study various characteristics in URL scan reports, such as the attack types of URLs, the stability and correlation of individual scanners and aggregated results in terms of detection and attack type classification, and lead/lag behavior. 

Recent studies on VT are limited in scale or diversity~\cite{zhu2020:labeldynamics,tops2021maat,peng2019blackbox,Kantchelian:2015:VTMalClassifier,salem2021maat,bouwman2022helping}. They often focused on one type of entity: malware binaries~\cite{zhu2020:labeldynamics,tops2021maat,Kantchelian:2015:VTMalClassifier,salem2021maat}, phishing URLs~\cite{peng2019blackbox}, or domains for COVID-19 related threats~\cite{bouwman2022helping}. Furthermore, while most previous works focused on malware binaries, it is more challenging to analyze URL reports than binaries as URLs themselves exhibit dynamic behaviors (i.e., although a URL does not change, its contents could change from time to time). Thus, it necessitates an independent analysis for URL reports to answer three key questions: (1) \textit{How do VT scanners behave for URLs?} (2) \textit{How to identify attack types of URLs from VT reports?} (3) \textit{Do VT scanners behave differently for URLs of different attack types?} 

The number of scanners that label a URL as malicious is called the number of positives in the VT ecosystem, which is often used to represent the level of maliciousness~\cite{www:vt_level_malicious}. A threshold-based approach has been commonly used to aggregate VT reports, i.e., if the positive count for a URL is above a threshold, it is considered malicious. 
%
But \textit{Which threshold should one use?} Setting the threshold too high may result in many false negatives, whereas a very low threshold (e.g., 1) would incur many false positives. So far, there is no principled way in the literature to set this threshold. Indeed, prior research assigns arbitrary thresholds such as 1~\cite{vt1:imc:2018,vt1:phishing:2018,vt1:www:2018}, 2~\cite{sharif2018impending}, 5~\cite{www:vt_level_malicious} to 10 or more~\cite{malwarevt:ccs18}, as they see fit. 

Our study shows that using a fixed threshold is sub-optimal and VT positive count for a URL does not necessarily reflect its maliciousness level, due to various reasons: (1) scanners could be highly correlated and should not be treated independently. The level of correlation is also different for different attack types of URLs; (2) some scanners could report the maliciousness of a URL much earlier than others, which means a threshold suitable at one time point may not be so anymore at another time period; (3) scanners often specialize in detecting URLs with different attack types such as phishing and malware hosting URLs; and (4) the quality and reliability of scanners could vary significantly. Some scanners change their labels on a URL multiple times in a short time period, suggesting they should not carry the same weight when building the ground truth of malicious URLs.



Strategies to mitigate attacks and/or the remedial actions in case of a compromise greatly depend on the types of attacks~\cite{choi2011detecting}. The issue is compounded as an overwhelming majority of cyber attacks are launched through compromising benign websites~\cite{compromised:usenix:2021}. For example, when a website is infected with malware, an initial remediation action is to check for the file integrity and malicious code injections whereas when a website is compromised with a phishing page, an initial task is to identify pages or folders that are created recently and contain login/payment forms. It is thus crucial to promptly and accurately determine the attack type of a malicious URL to build the corresponding ground truth ~\cite{feal2021blocklist}. 
The current practice is to rely on either blocklists or heuristics. Blocklist-based approaches utilize lists tailored to specific attack types (e.g., OpenPhish and Phishtank for phishing URLs~\cite{phishtank,openphish}, or URLhaus and Malware Domain List for malware URLs (i.e., those distributing malware) ~\cite{urlhaus, maldomlist}). However, each of these sources is either slow to update or has a low coverage of malicious URLs~\cite{phishingbl:2020,bouwman2022helping,li2019tealeave}. On the other hand, heuristic-based approaches, such as identifying input forms in a malicious URL to label it as phishing, tend to have both high false positives and false negatives as attackers increasingly adopt cloaking techniques to evade detection~\cite{Zhang2021CrawlPhishLA}. In light of this need, recent research utilized attack types labeled by individual VT scanners and the majority voting approach is employed to consider noises in VT reports~\cite{compromised:usenix:2021}. However, our analysis shows that conflicting attack labels frequently exist for individual scanners over time (i.e., temporal conflicts) and among multiple scanners at a given time point (i.e., cross-scanner conflicts). We also show that majority voting could assign incorrect attack type labels for the following reasons. First, highly correlated scanners would bias the final label if they are treated independently in majority voting. Second, some scanners give a generic ``malicious'' label for a URL which could be phishing or malware. The simple majority voting approach may not thus identify a specific attack type and lead to failure to mitigate the attack. 
In this paper, we propose a machine learning based approach to classify the attack types of malicious URLs into Phishing or Malware given their VT reports considering correlations between scanners. We show that the proposed model achieves high accuracy, and utilize it to perform an in-depth study of the behavior of VT scanners.

\heading{Our Contributions.} First, we perform a large-scale study of VT URL feed data spanning two years and measure various characteristics of VT reports (Section~\ref{sec:measurement}), including scanners' specialty, scanners' stability, attack type classification, scanner correlations, and lead/lag behavior. To the best of our knowledge, this is the first work of a large-scale in-depth analysis on VT URL scan reports and the first systematic comparative study for different attack types of URLs. 

Second, we propose a machine learning-based approach that takes scanners' correlations and specialties to identify  the attack type of malicious URLs and then apply the trained models to study the attack types reported in VT (Section~\ref{sec:attack_type_detection}). Our approach greatly outperforms a baseline majority voting approach in terms of accuracy (our approach: 97.47\% vs majority voting: 81.72\%).

Finally, we provide practical suggestions using VT to compile better malicious ground truth considering URL types and characteristics of scanners (Section~\ref{sec:discussion}).

\section{Our Key Findings}\label{sec:key_finding}
Before presenting the details of our study, we summarize our key observations in this section.

\heading {Observation 1. Conflicting Attack Type Labels.} We perform a systematic quantitative study and show that conflicting attack type labels are common in individual scanners temporally and across scanners (Section~\ref{sec:attack_type_analysis} and Section~\ref{sec:scanner_stability}). In general, we observe that phishing URLs have more conflicting labels than malware URLs. We emphasize that researchers need to consider such conflicts and identify attack types to collect reliable corresponding ground truth. In line with it, we propose a method to identify attack types given the conflicting labels (Section~\ref{sec:attack_type_detection}).

    
{\bf Observation 2. Scanners' Specialty and Detection Performance.} Scanners specialize in different attack types, and no scanner performs well for all types of URLs (Section~\ref{sec:scanner_accuracy}). For example, we observe AegisLab WebGuard performs well in detecting malware URLs, whereas Bitdefender performs well in detecting phishing URLs. Also, scanners often work poorly in the early reports when a URL first appears in VT. We observe that scanners reach the maximum F-1 score near the 5th day since the first appearance in VT. This indicates the need to evaluate scanners' reliability and quality depending on attack types and derive the optimal time period to collect ground truth. 

{\bf Observation 3. Scanners' Correlation on Detection and Attack Types Classification.} Some scanners are highly correlated in terms of their detection, attack type labels, their temporal label similarity, and trends of label patterns (Section~\ref{sec:correlation}). The set and number of highly correlated scanners are different depending on attack types. Also, we observe that scanners detecting phishing URLs are more correlated than those detecting malware URLs. Finally, fewer scanners correlate in terms of their labels on attack types than detection itself. Concretely, 27\% of scanners detecting phishing URLs and 5\% detecting malware URLs have a high correlation on co-detected URLs; while, only 3\% of scanners detecting phishing URLs and 3\%  detecting malware URLs have a high correlation on attack label assignment for given URLs. 

{\bf Observation 4. Lead/Lag Relationships among Scanners for Each Attack Type.} Lead/lag relationships exist among highly correlated scanners (Section~\ref{sec:lead_lag}). For example, Webroot and alphaMountain.ai have highly correlation for phishing URLs, while Webroot always detects earlier than alphaMountain.ai. Meanwhile, the set of leaders is different depending on the attack types (e.g., top 5 leaders are Sophos, OpenPhish, PhishLabs, Netcraft, and Segasec for phishing URLs; Kaspersky, Fortinet, Webroot, Sophos, Segasec for malware URLs). We recommend that researchers consider correlation and lead/lag relationships to choose a proper threshold and a set of scanners to form ground truth of URLs with a specific attack type.

\section{Data Collection and Preliminaries}\label{sec:data}
This section describes our data collection methodology, and the terminologies and notations used in the paper. The final dataset is summarized in Table~\ref{table:vtdata_summary} and Table~\ref{table:gtdata_summary}.

\begin{table}[tb]
\small
\begin{center}
\parbox{0.5\textwidth}{

\begin{tabular}{|c|c|c|}\hline
\textbf{URL Types}  & \textbf{\# of URLs} & \textbf{URL Collection Date}\\
\hline
VT General Feed         &  1577M  & 07/2019 - 01/2022\\
\hline
VT Fresh        & 224 M  & 07/2019 - 01/2022\\
\hline
\end{tabular}
\captionsetup{width=.98\linewidth}
\caption{Statistics for VT URL feeds with collection dates}~\label{table:vtdata_summary}
}
\end{center}
\end{table}
\begin{table}[tb]
\small

\begin{center}
\parbox{0.5\textwidth}{
\begin{tabular}{|p{1cm}|c|p{0.8cm}|p{2.5cm}|}\hline
\multicolumn{2}{|c|}{\textbf{URL Types}}  & \textbf{\# of URLs} & \textbf{URL Collection Date}\\ 
\hline
\multicolumn{2}{|c|}{Manual GT Benign} & 421 & 08/30/2020 - 09/03/2020, 10/10/2021\\
\hline
\multicolumn{2}{|c|}{Manual GT Malicious} & 352 & 08/30/2020 - 09/03/2020, 10/10/2021\\
\hline
Phishing & APWG Phishing & 9186 & 04/20/2021\\ \cline{2-4}
 & SiteAdvisor Phishing & 7000 & 06/05/2021\\\cline{2-4}
\hline
Malware & APWG Malware & 223 & 04/20/2021\\ \cline{2-4}
 & SiteAdvisor Malware & 6485 & 06/05/2021\\\cline{2-4}
\hline
\end{tabular}}
\captionsetup{width=.98\linewidth}
\caption{Statistics for ground truth URLs with collections dates (For all URLs, we use all scan reports between 07/2019 - 01/2022)}
\label{table:gtdata_summary}
\end{center}
\end{table}

\subsection{VT General Feed}\label{sec:vt_general_data}
We collected the scan reports of all URLs submitted to VT from July 2019 to Jan. 2022 through a subscription service to VT. There are nearly 5 million unique scan reports in total. However, interestingly, there are only about 500K and 350K newly observed FQDNs (Fully Qualified Domain Names) and apexes, respectively.

Each URL scan report presents the aggregated results of scanners.  We observe that each URL is not always scanned by the same set of scanners. We consider all scanners appearing in our dataset. In total, we have 95 scanners, which are listed in Appendix II. Between 1 and 2 million unique scan reports indicate that the scanned URL is detected by at least one scanner each day. Among them, we observe around 50K and
20K newly observed hosts and domains, respectively. 

In each scan report, we are specifically interested in the following fields: \emph{url}, \emph{scan\_date}, \emph{first\_seen}, \emph{scan\_id}, \emph{positives}, and \emph{scans}. The field \emph{scans} contains the name of the scanners, and two subfields: \emph{detected} (whether a URL is malicious or not according to the scanner), and \emph{result} (the attack type indicated by the scanner, such as malicious, phishing, malware, suspicious, mining, not recommended, and spam sites).  

The \emph{scan\_id} represents a unique scan. When a user queries a URL scanned before, VT may either rescan the URL and generate an updated scan report with a new \emph{scan\_id}, or simply return the previous scan report with the same \emph{scan\_id}. We observe that VT returns previous reports if the scan was done within a short time unless the user explicitly requests to rescan. As multiple reports with the same \emph{scan\_id} are duplicates leading to biased results, we only extract scan reports with a unique \emph{scan\_id}. 
The field \emph{positives} is the number of scanners that detect a particular URL as malicious. The detailed information for fields in a VT report is in ~\cite{virustotal}.

As our goal is to study the trends of scan reports of malicious URLs, we filter URLs that are never detected as malicious by any scanners during our study period. 


\subsection{VT Fresh URLs}\label{sec:vt_freshdata}

To better understand how VT reacts to URLs over time, we further compile VT fresh URLs among VT general feed, i.e., those URLs that are first scanned during our study period. Concretely, we extract URLs whose \emph{first\_seen} is the same as its \emph{scan\_date}. Then, we extract all the scan reports over two years for the fresh URLs. 

\subsection{Ground Truth Data Collection }\label{sec:gtdata}

\begin{figure}[tb]
\begin{center}
\parbox{0.5\textwidth}{
\centering 
\epsfig{file=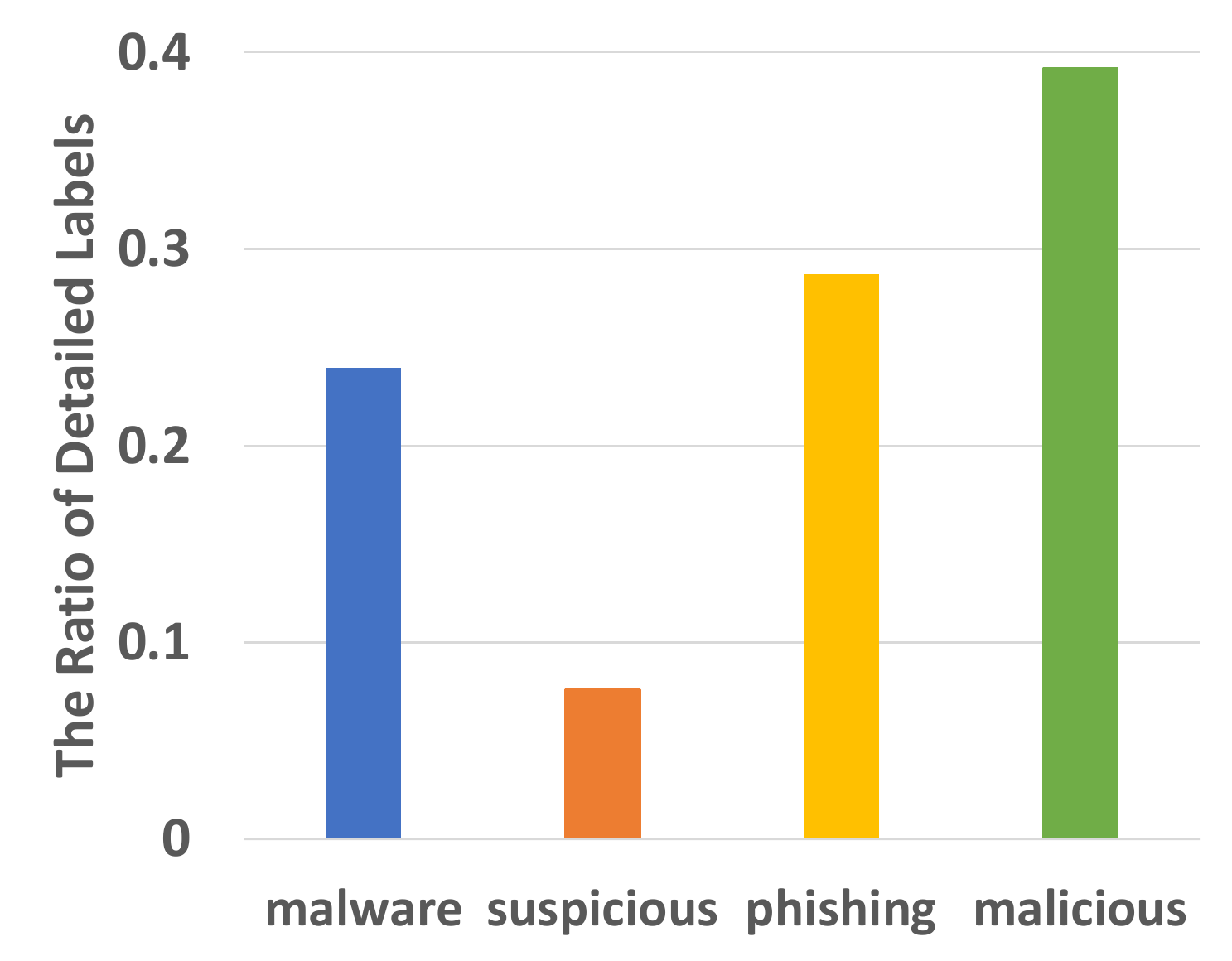,width=0.46\textwidth}
}
\captionsetup{width=.95\linewidth}
\caption{The ratio of top 4 detailed labels over the total number of VT reports for VT Fresh URLs}\label{fig:detail_label_stat}
\end{center}
\end{figure}

In Section~\ref{sec:measurement}, we analyze the characteristics of VT scan reports and individual VT scanners, respectively. To do so, we build ground truth datasets with multiple approaches: manual labelling (Section~\ref{sec:manual_data}) and two publicly available URL intelligence feeds not part of VT scanners (Section~\ref{sec:gt_public}). Figure~\ref{fig:detail_label_stat} shows that the non-generic attack types of most malicious URLs in VT are either phishing or malware. We thus focus on phishing and malware ground truth. For better confidence on the ground truth, labels for all URLs are manually verified using the rubric described in Section~\ref{sec:manual_data}.

In practice, it is crucial to detect short-lived malicious URLs as early as possible for their threats to be quickly contained. 
To fully understand the behavior of each scanner and URLs, we track scanner reports from the very first appearance of URLs in VT and study the dynamics of scanner responses over time. To do so, after collecting the ground truth datasets of URLs, we take two approaches: (1) submit the URLs to VT and request to rescan them periodically (\textbf{prospective study}); and (2) do a \textbf{retrospective study} to extract reports for the URLs from VT General Feed. A few factors need to be considered when selecting the proper time granularity for building the periodic reports. First, the status of malicious URLs changes rapidly. They could be taken down after attacks~\cite{choo2019devicewatch}, cleaned after being compromised~\cite{compromised:usenix:2021}, or re-registered after take-down to reuse for new attacks~\cite{reg:2017:RAID}. Recent research suggests that only a few malicious URLs have a lifetime of more than a month ~\cite{kim2020anatomy}, while most malicious URLs have a few days or even a few hours ~\cite{kim2020anatomy,oest2020phishtime}. Second, it has been shown that even though a scanner may update its malicious URL list shortly after detection of new malicious URLs~\cite{oest2020sunrise}, VT does not necessarily promptly update the scanner's result in its database~\cite{peng2019blackbox}. We collect periodic reports daily and hourly based on these observations and our empirical analysis. We will discuss each dataset in the following sections and their possible limitations in Section~\ref{sec:discussion}.

\subsubsection{Manually Labeled URLs (Manual GT URLs)}\label{sec:manual_data}
We pick 3800 URLs from VT Fresh URLs using a stratified sampling based approach~\cite{bennett2010online,katariya2012active}, which are then manually labeled by 5 domain experts. The detailed process is described in the Appendix and the dataset is summarized in Table~\ref{table:gtdata_summary}. In brief, the experts applied rules including but not limited to the following. A URL is labeled as a malware URL if a malware file is located in the URL, as a phishing URL if it is a squatting domain~\cite{moubayed2018dnssquatting} or unknown URLs mimicking popular URLs (e.g., \url{malicious.com} with the login image of \url{paypal.com})~\cite{peng2019blackbox}, and as a benign URL if the URL does not have any malicious signals and is operational for at least 3 months. For a better confidence on labeling, all URLs are labeled by two experts and we exclude URLs with conflicting labels. 773 out of the 3800 URLs are labeled while the remaining 3027 URLs cannot be labeled as it
has neither malicious nor benign signals, or it has conflicts between two experts. While we observe that most malicious URLs in this set have phishing signals, we only use binary labels for this dataset: benign or malicious. We use benign URLs in this set as benign ground truth for all analyses. 



\subsubsection{Publicly Available Intelligence Sources}\label{sec:gt_public} 
We additionally collect malicious URLs using two publicly available intelligence sources - APWG and SiteAdvisor. We specifically choose these two because they are not part of VT scanners. These two sources are further manually verified to identify the attack types of URLs. The manual annotation process uses the same rubric as in Section~\ref{sec:manual_data}.

\paragraph{3.3.2.1. Anti-Phishing Working Group (APWG) URLs}\label{sec:apwg_data}
APWG is a community-based service where the URLs are labeled by domain experts from multiple institutions~\cite{APWG}. We download the latest 10K URLs from APWG and filter out invalid URLs (e.g., malformed URLs) and non-fresh URLs (i.e., the URL's first appearance in VT (\emph{first-seen}) is older than the studying period). Since APWG does not provide the attack types of URLs, they are labeled manually using the above-mentioned rubric. 

\paragraph{3.3.2.2. McAfee SiteAdvisor URLs}\label{sec:siteadvisor_data}
We collect 7K and 6.5K phishing and malware URLs respectively as follows. We first collect random samples of URLs from VT Fresh URLs having at least one phishing or malware label. We then employ SiteAdvisor (SA) URL report to assist in labeling ground truth. In addition to detailed comments, SA reports~\cite{siteadvisor} include an attack category and one of the four risk levels: unverified, low, medium and high risk. We extract the SA medium and high risk URLs 
and utilize the detailed threat reports as a guide along with the above-mentioned rubric to manually label the attack types of URLs as phishing or malware. 

\subsection{Terminologies and Notations}\label{sec:notation}





We sort the scan reports by time for each URL and represent the data per scanner as a time series, i.e., a sequence of chronologically ordered data points. Each data point corresponds to the scanner's label of the URL for a given time frame. To see the long-term trend of scanners and URLs, we present results using the daily time granularity throughout the paper. 

We use two types of labels for each time frame: a binary label (\emph{detected} field in VT reports) and a detailed label (\emph{result} field in VT reports). A binary label represents whether the scanner detected the URL as malicious or not, encoded as 1 or 0; a detailed label identifies the attack label assigned by scanners such as  ``malware site'' and ``phishing site''. 

If there are multiple scan reports in a given time frame (i.e., a day), we select the highest binary label given by each scanner as the binary label of the scanner. 
We observe that although a scanner may have both 0 and 1 binary labels within a day, reports with ``1'' as binary labels have a single detailed label per scanner within a day. We thus use it as the detailed label for a given time frame. 


For a given scanner $s$ and a URL $u$, its time series is represented 
as a binary sequence $BL_{s,u}=[bl_{t_1},bl_{t_2},..bl_{t_n}]$ or a detailed label sequence $DL_{s,u}=[dl_{t_1},dl_{t_2},..dl_{t_n}]$ where $t_i$ is the $i^{th}$ time frame, $bl_{t_i} \in \lbrace 0 , 1 \rbrace$, and $dl_{t_i} \in \lbrace$ 0, ``phishing sites'', ``malicious sites'', ``malware sites'', ``suspicious sites'', ``spam sites'', ``mining sites'', ``not recommended sites'' $\rbrace$.

\section{Measurement Study on VirusTotal}\label{sec:measurement}
\subsection{Attack Type Analysis}\label{sec:attack_type_analysis}
Recall that VT scanners assign a detailed label for a malicious URL they detect. We observe that the detailed labels assigned by multiple scanners in the same VT report often conflict.
For example, scanners may give different labels depending on their detection method (e.g., one says ``malware'' and another says ``phishing'' for a URL). This section analyzes the statistics of attack types for each set of URLs based on the detailed labels. 

\begin{figure}[htbp]
\parbox{1.0\textwidth}{
\centering
\subfigure[The CDF of the number of detailed labels]{
\psfig{file=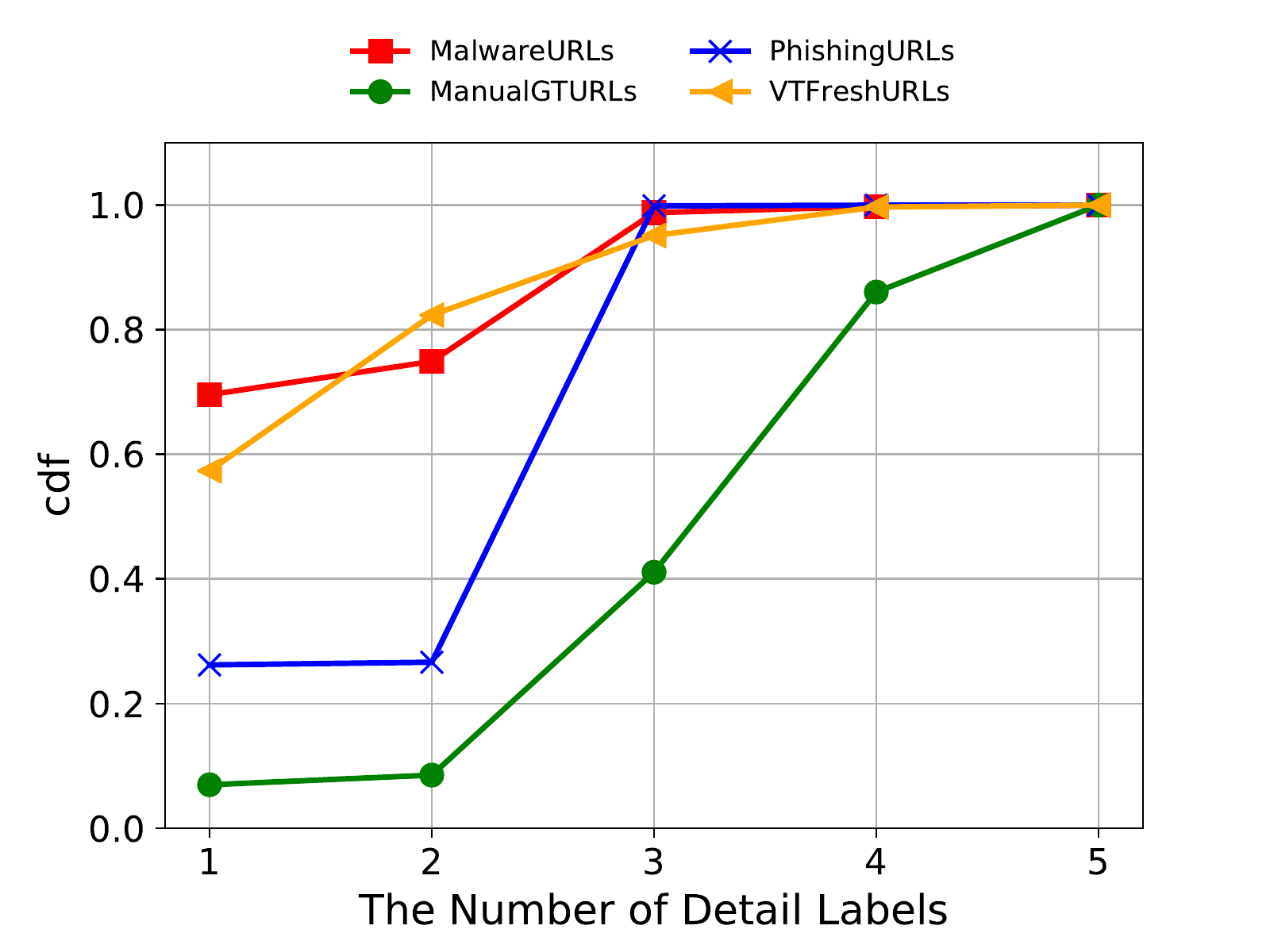,width=0.45\textwidth}\label{fig:attacktype_cdf}}
\subfigure[The ratio of top 4 detailed labels over the total number of scan results ]{
\psfig{file=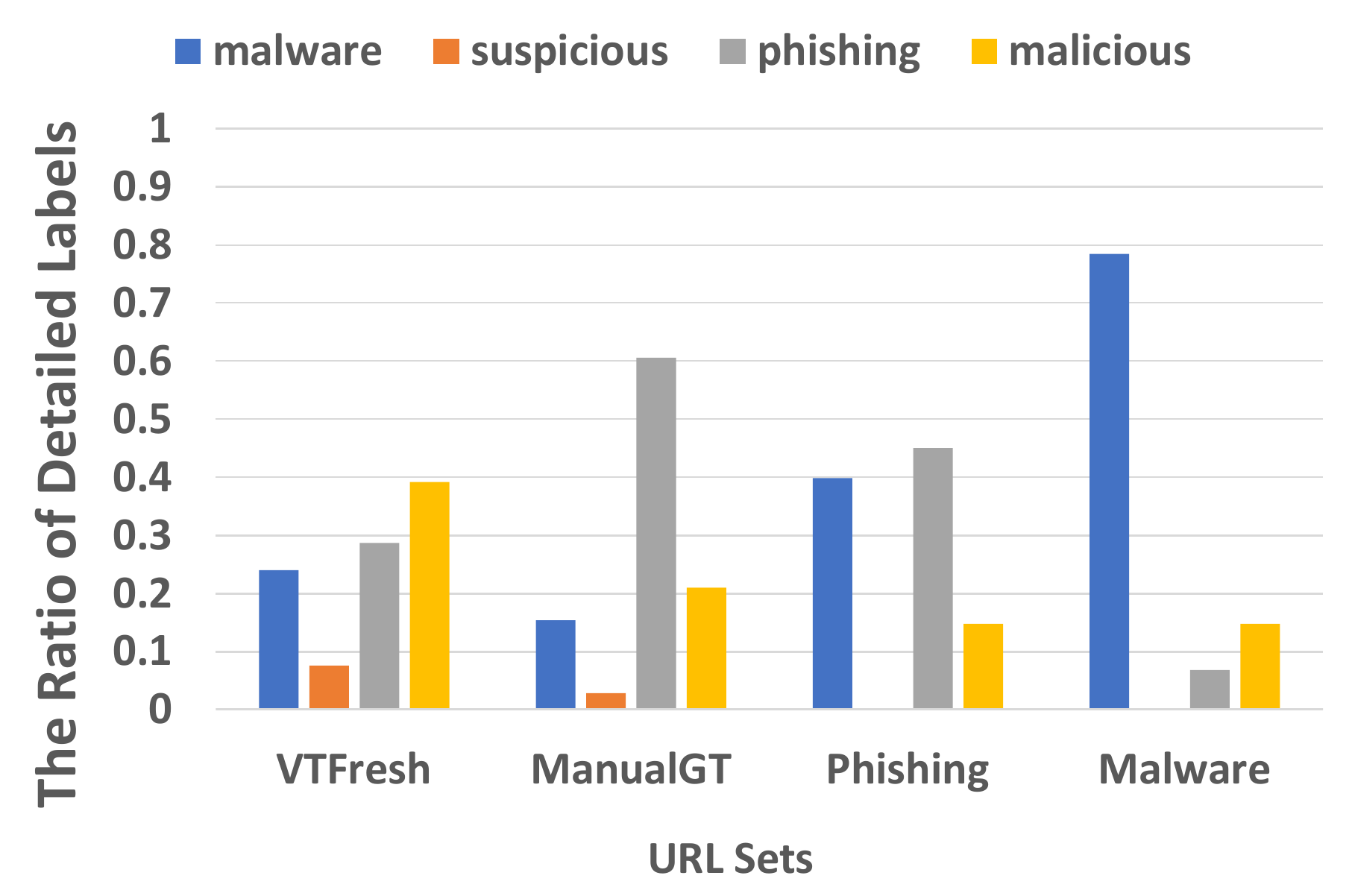,width=0.45\textwidth}\label{fig:attacktype_ratio}}
\captionsetup{width=.98\linewidth}
\caption{The statistics of detailed labels (attack types) for each URL set
}\label{fig:attack_type_analysis}}
\end{figure}

Figure~\ref{fig:attacktype_cdf} shows the CDFs of the number of the detailed labels for each URL set where the x-axis presents the number of detailed labels for each URL, and the y-axis presents the CDF (i.e., the portion of URLs). Figure ~\ref{fig:attacktype_ratio} shows the ratio of the top 4 detailed labels over the total number of scan results. Each bar presents a detailed label, the x-axis presents URL sets, and the y-axis presents the ratio of each detailed label. Figure~\ref{fig:attacktype_cdf} clearly shows the different trends. That is, phishing URLs tend to have more labels than malware URLs. 75\% of phishing URLs have 3 or more but only 25\% of the malware URLs have 3 or more labels. Further, Figure~\ref{fig:attacktype_ratio} shows that while 78.5\% of labels for malware URLs are malware, only 45\% of labels for phishing URLs are phishing.

We observe multiple scenarios leading to multiple detailed labels for a URL. First, different scanners often assign different detailed labels to the same URL. We observe that 84.3\% of URLs with multi-labels are due to this scenario. For example, \url{http://faceasdasdasd.000webhostapp.com/} is always marked as ``phishing'' by AegisLab WebGuard, Fortinet, Kaspersky, Phishtank, Avira, CLEAN MX, Phishing Database, ESET, OpenPhish, G-Data, Emsisoft, and Google Safebrowsing; as ``malware'' by Sophos, BitDefender, and SCUMWARE.org; and as ``malicious'' by AlienVault, CRDF, Netcraft, CyRadar, and Forcepoint ThreatSeeker. 

Second, some scanners change their detailed labels for the same URL. We observe that 15.7\% of URLs with multi-labels are due to such scanners switching their detailed labels in a short time. In Section~\ref{sec:scanner_stability}, we will show that 50\% of scanners have URLs for which they keep changing the detailed labels. One may choose a highly reputable scanner. However, we later show that scanners considered highly reputable in the literature also often change their detailed labels over time. 


\obssum{} There largely exist conflicting detailed labels for given URLs. Furthermore, we observe that different URL types have different trends in the number of detailed labels. In general, phishing URLs tend to have more conflicting labels than malware URLs. We observe two scenarios leading to conflicting labels: individual scanners' temporal conflict and cross-scanner conflicts.  Given such conflicting labels, assigning one type of attack would  be apparently challenging. As we will show in Section~\ref{sec:attack_type_detection}, majority voting to decide an attack type will result in high false positives and negatives. We analyze individual scanners' behavior in assigning attack types in more detail in Section~\ref{sec:scanner_stability} and propose a method to assign a final attack type given such conflicts in Section~\ref{sec:attack_type_detection}.

\eat{
\begin{figure}[t]
\centering
    \includegraphics[width=0.4\textwidth,height=0.17\textheight]{figure/url_label_fresh_cdf_new.pdf}
\caption{The CDF of the number of detailed labels (attack types) for each URL set}
\label{fig:attacktype_cdf}
\end{figure}
\begin{figure}[t]
\centering
    \includegraphics[width=0.4\textwidth,height=0.17\textheight]{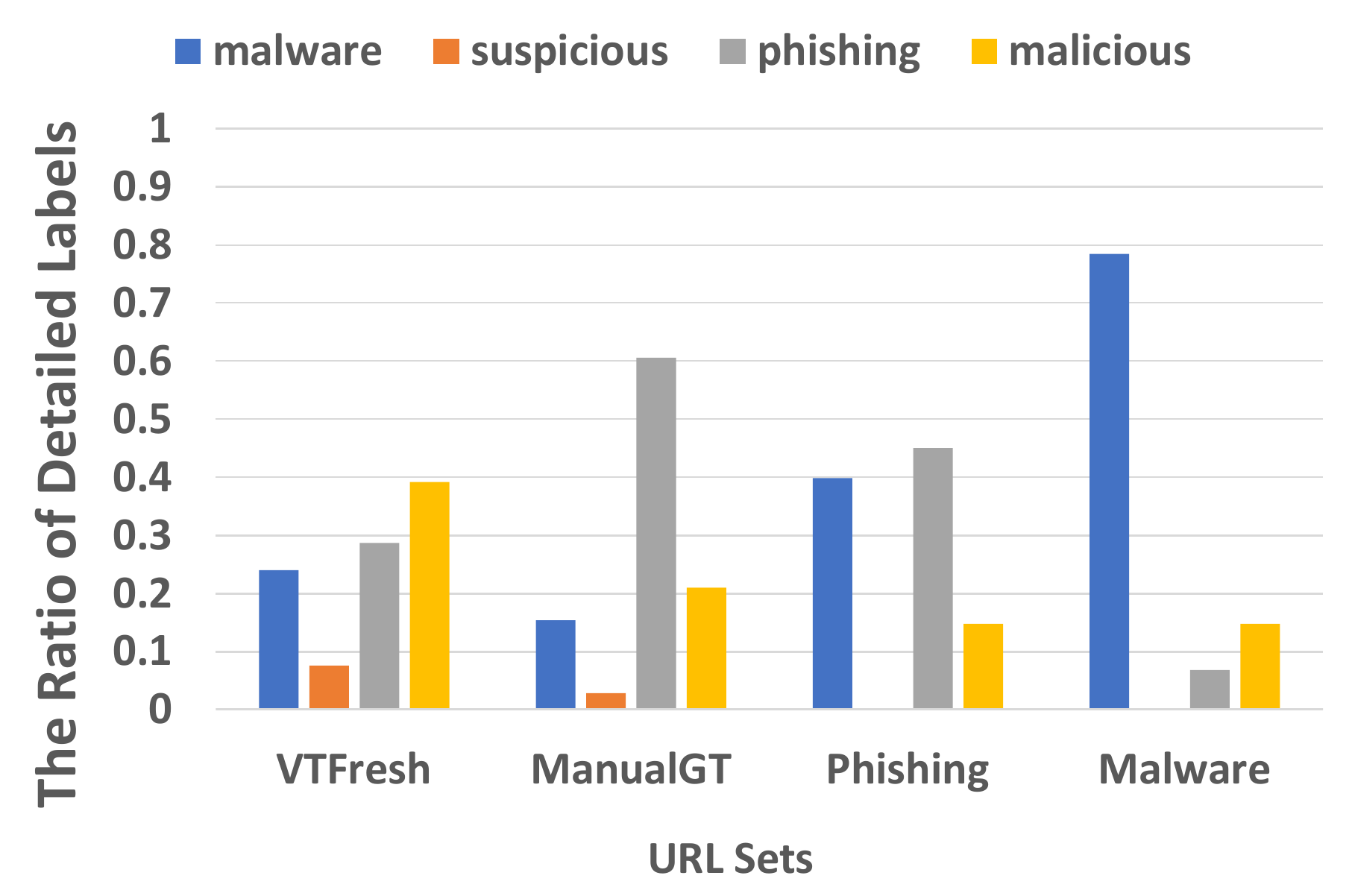}
\caption{The ratio of top 4 detailed labels over the total number of scan reports for each URL set}
\label{fig:attacktype_ratio}
\end{figure}
}


\begin{figure*}[htbp]
\begin{center}
\parbox{1.0\textwidth}{
\centering
\subfigure[Manual GT]{
\psfig{file=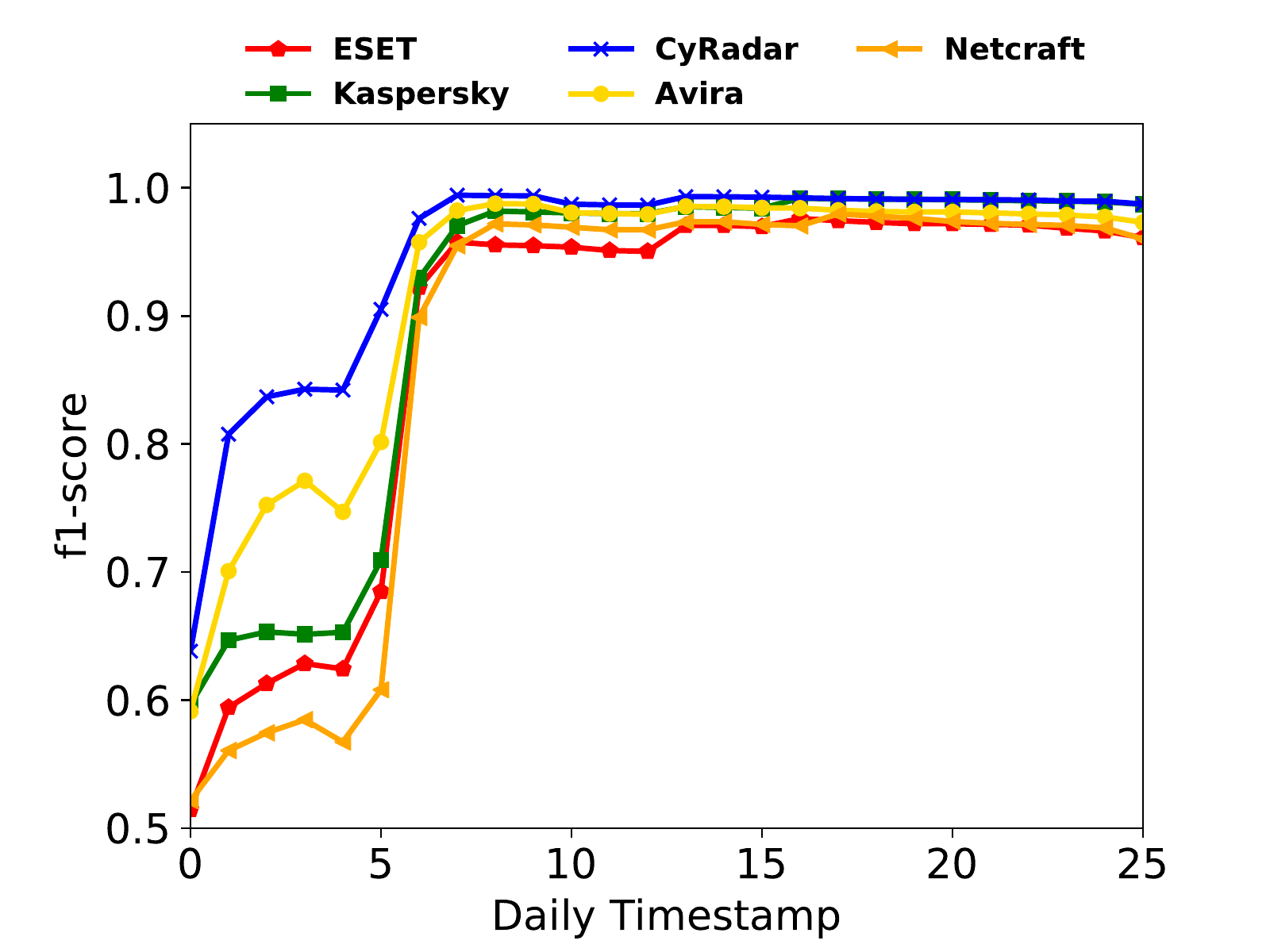,width=0.323\textwidth}\label{manual_eachengine_daily_accuracy}}
\subfigure[Phishing]{
\psfig{file=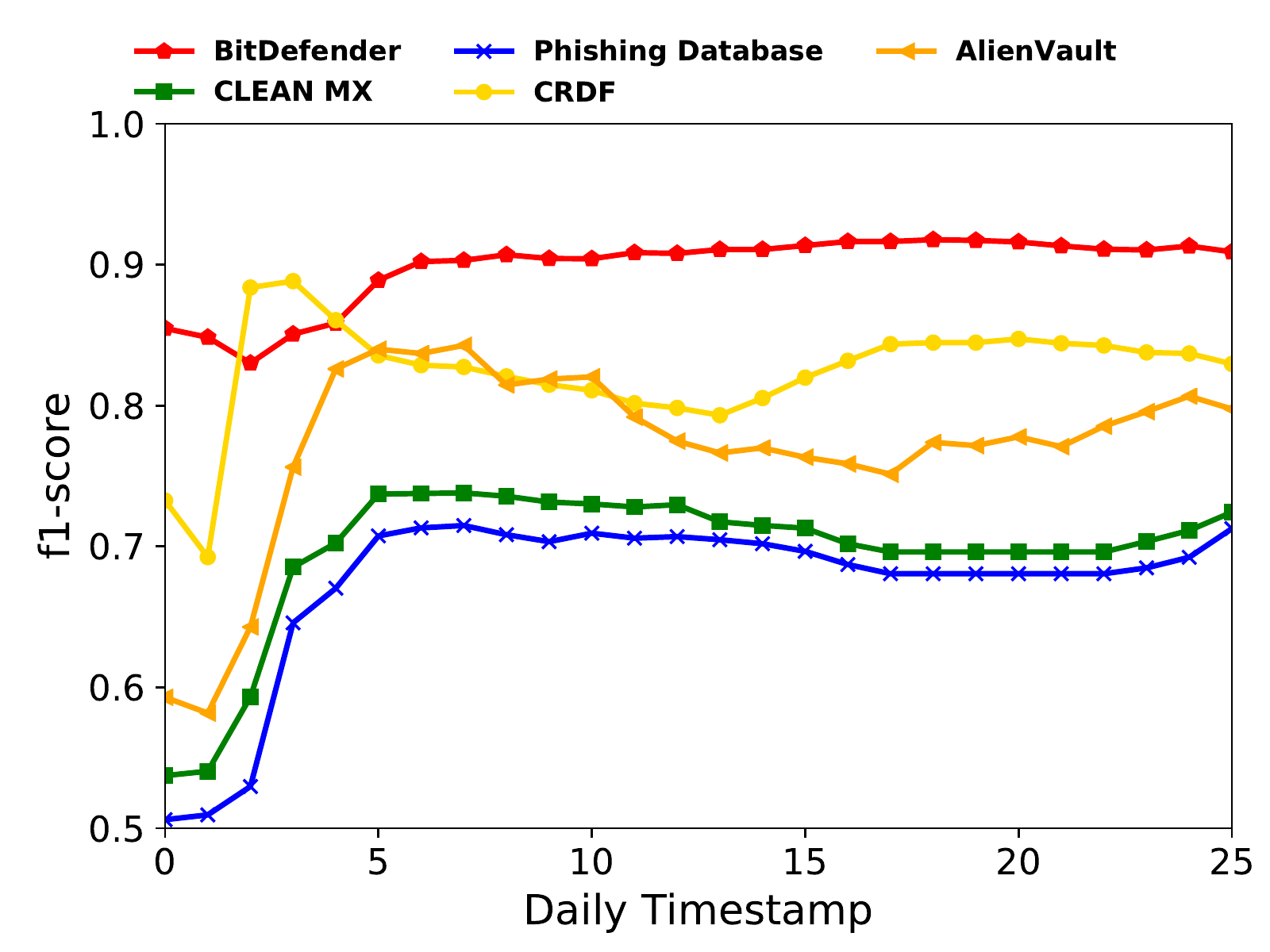,width=0.323\textwidth}\label{fig:phishing_f1}}
\subfigure[Malware]{
\psfig{file=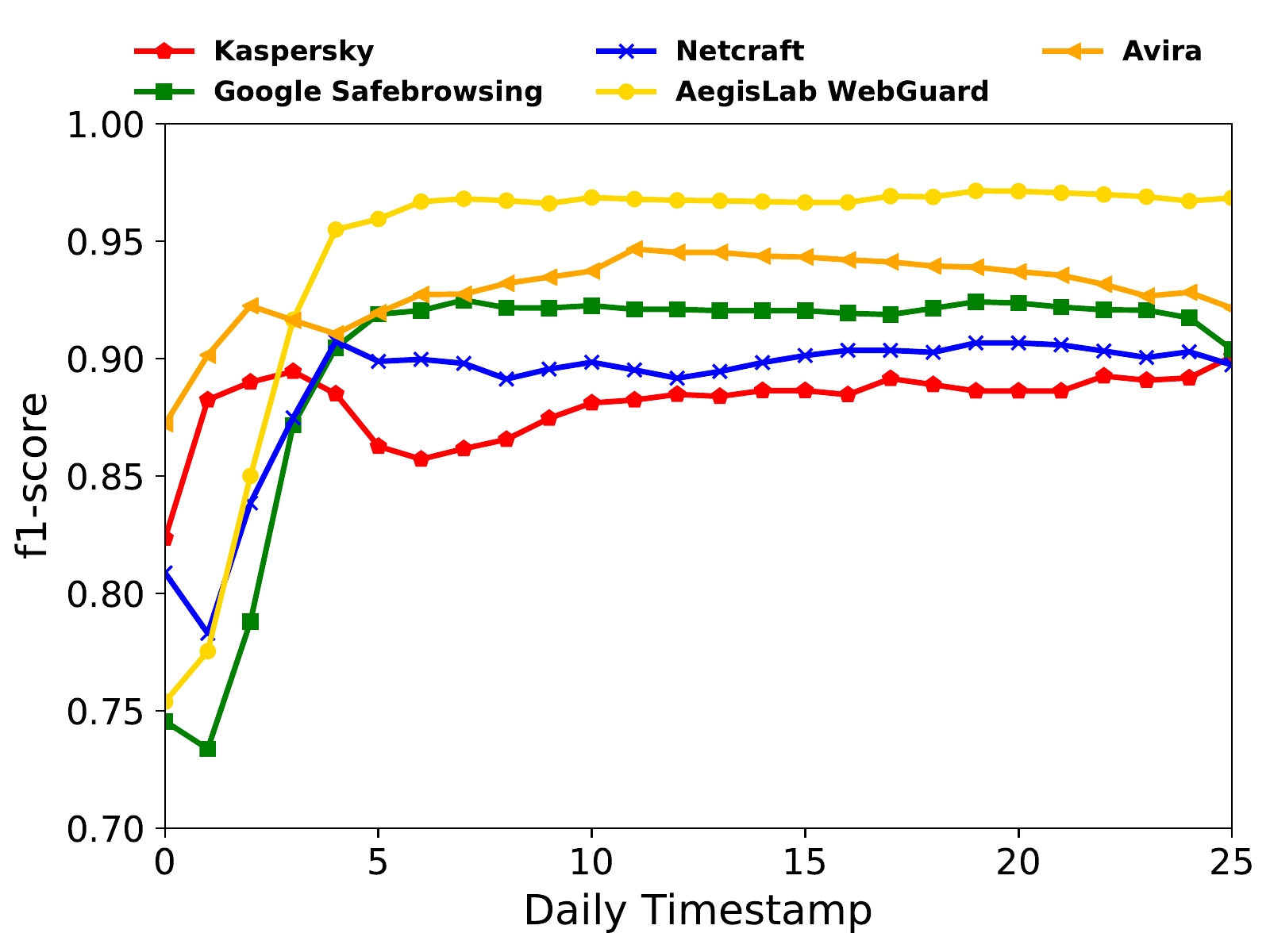,width=0.323\textwidth}\label{fig:malware_f1}}
}

\caption{F-1 score trends of top 5 scanners over daily timestamp for each ground truth data }\label{fig:gt_engine_daily_accuracy}
\end{center}
\end{figure*}
\subsection{Scanners' Specialties - Detection Accuracy}~\label{sec:scanner_accuracy} 

In this section, we study the specialties of scanners by measuring scanners' detection accuracy. To do so, we focus on whether the scanner detected the URL or not at each timepoint and thus we use the binary label sequence $BL_{s,u}=[bl_{t_1},bl_{t_2},..bl_{t_n}]$ defined in Section~\ref{sec:notation}. We measure the F-1 score at each time point. We use F-1 score ($\frac{2 \times \text{precision} \times \text{recall}}{\text{precision + recall}}$) as it accommodates the unbalanced datasets ~\cite{f1auc}. 

We rank the scanners by the maximum F-1 score. We observe that the maximum scores of 58\% of scanners for malware URLs, 37.5\% for phishing URLs, and 25\% for manual GT URLs are less than 0.5. Figure~\ref{fig:gt_engine_daily_accuracy} shows each scanner's F-1 score (the y-axis) trends over daily timestamps (the x-axis). Day 0 means the first appearance in VT. We only show top 5 scanners based on their maximum F-1 score for clarity. 

The figure shows a few interesting observations. First, no scanner performs well for all URL types and thus the top 5 scanners are different for different URL types. For example, BitDefender works well for phishing URLs but poorly on malware URLs. Second, scanners often perform poorly in the early reports in VT. In general, we observe that top 5 scanners reach the maximum F-1 score near the 5th day since the first appearance in VT. Third, certain scanners do not change their label once they detect certain types of URLs, resulting in continuously high F-1 scores (e.g., ESET, CyRadar, Netcraft, Kaspersky, Avira in Figure~\ref{manual_eachengine_daily_accuracy}, BitDefender in Figure~\ref{fig:phishing_f1}, and AegisLab WebGuard in Figure~\ref{fig:malware_f1}). Finally, certain scanners quickly reach their maximum F-1 score, and then the score continuously decreases (e.g., CRDF in Figure~\ref{fig:phishing_f1}). 

Note that a scanner consistently having high F-1 scores is not necessarily a good scanner, as the status of URLs can change (e.g., compromised and cleaned). Indeed, we observe scanners not changing their decision about URLs that are once detected then become NX (non existent). For example, \url{http://bstange.alinaalexandrovacademy.ro/} 
becomes NX URL but scanners such as Fortinet and Webroot still  mark it as ``phishing'' or malicious.

\obssum{} Scanners specialize in different attack types and highly accurate scanners are different for different URL types. Threshold-based approaches without considering such specialties may result in less accurate groundtruth. Further, scanners perform poorly in the early reports. This suggests that researchers need to evaluate scanners' reliability depending on the attack types of URLs and derive the optimal time to collect ground truth. 

\subsection{Scanners' Stability on Binary and Detail Labels}\label{sec:scanner_stability}


Essentially, the F-1 score changes over time because scanners change their labels for URLs. In this section, we thus measure the stability of binary and detailed labels of scanners. As malicious URLs are often short-lived, the label changes over a long period may be natural due to external dynamics on URLs (e.g., a URL is used once for phishing, and later for malware). Based on the lifetime analysis of malicious URLs including phishing and malware in the previous research~\cite{kim2020anatomy}, we consider a month for this analysis. 

Inspired by ~\cite{tops2021maat}, we measure the stability of scanners' labels for malicious URLs by two certainty scores: binary and detailed label certainty scores. A binary label certainty means how \textit{certain} a scanner is about its detection; a detailed label certainty means how \textit{certain} a scanner is about its detailed label (i.e., an attack type).

A \textbf{binary label certainty} of scanner $s$ for URL $u$ measures how much time $s$ labels $u$ as malicious over time. For example, let us assume that $s$ has reports for two URLs $u_1$ and $u_2$ where $s$'s binary sequences are $BL_{s,u_1} = [0,\textbf{1},\textbf{1},0]$ and $BL_{s,u_2} = [0,\textbf{1},\textbf{1},\textbf{1}]$, respectively. Then, the binary label certainties of $s$ for $u_1$ and $u_2$ is $certainty_b(s, u_1) = 0.5$ and $certainty_b(s, u_2) = 0.75$, respectively. Then, the binary label certainty score, \emph{BLCertainty}, of $s$ is computed as the average of binary label certainties for all URLs. That is, BLCertainty(s) = $(0.5 + 0.75)/2 = 0.625$.

A \textbf{detailed label certainty }of scanner $s$ for URL $u$ measures whether $s$ constantly gives the same detailed label over time. Essentially, the most \emph{certain} label of $s$ for $u$ will be the most common label given by $s$ to $u$. We thus extract $s$'s most common label for $u$ and compute a detailed label certainty as the ratio of occurrences of most common labels over time. For example, let us assume $s$'s detailed label sequence for $u_2$ is $DL_{s,u_2} = [$0, phishing, \textbf{malware}, \textbf{malware}$]$. Its most common label is ``malware'' and it appears twice in 4 time periods, and thus the detailed label certainty $certainty_d(s,u_2)$ is $2/4=0.5$. If $s$ always gives the same detailed label, $certainty_d(s,u)$ will be the same as $certainty_b(s,u)$. Similar to BLCertainty, the detailed label certainty score, \emph{DLCertainty}, of $s$ is computed as the average of detailed label certainties for all URLs. 
\begin{figure}[htbp]
\parbox{1.0\textwidth}{
\subfigure[Phishing]{
\psfig{file=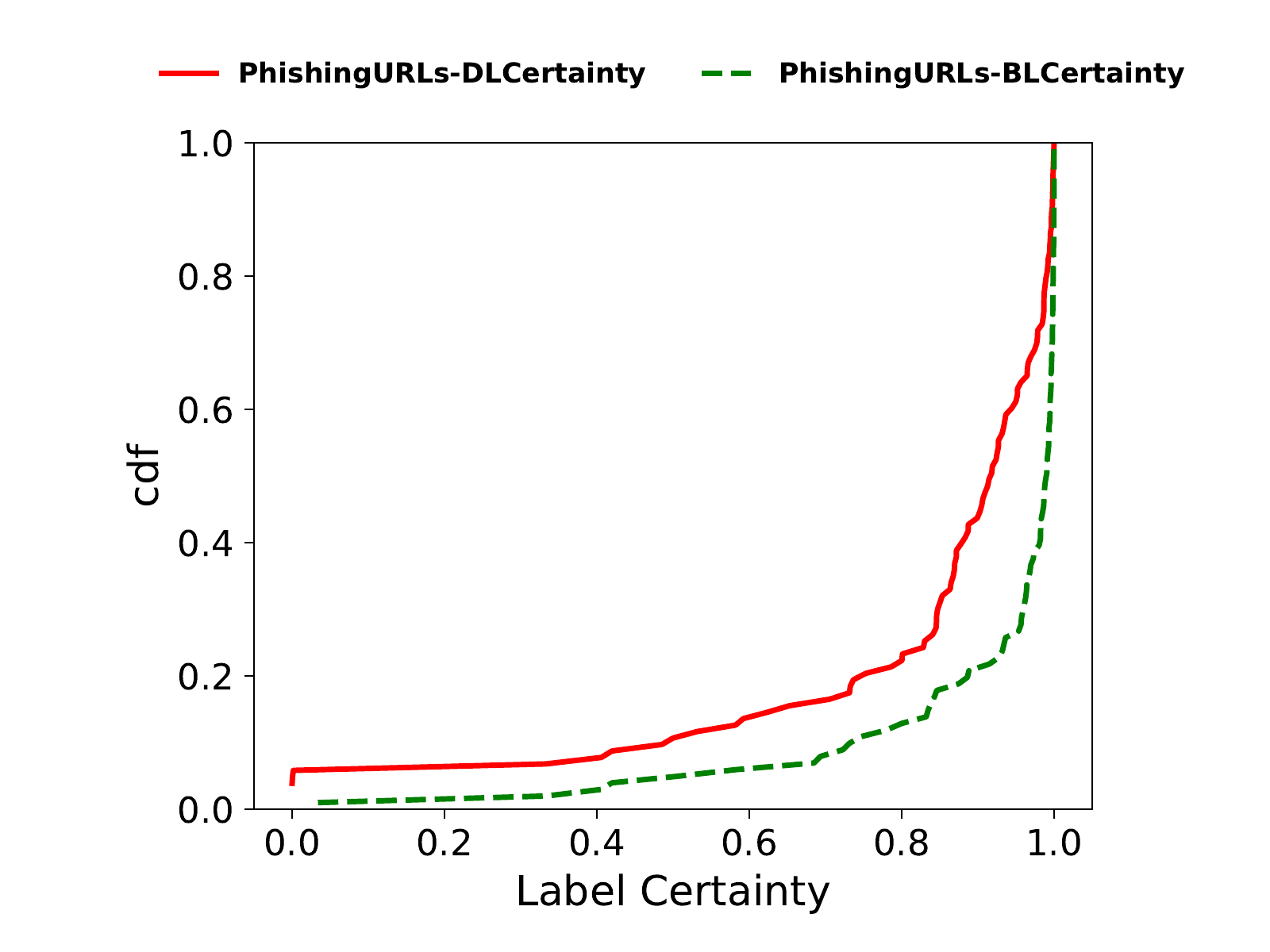,width=0.45\textwidth}\label{phishing_certainty}}
\subfigure[Malware]{
\psfig{file=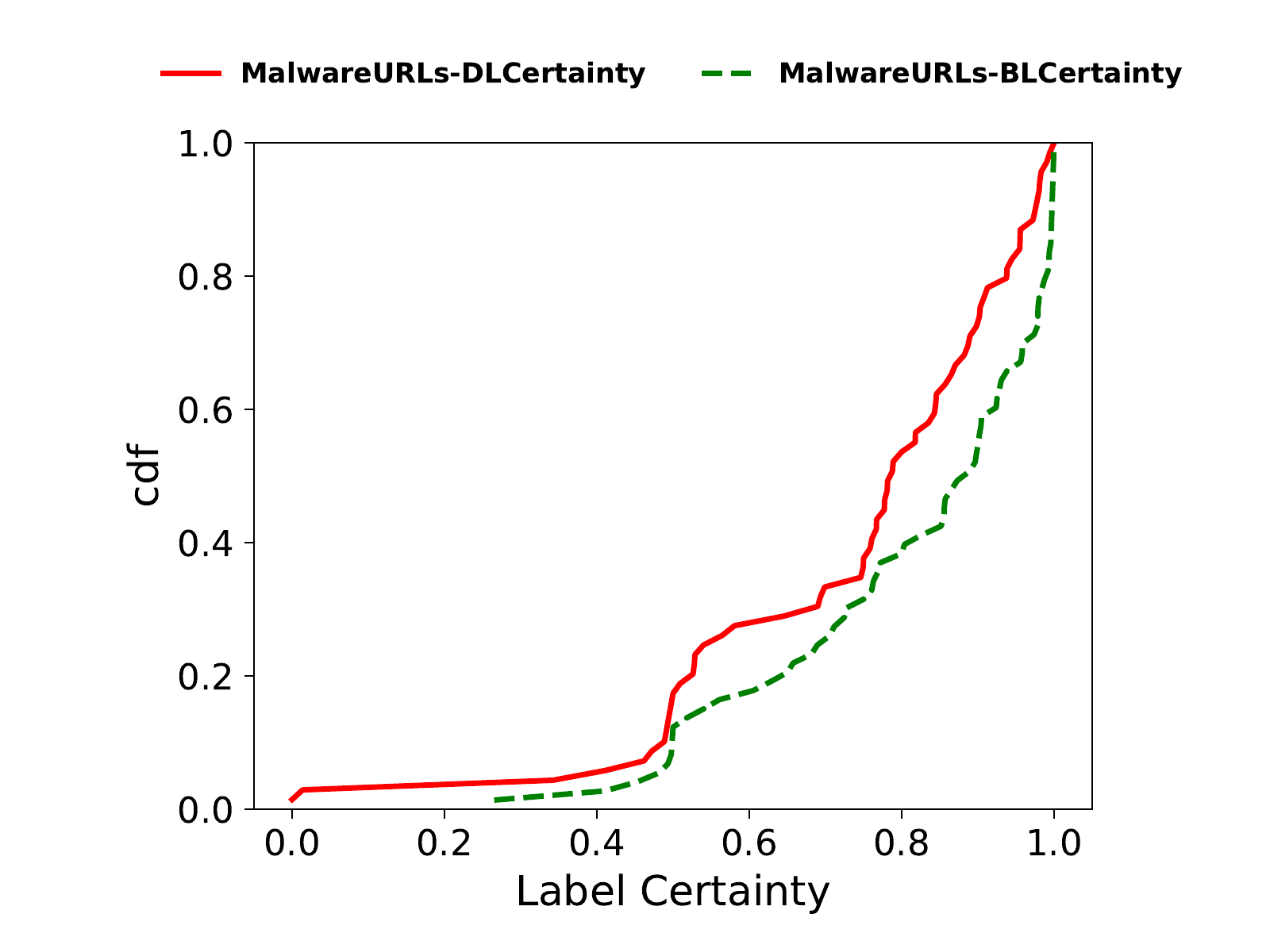,width=0.45\textwidth}\label{malware_certainty}}

\captionsetup{width=.98\linewidth}
\caption{Scanner label certainty scores for phishing and malware URLs (other sets are in the Appendix (Figure~\ref{fig:certainty_appendix}))
}\label{fig:engine_stability}}
\end{figure}
Figure~\ref{fig:engine_stability} represents the CDFs of two certainty scores of all scanners for different types of URLs. The x-axis represents the label certainty score and the y-axis represents the CDF (i.e., the portion of scanners). Essentially, the line in the left side means that there are more scanners with lower label certainty scores. In general, we observe that scanners have lower DLCertainty than BLCertainty. This means that although a scanner may have relatively stable binary labels for URLs, it changes detailed labels (i.e., assign multiple attack types to a URL) over time. 

To study more on scanners' detailed label stability, we further measure the number of detailed labels for each URL per scanner. Figure~\ref{fig:scanner_label_ratio} shows the distribution for the number of detailed labels per scanner over a month. The x-axis represents the set of scanners. Each bar represents the number of detailed labels. The y-axis represents the ratio of URLs that scanners assign the corresponding number of detailed labels. We only show scanners having URLs with more than 1 label in the figure.
\eat{
\begin{figure*}[htbp]
\begin{center}
\parbox{1.0\textwidth}{
\centering
\subfigure[VT Fresh]{
\psfig{file=figure/vtfreshlabel_dist_sorted.pdf,width=0.45\textwidth,height=0.2\textheight}\label{allengine_slabel}}
\subfigure[Manual GT Malicious]{
\psfig{file=figure/manuallabel_dist_sorted.pdf,width=0.45\textwidth,height=0.2\textheight}\label{manualgt_slabel}}
\subfigure[Phishing (PT Verified + OpenPhish)]{
\psfig{file=figure/phishinglabel_dist_sorted.pdf,width=0.45\textwidth,height=0.2\textheight}\label{phishing_slabel}}
\subfigure[Malware (AlienVault + URLhaus)]{
\psfig{file=figure/malwarelabel_dist_sorted.pdf,width=0.45\textwidth,height=0.2\textheight}\label{malware_slabel}}
\vspace{-.2in}
}
\caption{The distribution for the number of detailed labels per scanner (Scanners having URLs with more than 1 label are shown) }\label{fig:scanner_label_ratio}
\end{center}
\end{figure*}
}

\begin{figure*}[htbp]
\begin{center}
\parbox{1.0\textwidth}{
\centering
\subfigure[VT Fresh]{
\psfig{file=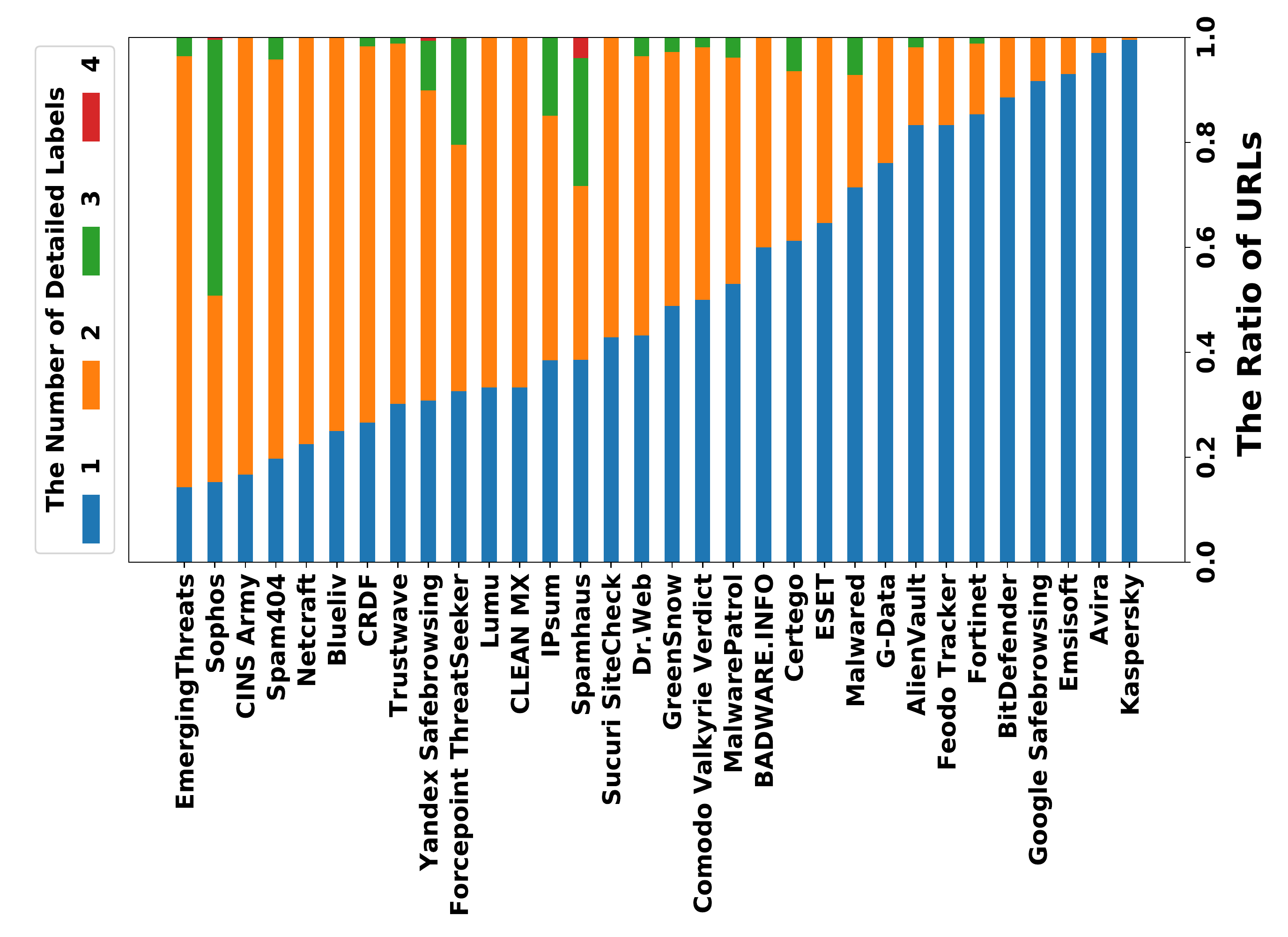,width=0.47\textwidth,height=0.2\textheight}\label{allengine_slabel}}
\subfigure[Manual GT Malicious]{
\psfig{file=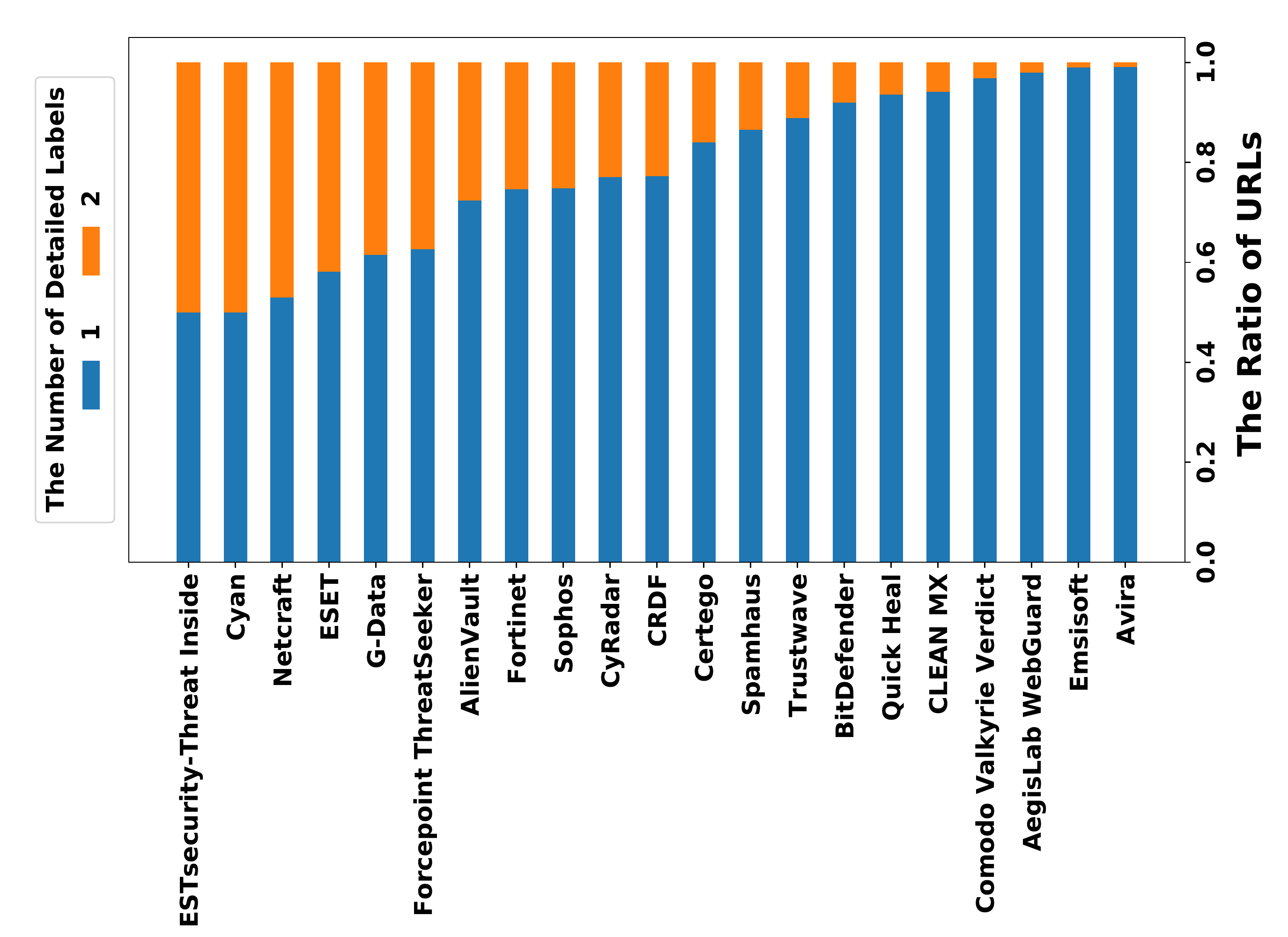,width=0.47\textwidth,height=0.2\textheight}\label{manualgt_slabel}}
\subfigure[Phishing]{
\psfig{file=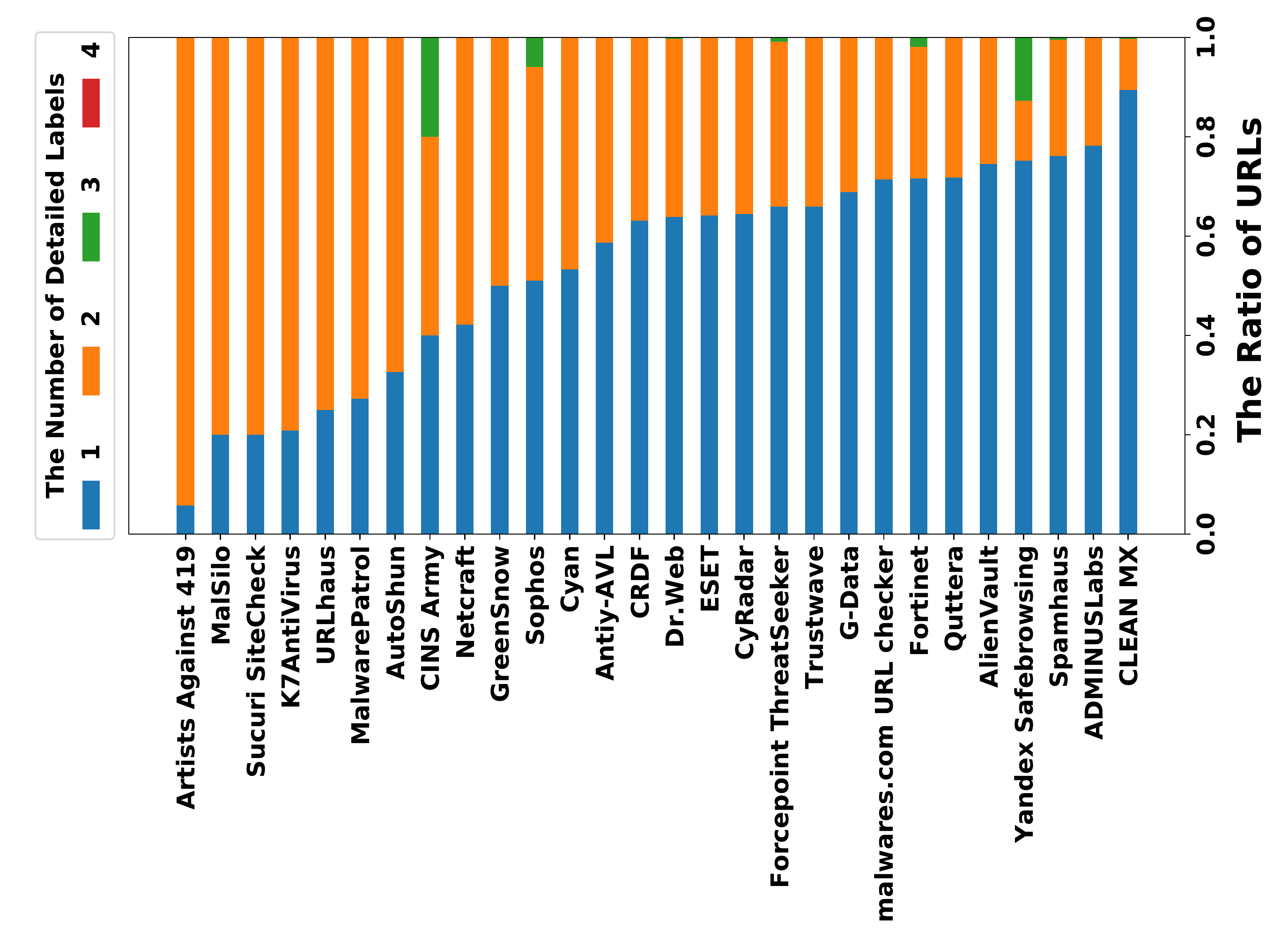,width=0.47\textwidth,height=0.2\textheight}\label{phishing_slabel}}
\subfigure[Malware]{
\psfig{file=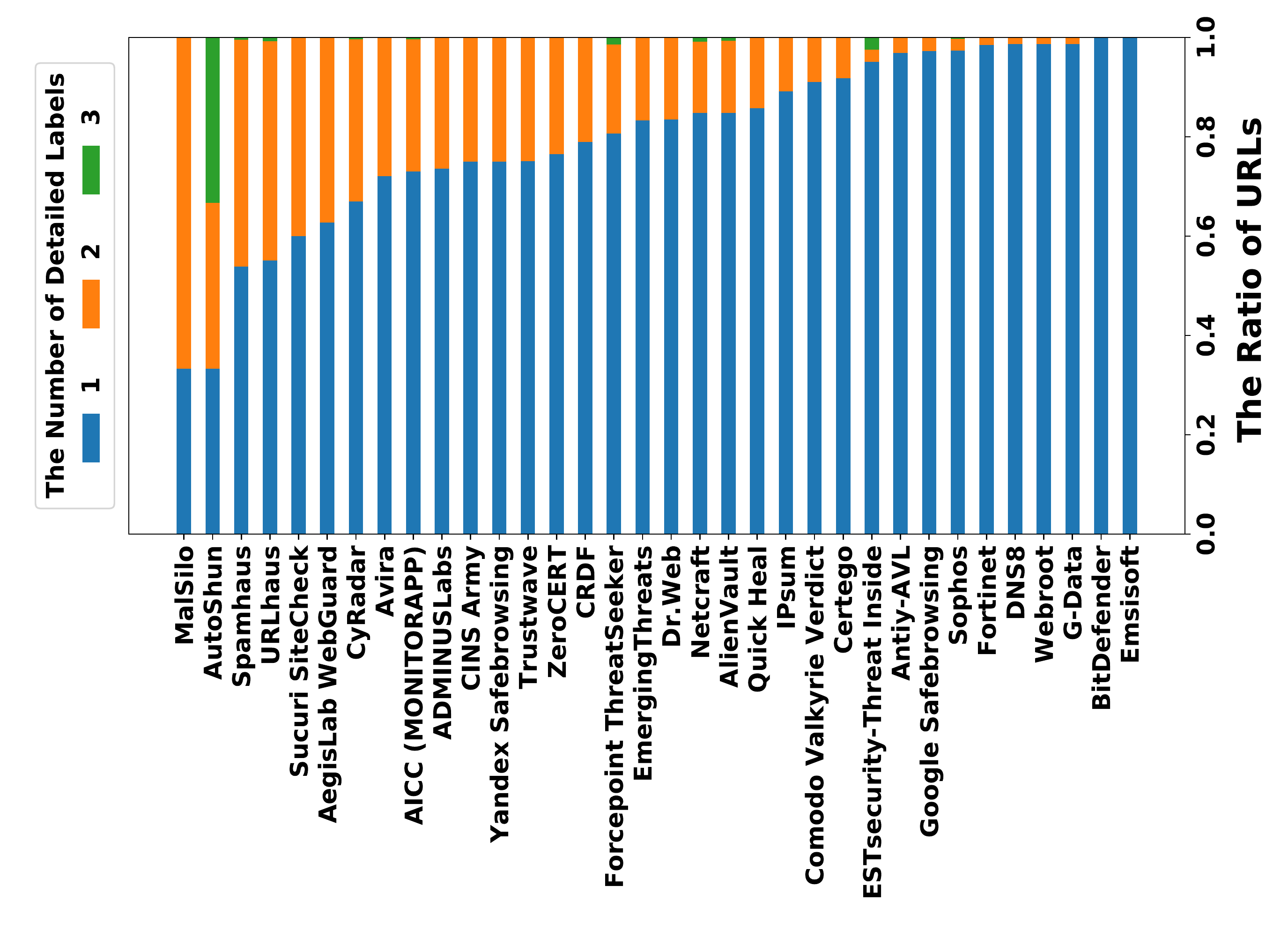,width=0.47\textwidth,height=0.2\textheight}\label{malware_slabel}}
}
\vspace{-.1in}
\caption{The distribution for the number of detailed labels per scanner (only scanners having URLs with more than 1 label are shown) }\label{fig:scanner_label_ratio}
\end{center}
\end{figure*}

The figure shows that 49 scanners assign multiple attack types to at least one of the URL sets. We observe that some scanners even assign 4 attack types to the same URL. For example, Spamhaus assigns 4 attack types to 3.9\% of VT Fresh URLs (Figure~\ref{allengine_slabel}) and a few phishing URLs (not visible in the figure) (Figure~\ref{phishing_slabel}). Figure~\ref{fig:scanner_label_ratio} shows that scanners considered highly reputable in the literature such as Sophos, Bitdefender, and Kaspersky ~\cite{zhu2020:labeldynamics,thomas2015ad,arp2014drebin} also assign multiple attack types to given URLs. For example, Sophos assigns at least 2 attack types to 85\% of VT Fresh URLs (Figure~\ref{allengine_slabel}), 50\% of phishing URLs (Figure~\ref{phishing_slabel}), and 2\% of malware URLs (Figure~\ref{malware_slabel}). This suggests that although using only highly reputable scanners may increase detection accuracy, assigning an attack type to a URL would still be challenging. 


Interestingly, we observe different behavior of scanners that assign multiple types of attacks to the same URL. Specifically, 53 \% of scanners constantly change their detailed label from one to another in the beginning, and then they stabilize with one type of attack. For example, Sophos switches its label every day for \url{http://jp-billverify.com} between ``malware'' and ``phishing''; then later it stabilizes as ``phishing''. Meanwhile, 47 \% of scanners never stabilize their labels. For example, Fortinet keeps changing its label between ``phishing'' and ``malware'' for \url{http://wikiarch.cz/wiki/nabidka-projektovych-praci?rev=326}. 

\obssum{} Scanners often change their binary and detailed labels for the same set of URLs. Moreover, scanners are less ``certain'' about the attack types (DLCertainty) than the maliciousness itself (BLCertainty) leading to challenges in deciding an attack type for given URLs. Given these different behaviors of scanners, we propose a method to assign a final attack type to each URL at a given time point in Section~\ref{sec:attack_type_detection}.

\subsection{Scanners' Correlation on Binary and Detailed Labels}~\label{sec:correlation}

One may take scanners consistently having high F-1 scores as reputable for each attack type and choose thresholds considering only such reputable scanners~\cite{zhu2020:labeldynamics}. However, this section shows there exist highly correlated scanners in terms of both binary and detailed labels that may degrade threshold-based approaches for detection and produce a bias for a majority voting based approach for attack type detection. We analyze the pairwise correlation among scanners using two similarity measures: Jaccard similarity ~\cite{jaccard1912distribution} and dynamic time warping (DTW)~\cite{berndt1994dtw}. 

\heading{Scanners' Co-labeled URL Similarity.} To measure the similarity in terms of co-labeled URLs, we employ Jaccard similarity for binary and detailed labels at each time point as well as over time. Specifically, we measure Jaccard similarity for binary labels by the number of co-detected URLs over the total number of URLs; Jaccard similarity for detailed labels by the number of URLs having the same detailed labels over the total number of URLs. For example, when the set of total URLs is ${u_1,u_2,u_3,u_4,u_5}$, scanner $s_1$ detected ${u_1,u_2,u_3}$, and scanner $s_2$ detected $u_2,u_3,u_4,u_5$, then the Jaccard similarity for a binary label is 2/5. Although $s_1$ and $s_2$ co-detected $u_2$ and $u_3$, $s_1$ and $s_2$ may have different detailed labels. And if $dl(s_1,u_2)$=``malware'', $dl(s_1,u_3)$ = ``malware'', $dl(s_2,u_2)$=``malware'', and $dl(s_2,u_3)$ = ``phishing'', the Jaccard similarity for a detailed label is 1/5 due to their different detailed labels for $u_3$.

\begin{figure}[htbp]
\parbox{1.0\textwidth}{
\centering
\subfigure[Scanner binary label similarity Frobenius norm]{
\psfig{file=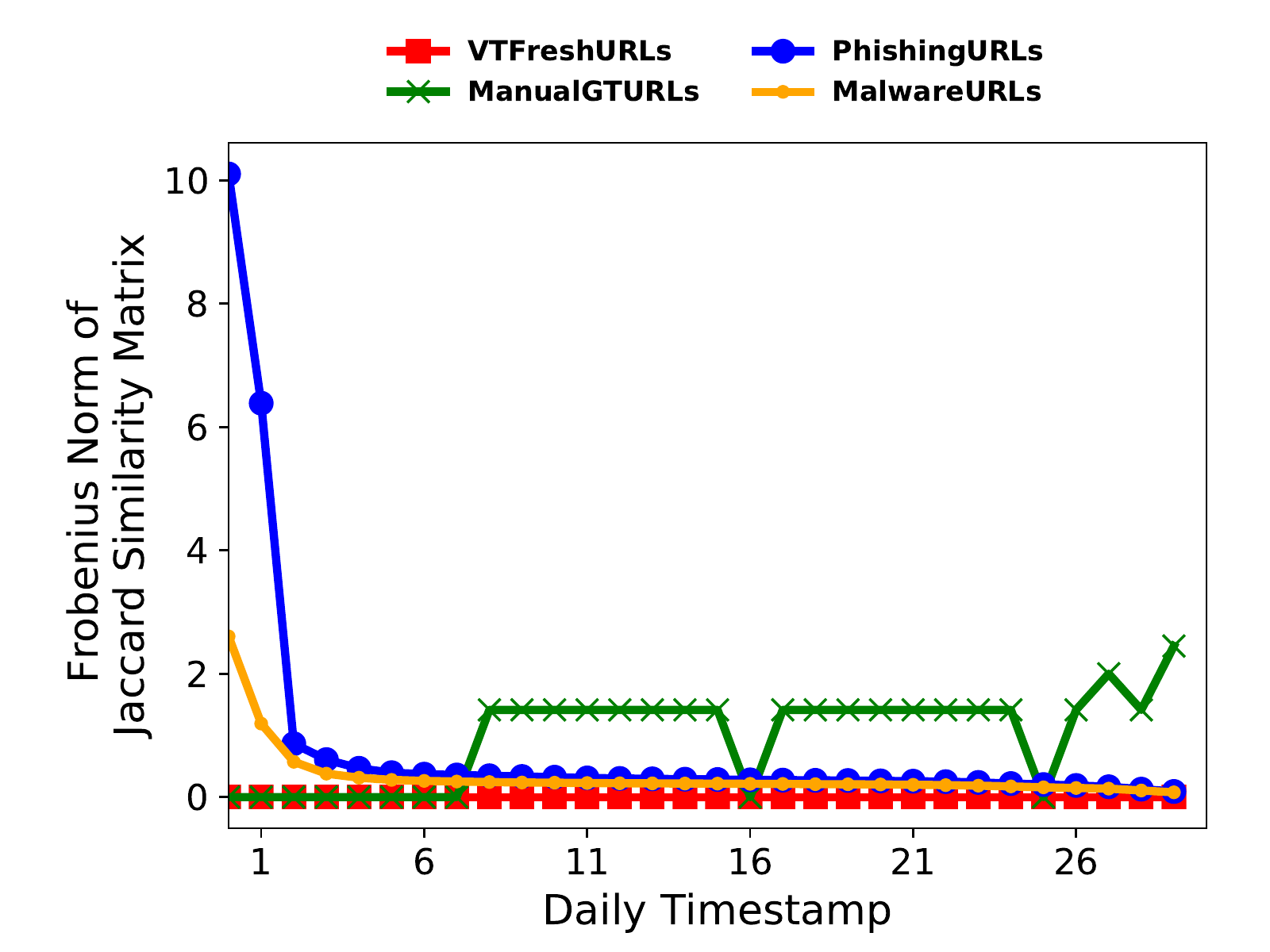,width=0.45\textwidth}\label{binary_frob}}
\subfigure[Scanner detail label similarity Frobenius norm]{
\psfig{file=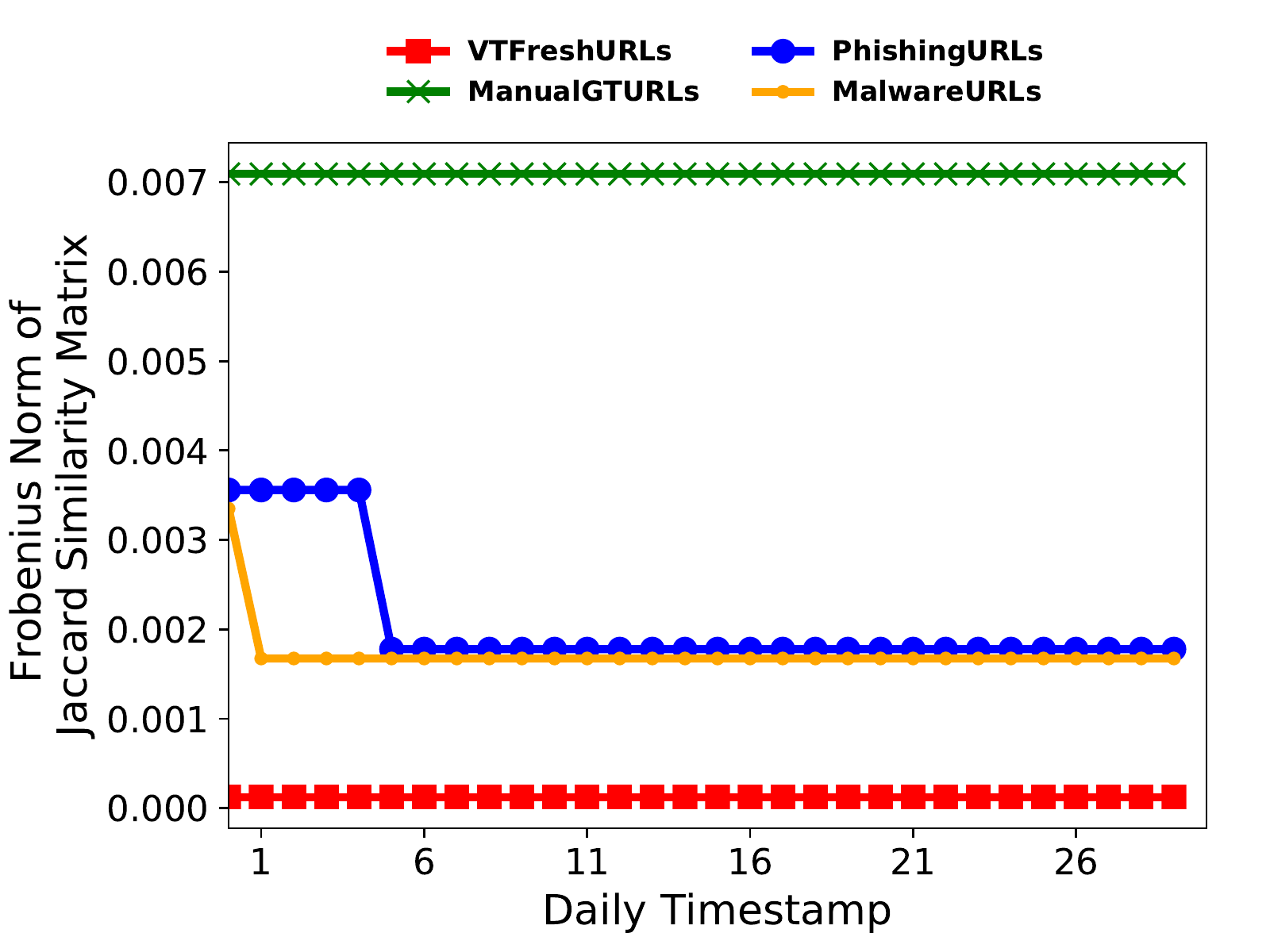,width=0.45\textwidth}\label{detail_frob}}
\captionsetup{width=.95\linewidth}
\caption{Frobenius norm of Jaccard similarity of scanner's binary/detail labels over time}\label{fig:label_jaccard_similarity_trend}}
\end{figure}

We present the heatmaps of pairwise Jaccard’s similarity of binary and detailed labels over all periods in Appendix V. Scanners co-detecting URLs with at least one scanner are shown in the heatmaps. A darker cell in heatmap means high similarity, while a lighter cell means low similarity. We also compute the Frobenius norm ~\cite{horn2012matrix} of the pairwise Jaccard similarity matrix at each time point to measure if the similarity is consistent over time. Figure~\ref{fig:label_jaccard_similarity_trend} shows how the Frobenius norm  changes over time where the x-axis presents daily timestamps and the y-axis presents the norm at the timepoint. The larger norm indicates that there are more highly similar scanners in terms of detection (binary labels) or attack type assignment (detailed labels). 

In general, we observe that more scanners have high Jaccard similarity for phishing URLs (more darker cells in heatmaps and the larger norm in Figure~\ref{fig:label_jaccard_similarity_trend}) than for malware URLs. Also, the Jaccard similarity of detailed labels is lower in general (lighter in heatmaps and the lower norm in Figure~\ref{fig:label_jaccard_similarity_trend}). Meanwhile, we observe a few scanners having high Jaccard similarity for detailed labels (the darkest) such as ESTsecurity and Scantitan for phishing URLs. 

Figure~\ref{binary_frob} shows that for phishing URLs, there are more scanners having high similarity for binary labels in the beginning, then continuously the norm decreases over time. One possible reason is that shortly after detecting the phishing URLs, some scanners gradually change their label to benign, resulting in less similarity. Furthermore, while there are fewer scanners having high similarity for malware URLs, the norm is relatively consistent over time. 

Scanners may have high similarities due to multiple reasons. If a scanner copies others directly (e.g., a scanner uses a blacklist provided by another scanner), the simple threshold-based approaches will be biased and unreliable. Meanwhile, scanners having high similarity, albeit their independent methods, may indicate high confidence of detection, so that the higher positive counts provide stronger signals. 

We also observe fewer scanners having high similarity for detailed labels (and thus low norm such as 0.0036 compared to norm of 10 for binary labels) and the consistent norm. Scanners having high binary label similarity yet low detailed label similarity suggest that such scanners may have independent approaches (inspecting different signals from URLs) and thus one may treat the positive counts from such scanners as the level of maliciousness. Meanwhile, scanners having both high binary and detailed label similarities suggest high correlations, and thus one may penalize the count accordingly.

\heading{Scanners' Labeling Trend Similarity.} If one scanner copies another, or two scanners share similar (if not the same) features, their label trends should be similar. If one scanner copies another, the copied version's detection would be delayed with the same label trend. We thus further compare the patterns of scanners' labels. To measure the similarity of scanners' binary labels' patterns, we employ dynamic time warping (DTW) distance~\cite{berndt1994dtw} that computes the similarity between two temporal sequences. Essentially, DTW distance can measure if the evolution of labels are similar regardless of their speed. To get the final DTW distance between two scanners, we measure the DTW distance of all pair sequences for co-detected URLs and then compute the average.

We run a hierarchical clustering algorithm based on DTW distance and cut the dendrograms (Figure~\ref{fig:engine_clustering_dendro} in Appendix) by the level. Figure~\ref{fig:engine_clustering} shows the resulting clusters for phishing and malware URLs. Note that we do not consider similar when both scanners do not detect the URL at all over time. 
\begin{figure}[htbp]
\begin{center}
\parbox{1.0\textwidth}{
\centering
\subfigure[Phishing]{
\psfig{file=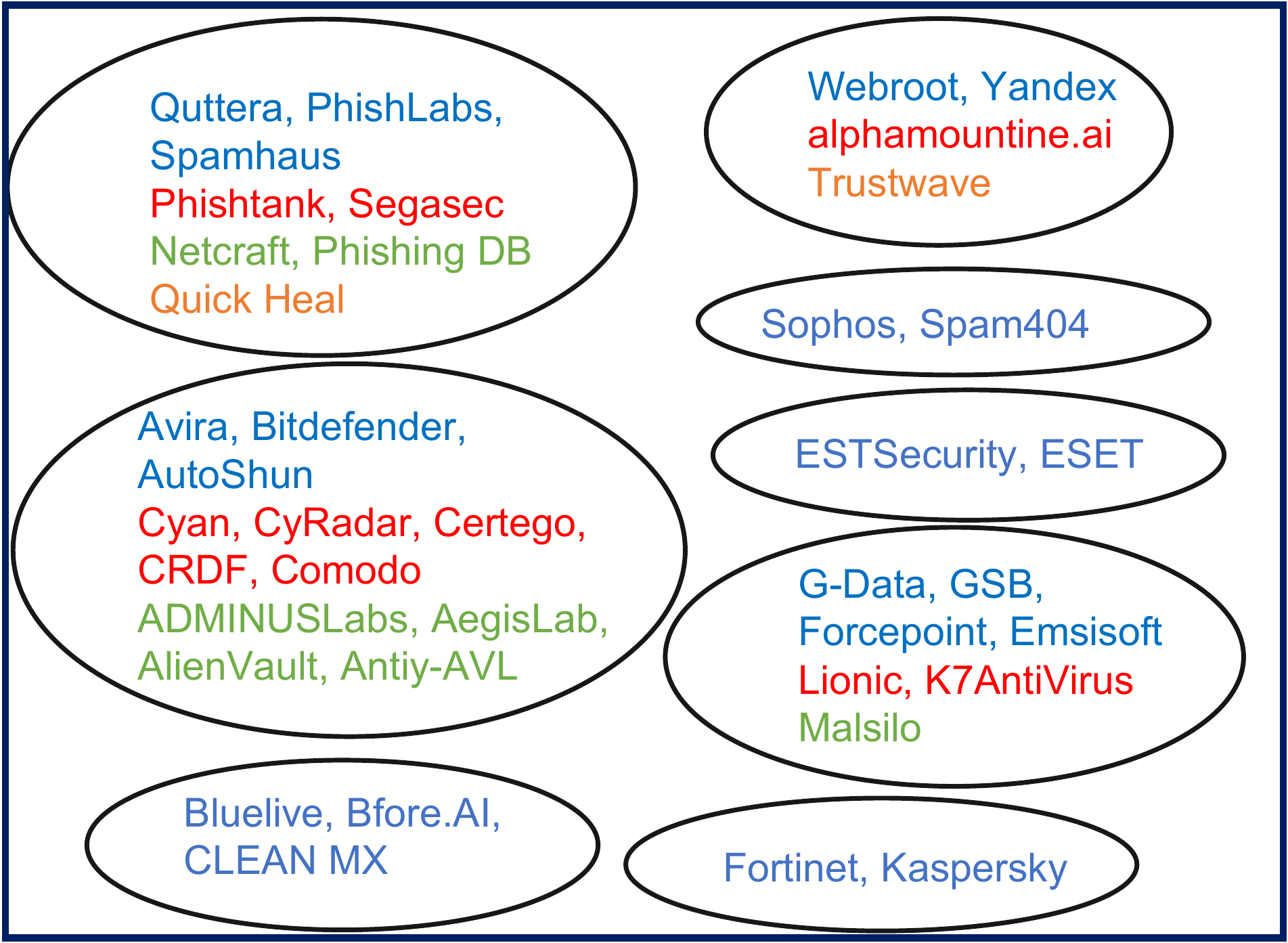,width=0.45\textwidth}\label{fig:phishing_clustering}}
\subfigure[Malware]{
\psfig{file=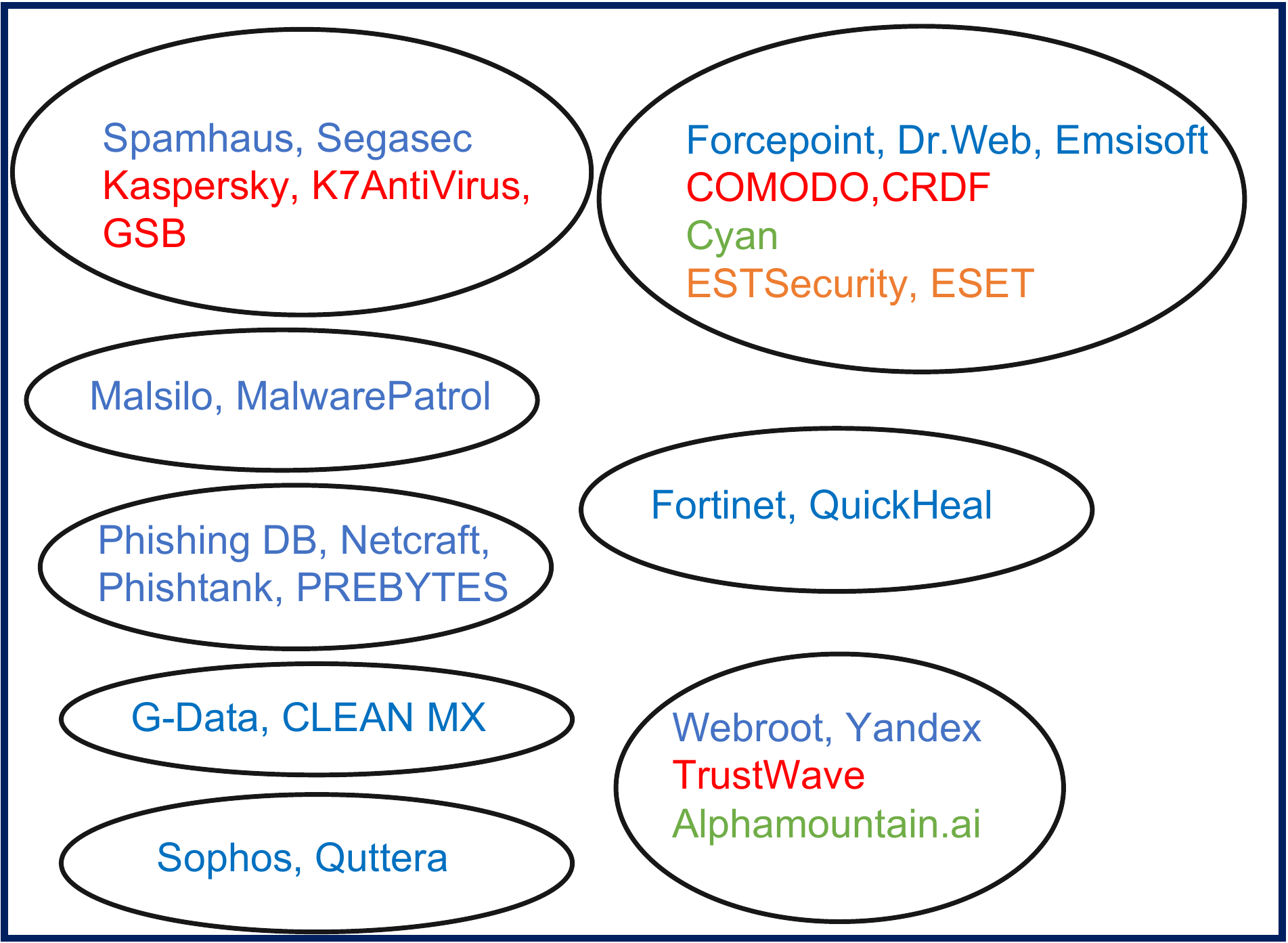,width=0.45\textwidth}\label{fig:malware_clustering}}
}

\captionsetup{width=.95\linewidth}
\caption{Scanner clustering using DTW distance}\label{fig:engine_clustering}
\end{center}
\end{figure}

The figure suggests multiple observations. First, different clusters are built for different types of URLs. For example, for phishing URLs (Figure~\ref{fig:phishing_clustering}), G-Data is closely clustered with Google Safe Browsing (GSB), but it is closely clustered with CLEAN MX for malware URLs (Figure~\ref{fig:malware_clustering}).

The phishing scanners (e.g., PhishTank, PhishLabs, and Phishing Database) clustered for phishing URLs (Figure~\ref{fig:phishing_clustering}) and the malware scanners (e.g., MalwarePatrol and Malsilo)  clustered for malware URLs Figure~\ref{fig:malware_clustering}) confirm that our clustering method indeed captures meaningful clusters. Meanwhile, such clustered scanners (i.e., having highly similar trends of binary labels) suggest that some scanners may not be independent (e.g., one may copy another's labels and do delayed detection compared to another with a similar labeling trend) for a URL. 



\obssum{} We observe highly correlated scanners in terms of their temporal similarity (Figure~\ref{fig:label_jaccard_similarity_trend}) and overall similarity on the trend of their label patterns (Figure~\ref{fig:engine_clustering}). One may prefer scanners always detecting URLs earlier than others among those highly correlated ones. In the next section, we thus analyze if lead/lag relationships between scanners exist.


\subsection{Lead Lag Analysis}\label{sec:lead_lag}

As malicious URLs are often short-lived, it is crucial to detect URLs as early as possible. In Section~\ref{sec:correlation}, we observe highly correlated scanners in terms of the co-detected URLs (Jaccard similarity) and the patterns of binary label trends (DTW distance). In this section, we analyze if there is any lead/lag relationship among those correlated scanners. For example, if scanners $s_1$ and $s_2$ detect the same set of URLs yet the $s_1$ always detects URLs earlier than $s_2$, we may fairly say $s_1$ is a \emph{leader} and $s_2$ is a lagger. We thus compare the first detection time of two scanners for co-detected URLs.

\begin{figure*}[htbp]
\begin{center}
\parbox{0.9\textwidth}{
\centering
\subfigure[Phishing]{
\psfig{file=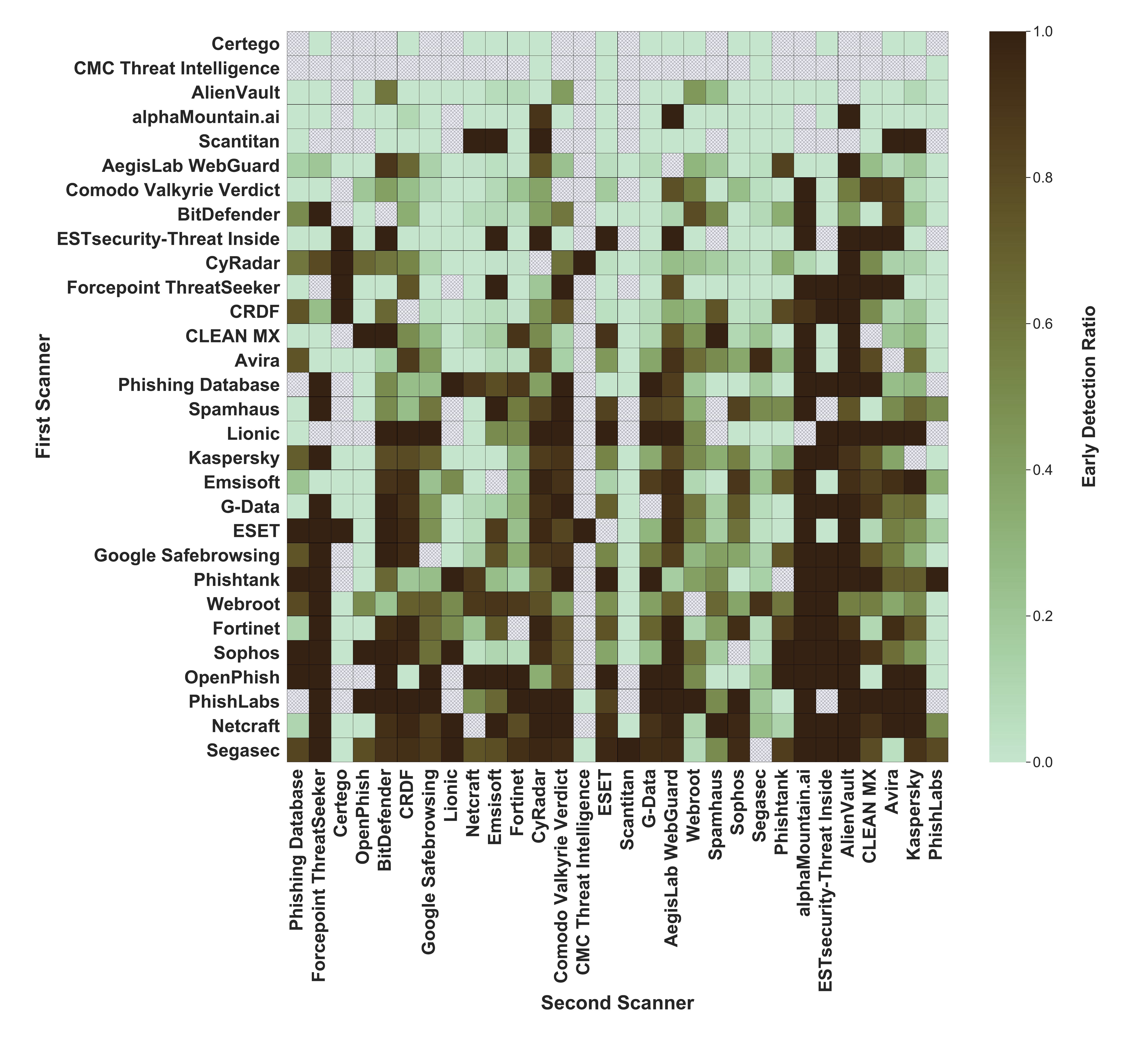,width=0.44\textwidth}\label{fig:phishing_early_heatmap}}
\subfigure[Malware]{
\psfig{file=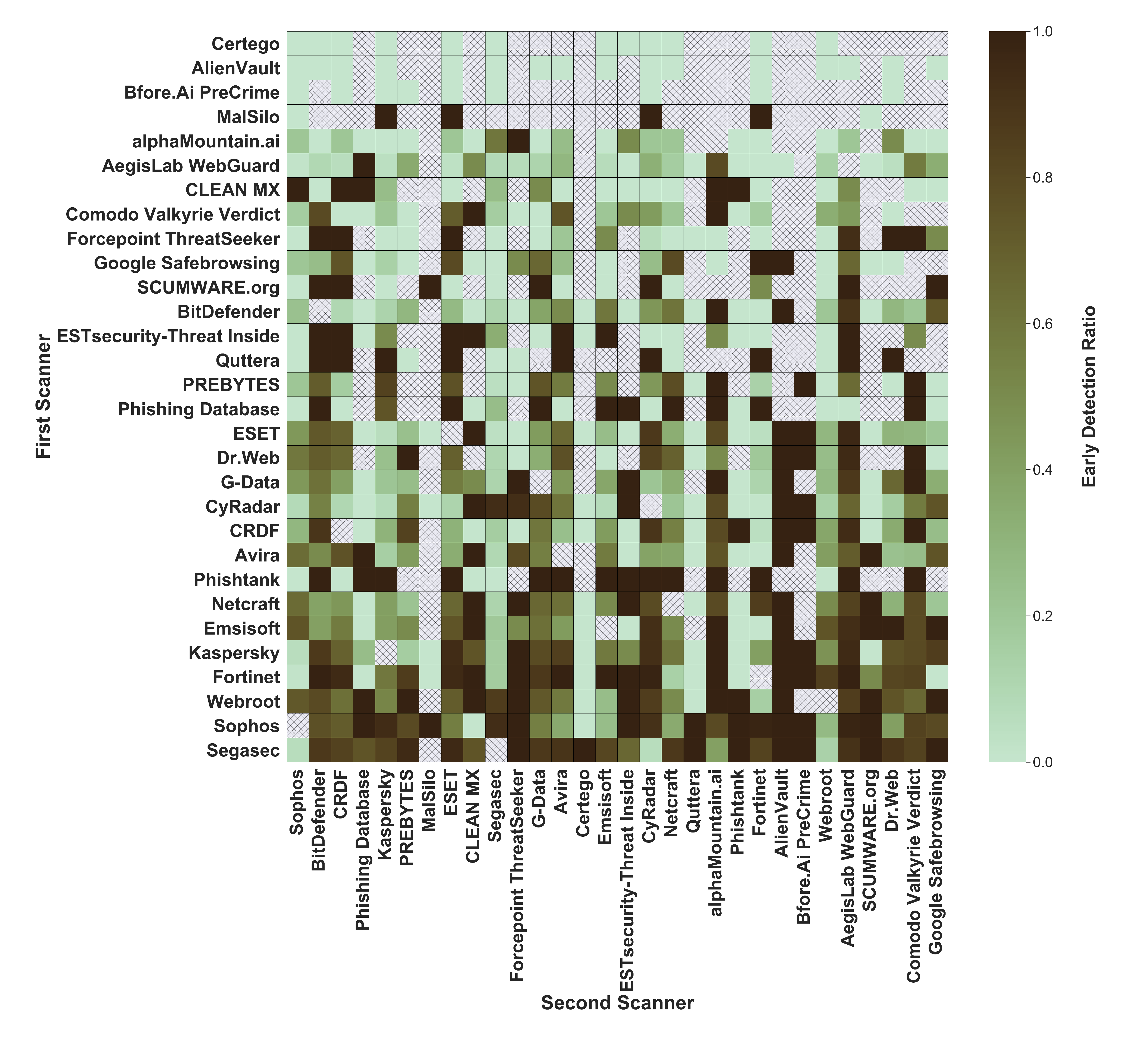,width=0.44\textwidth}\label{fig:malware_early_heatmap}}
}

\caption{Early detection ratio of y-axis scanner being earlier than x-axis scanner (sorted by the darkness of rows)}\label{fig:early_detectratio_heatmap}
\end{center}
\end{figure*}
Figure~\ref{fig:early_detectratio_heatmap} presents the pairwise early detection ratio matrices for phishing and malware URLs measured by the number of URLs that the first scanner (the y-axis) detected earlier than the second scanner over the total number of co-detected URLs. Matrices for VT Fresh and manual GT URLs are given in Figure~\ref{fig:early_detectratio_appendix} in the Appendix. The matrix is sorted so that the darkest row is at the bottom. If there are no co-detected URLs, it is marked as \emph{xxx}. A scanner that does not co-detect URLs with any scanner will not appear in the matrix. Essentially, a completely dark row means that the corresponding row scanner always detects earlier than other scanners; a completely dark column indicates that the corresponding column scanner always detects later than other scanners. 

From Figure~\ref{fig:early_detectratio_heatmap}, first, we observe scanners detect relatively earlier than others (e.g., Segasec) and scanners detect relatively later than others (e.g., alphaMountain.ai). Second, we observe closely clustered scanners (i.e., the label trend is highly similar) where one always detects URLs earlier than another for a specific type of URLs. For example, while Webroot and alphaMountain.ai have similar labeling patterns (and thus closely clustered) for phishing URLs (Figure~\ref{fig:phishing_clustering}), Webroot always detects URLs earlier than alphaMountain.ai (Figure~\ref{fig:phishing_early_heatmap}). Then, one may prefer Webroot over alphaMountain.ai for phishing URLs. 

While MalSilo do not co-detect many URLs with other scanners (i.e., most cells are \emph{xxx}), it mostly detects earlier than other scanners among those co-detected URLs. This suggests that MalSilo may employ an independent method that can compensate for other scanners' detection. 

Meanwhile, we observe there are more scanners detecting the same set of phishing URLs than those detecting the same set of malware URLs (i.e., Figure~\ref{fig:phishing_early_heatmap} has fewer cells with \emph{xxx} than Figure~\ref{fig:malware_early_heatmap}). Further, more lead/lag relationships exist in phishing URLs than malware URLs (i.e., Figure~\ref{fig:phishing_early_heatmap} has more darker cells than Figure~\ref{fig:malware_early_heatmap}). This means that the approaches detecting malware URLs are more likely to be independent of other scanners than approaches detecting phishing URLs.


\obssum{} We observe that there exist lead/lag relationships among scanners that have similar label patterns. On the other hand, more scanners are correlated to detect phishing URLs than malware URLs. Along with the results in Section~\ref{sec:correlation}, one may consider leading scanners' results while penalizing the positive counts. Furthermore, the positive counts for malware URLs and phishing URLs should be treated differently, given the higher correlation in phishing URLs.

\section{Attack Type Detection}\label{sec:attack_type_detection}

Given a malicious URL, identifying if it is involved in a phishing or a malware attack is quite important in practice as, for example, these two attacks require different mitigation actions and malicious URLs are aggregated to threat specific feeds in practice~\cite{feal2021blocklist}. As examined in Section~\ref{sec:measurement}, VT detailed labels are noisy since the scanners often \eat{that detect a URL} do not agree on a single attack type label. Hence, the simple majority voting based approach, our baseline, is sub-optimal (see Table~\ref{tab:attackresults}). Instead, one needs an approach to account for the dependencies and the varying expertise of scanners. Towards this end, one approach is to learn a set of latent variables for each scanner from a large corpus of historical VT reports, capturing the scanner dependencies and expertise. Utilizing these latent variables along with other commonly available features from prior work, we \eat{then} construct a supervised learner to classify malicious URLs, which \eat{
and show that our approach }indeed achieves 10-45\% higher classification performance compared to the baseline. 



\begin{figure}
\centering
  \includegraphics[width=1.0\textwidth]{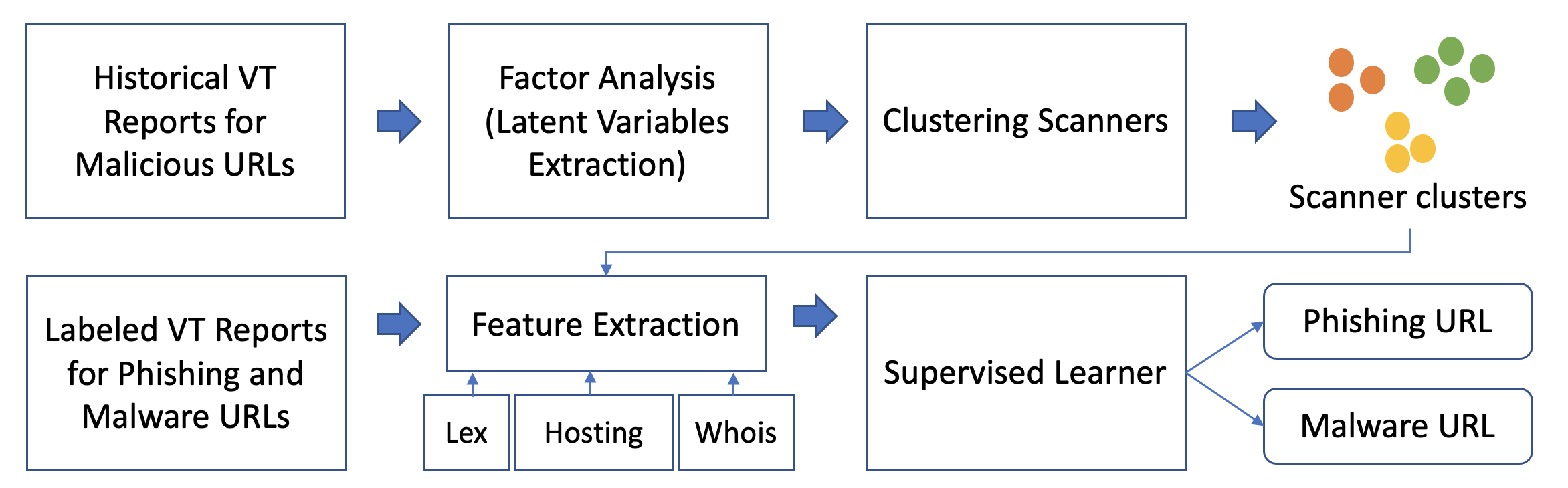}
  \captionsetup{width=.95\linewidth}
  \caption{Overall workflow of classifying malicious URLs as phishing or malware}
  \label{fig:pipeline}
  
\end{figure}

\textbf{Our Approach}. Figure~\ref{fig:pipeline} shows the overall classification pipeline of our approach. Our analysis in Section~\ref{sec:correlation} shows that scanners are highly correlated in terms of both detailed labels as well as binary labels. Further, scanners detecting phishing and malware URLs form distinct clusters. Motivated by these observations, first, we cluster similar scanners together based on the latent variables we derive. Along with the VT cluster features, we then utilize three groups of features: lexical, hosting,  and WHOIS. Lexical features refer to the textual features related to  URLs~\cite{blacklist:2009}. It is more likely for phishing URLs to have lexical features impersonating popular brands compared to malware ones. Hosting features, capturing the differences in the hosting infrastructures utilized for these two types of attacks, extract attributes related to the IP addresses where URLs are hosted~\cite{blacklist:2009}. WHOIS features are extracted from WHOIS registration records for each domain~\cite{predictive:leet:2010}. VT cluster features include scanner attack labels and the features derived from scanner clusters. Latent scanner features are derived from the factor analysis on randomly selected 20K recent historical VT reports with at least two positives. 
We use the detailed and binary labels of the scanners in each report as input features to the factor analysis. Our intuition is that these features capture scanner dependencies and varying degrees of expertise. We take the top 5 factors and cluster scanners into multiple groups. We vary the number of clusters from 5 to 20 and identify that 15 clusters produces the best downstream performance. 
Utilizing these clusters, we extract VT cluster features 
taking the scanner cluster assignment as the input and computing adjusted phishing and malware label proportions for each malicious URL. We observe that the adjusted label proportions perform better in the downstream classification task compared to the raw label proportions. A key reason for the significant performance gain is due to, as we have shown earlier, dependencies among scanners and highly correlated results at times. The adjusted label proportions consider these dependencies and compute more discriminative features to differentiate between phishing and malware URLs.

\begin{figure}[!htbp]
\parbox{1.0\textwidth}{
\centering
\subfigure[ROC Curve for Phishing]{
\psfig{file=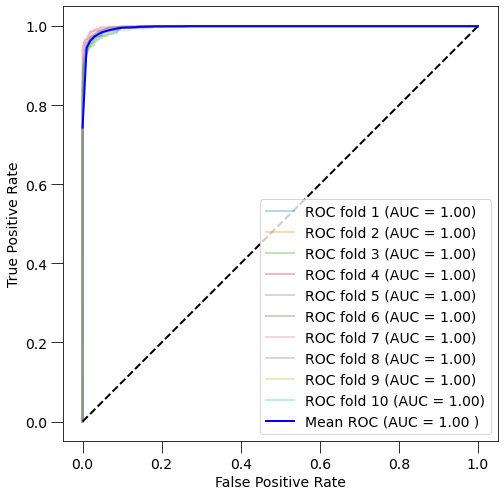,width=0.45\textwidth}\label{phishing_roc}}
\subfigure[ROC Curve for Malware]{
\psfig{file=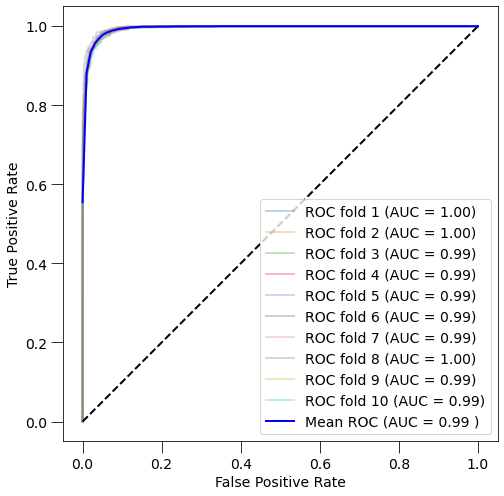,width=0.45\textwidth}\label{malware_roc}}
\eat{
\subfigure[ROC Curve for Malicious]{
\psfig{file=figure/attack_malicious.png,width=0.31\textwidth}\label{malicious_roc}}}
\captionsetup{width=.95\linewidth}
\caption{ROC Curves for Attack Types
}\label{fig:attack_roc}}
\vspace{-3mm}
\end{figure}

\textbf{Model Training and Testing}. To verify our approach, we use balanced datasets from each class - phishing and malware URLs - described in Section~\ref{sec:siteadvisor_data} so that 6,485 URLs from each class is used. We use the simple majority voting as a baseline model to compare with our approach. We train XGBoost, Random Forest (RF), Support Vector Machine, K Nearest Neighbor, Decision Tree, Naive Bayes, Logistic Regression and Linear Discriminant Analysis. 
RF yields the best result and hence all the experiments are performed with RF. Randomized search based hyperparameter optimization identifies the optimal maximum depth to be 250, a maximum number of features to be 55, the number of estimators to be 200. We utilize 80-20 train-test split and Figure~\ref{fig:attack_roc} shows the ROC curve for the two classes. Table~\ref{tab:attackresults} shows the offline performance metrics for the baseline model and our approach. While the baseline model performs poorly, our model achieves a high accuracy, precision, recall and a low false positive rate for each class in general. We attribute the performance improvement to the inclusion of latent scanner features along with lexical, hosting and WHOIS features.



\begin{table*}[!th]
\caption{Attack type classification performance of the baseline and our approach} 
\label{tab:attackresults}
\centering
\begin{tabular}{| p{1in} || p{0.4in} | p{0.4in} | p{0.4in} | p{0.4in} | p{0.4in} | p{0.4in} | p{0.4in} | p{0.4in} |}
\hline
 &
\multicolumn{4}{|c|}{\textbf{Baseline (Simple Majority Voting)}} &
\multicolumn{4}{|c|}{\textbf{Our Approach}}\\
\hline
\textbf{Type} & \textbf{Acc.} & \textbf{Prec.} & \textbf{Rec.} & \textbf{FPR} & \textbf{Acc.} & \textbf{Prec.} & \textbf{Rec.} & \textbf{FPR}\\
\hline
\hline
 Phishing & 81.72 & 69.90 & 92.98 & 25.43 & \textbf{97.47} & \textbf{95.45} & \textbf{96.91} & \textbf{2.3} \\ 
 \hline
 Malware & 70.10 & 51.59 & 19.93 & 8.12 & \textbf{96.34} & \textbf{95.93} & \textbf{93.30} & \textbf{2.1} \\
 \hline
 \eat{Malicious & 59.37 & 22.14 & 12.55 & 19.72 & 94.83 & 92.20 & 92.31 & 3.9 \\
 \hline}
\end{tabular}
\end{table*}


\textbf{Ablation Analysis}. As shown in Table~\ref{tab:ablation}, we analyze the performance with respect to different feature categories. We experiment lexical, hosting and WHOIS features separately along with VT cluster features.  While the performance improves around 1\% in each of these scenarios compared to only utilizing VT cluster features, 2-3\% improvement when all feature categories are considered suggesting each feature categories help learning different aspects of the phishing and malware URLs. Often times, collecting WHOIS records for domains are quite challenging. In such a situation, we recommend utilizing only lexical and hosting features with only slightly dropped performance (0.3\%) compared to having WHOIS features.

\begin{table*}[!th]
\caption{Performance of the attack type classification for different feature categories} 
\label{tab:ablation}
\centering
\begin{tabular}{| p{2in} || p{0.4in} | p{0.4in} | p{0.4in} | p{0.4in} | p{0.4in} | p{0.4in} | p{0.4in} | p{0.4in} |}
\hline
 &
\multicolumn{4}{|c|}{\textbf{Phishing}} &
\multicolumn{4}{|c|}{\textbf{Malware}}\\
\hline
\textbf{Feature Sets} & \textbf{Acc.} & \textbf{Prec.} & \textbf{Rec.} & \textbf{FPR} & \textbf{Acc.} & \textbf{Prec.} & \textbf{Rec.} & \textbf{FPR}\\
\hline
\hline
 VT cluster labels & 95.49 & 92.83 & 93.60 & 3.6 & 94.42 & 93.09 & 90.47 & 3.5 \\ 
 \hline
 VT cluster labels + lexical & 96.38 & 93.82 & 95.25 & 3.1 & 95.26 & 93.84 & 92.10 & 2.6 \\
 \hline
 VT cluster labels + hosting & 96.98 & 94.73 & 96.17 & 2.6 & 95.27 & 94.25 & 91.83 & 2.9 \\
 \hline
 VT cluster labels + whois & 96.91 & 94.91 & 95.78 & 2.5 & 95.71 & 94.85 & 92.48 & 2.6 \\
  \hline
 VT cluster labels + lexical + hosting & 97.21 & 95.24 & 96.35 & 2.4 & 95.67 & 94.65 & 92.53 & 2.7 \\
 \hline
 All (Our approach) & \textbf{97.47} & \textbf{95.45} & \textbf{96.91} & \textbf{2.3} & \textbf{96.34} & \textbf{95.93} & \textbf{93.30} & \textbf{2.1} \\
 \hline
\end{tabular}
\end{table*}

\begin{figure}[t]
\centering
    \includegraphics[width=.9\textwidth]{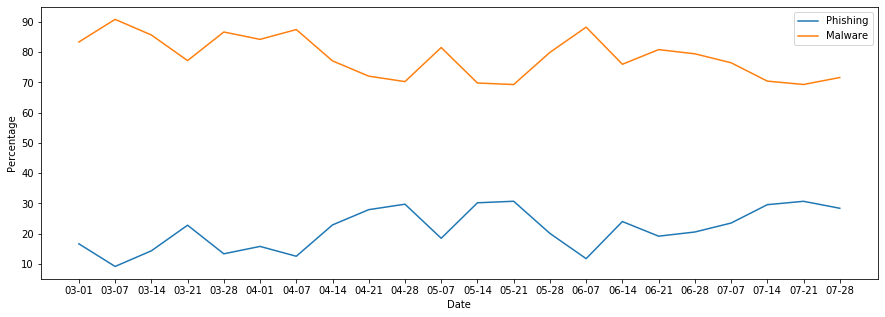}
   
    \captionsetup{width=.95\linewidth}
\caption{Attack types proportions observed in VT General feed over time}
\label{fig:attacktypecount}

\end{figure}

\textbf{Longitudinal Results}. We apply our trained classifier on a random sample of 56,138 VT malicious URL reports across 4 months during our study period starting from March 2021. Our predictions show that 11,922 and 44,216 are phishing, and malware  respectively. Figure~\ref{fig:attacktypecount} shows the weekly percentage of these two types of attacks over the 4 month period. The relative proportions of these attacks have been quite stable in this quarter and the malware URLs consistently dominate phishing URLs observed in VT over time. 

\section{Discussion}\label{sec:discussion}
\heading{Analysis Summary and Recommendation} Our measurement studies show that one important consideration to build a high quality ground truth is to identify attack types of URLs. 
Different scanners specialize in different attack types and the best thresholds for different attack types also vary due to the level of scanners' correlations for each attack type. We observe that in terms of detection, scanners detecting phishing URLs are more correlated than those detecting malware URLs. It is prudent to identify attack types and set different thresholds depending on the attack types. Instead of using a fixed threshold for all types of URLs, we suggest to penalize the count to total proportional to correlation coefficients to obtain a more reflective positive count for each attack type. Specifically, we recommend using a higher threshold to compile phishing URL ground truth compared to malware URL ground truth given higher correlation between scanners detecting phishing URLs. Furthermore, we show that a simple majority voting approach to determine the attack type works poorly. We observe that phishing URLs have more conflicting attack labels than malware URLs and the false positive rate can reach 25.43 \% for phishing URLs using a majority voting approach. We recommend that researchers employ our proposed approach to efficiently and effectively identify an attack type with high accuracy.







\heading{Limitation on Ground truth} It is often challenging to collect a large-scale URL ground truth~\cite{feal2021blocklist}. Despite our best efforts to cover various types of attacks in ground truth, our dataset may still have some limitation. First, 2 external sources in our ground truth dataset may have certain bias on their URL lists. However, for better confidence on the ground truth considering noises in 2 external sources, we take a conservative approach that we do additional manual verification and URLs with conflicting labels from domain experts are not considered as ground truth. Moreover, as we collected URLs from either VT itself or identified URLs from 2 external sources, we are not able to analyze VT's behavior throughout the life time of URLs (e.g., creation and take-down). It would be interesting to see such trend, we leave as future work. Second, while it is relatively easy to judge if a URL is phishing/non-phishing and malware/non-malware, it is hard to judge if the URL is benign by human experts. Accordingly, our manual GT benign URLs can be biased towards popular domains. In a future work, we will study VT scanners' behavior with less popular benign URLs.




\vspace{-2mm}

\section{Related Work}\label{sec:related}

\heading{VT as Ground Truth.} VT has been used to build a groundtruth in various domains including malware files~\cite{vt1:thomas2016investigating,vt1:wong2018tackling,korczynski2017capturing,duan2018things,yang2017malware,kwon2015dropper} and IP/URLs~\cite{li2019tealeave,sharif2018impending,vt1:phishing:2018,vt1:www:2018,vt3:oprea2018made,vt1:zuo2017smartgen} detection. In doing so, the most common approach is unweighted threshold-based methods, which employ a heuristically chosen number of detecting scanners with which the entity is labelled as malicious. While there is no consensus on such a number~\cite{zhu2020:labeldynamics,vtuse2:imc:2018}, surprisingly, small thresholds such as 1 or 2 have been widely used in the literature~\cite{vt1:www:2018,vt1:phishing:2018,vtuse2:imc:2018,vt1:thomas2016investigating,vt1:wong2018tackling,sharif2018impending,vt1:zuo2017smartgen,zhu2020:labeldynamics}. Apparently, such small thresholds will lead to high false positives depending on the quality of scanners~\cite{vt3:oprea2018made}. A few papers set aggressive thresholds~\cite{korczynski2017capturing,duan2018things,yang2017malware}, which may result in low coverage~\cite{choo2019devicewatch,peng2019blackbox,zhu2020:labeldynamics}. A few studies such as ~\cite{www:vt_level_malicious,choo2019devicewatch} treat the number of detecting scanners as the level of maliciousness. However, we show that the absolute number does not necessarily mean the level of maliciousness due to high correlations between scanners.

\heading{Threat Intelligence Aggregation.} Recently, there have been researches to measure the qualities of multiple threat intelligence sources including VT ~\cite{qiang:2018quality,du2018statistical,charlton2018measuring,zhu2020:labeldynamics,peng2019blackbox,Kuhrer:2014:blacklists,oest2020phishtime,li2019tealeave,feal2021blocklist,sakib2020maximizing,salem2021maat,tops2021maat,bouwman2022helping}. K{\"u}hrer \emph{et al.} ~\cite{Kuhrer:2014:blacklists}, Charlton \emph{et al.} ~\cite{charlton2018measuring}, and  Bouwman \emph{et al.} ~\cite{bouwman2022helping} evaluated the effectiveness of malware blacklists, VT malware file scanners, and COVID-19 Cyber Threat Coalition, respectively. Zhu \emph{et al.} analyzed the stability of VT malware file scanners in terms of the detection label dynamics and the dependency among scanners~\cite{zhu2020:labeldynamics}. Adam \emph{et al.} measured the effectiveness of three browser blacklists in terms of speed and coverage and investigated the evasion techniques of phishing websites against those blacklists ~\cite{oest2020phishtime}. Peng \emph{et al.} measured the detection accuracy of VT phishing URL scanners for IRS/paypal phishing URLs and compared with each vendor's own API ~\cite{peng2019blackbox}. While these researches provided insights about detection qualities of intelligent sources, measurement was done on limited datasets in terms of diversity and scale. Concretely, not only did each of them focus on one type of entity such as malware files ~\cite{zhu2020:labeldynamics,du2018statistical,charlton2018measuring,Kantchelian:2015:VTMalClassifier, salem2021maat,tops2021maat}, phishing URLs ~\cite{peng2019blackbox,oest2020phishtime}, malware URLs ~\cite{Vasek:2013:malwareurls}, and domains from COVID-19 related threat intelligence sources~\cite{bouwman2022helping}; but also most studies were done on a small-scale dataset~\cite{peng2019blackbox} or a single snapshot of reports~\cite{hurier2017euphony,sebastian2016avclass,sakib2020maximizing,mohaisen2014avmeter,charlton2018measuring}. In contrast, we provide a large-scale longitudinal analysis for various types of URLs. In doing so, we analyze the specialty of scanners and their correlations for different attack types, and propose a method to identify attack types of URLs.

A few studies proposed better ways to aggregate different sources considering qualities~\cite{ramanathan2020blag,Kantchelian:2015:VTMalClassifier,sebastian2016avclass,hurier2017euphony,mohaisen2014avmeter,sakib2020maximizing,siraj2022}. Kantchelian \emph{et al.} proposed two machine learning models~\cite{Kantchelian:2015:VTMalClassifier} with the assumption that scanners are independent. However, we show that some scanners are highly correlated and cannot be considered independent. Sakib \emph{et al.}~\cite{sakib2020maximizing} and Thirumuruganathan \emph{et al.}~\cite{siraj2022} proposed ways to optimally combine malware scanners and general threat intelligence sources, respectively, with consideration of dependencies between scanners. However, the approach in ~\cite{sakib2020maximizing} assumed that detection probabilities of scanners are available, which is limited because detection probabilities are not given when the ground truth is built depending solely on VT. We also show that the detection probability of each scanner can vary over time and for different attack types due to its detecting specialty. In ~\cite{siraj2022}, Thirumuruganathan \emph{et al.} proposed an approach to cluster URLs into benign or malicious by integrating noisy scan reports without such assumption. However, they did not provide a systematic quantitative study on the characteristics of scan reports, which is one of our main focuses. Furthermore, we show that the attack types of malicious URLs is an important factor in such characteristics and propose a method to classify the attack types.

\section{Conclusions}\label{sec:conclusions}
In this paper, we provide a large-scale analysis of VT URL scan reports spanning over two years. We show that using a fixed threshold to determine maliciousness of URL and using a majority voting to classify attack types are limited due to multiple factors including conflicts between scanners, and the specialty, stability, correlation, and lead/lag behavior of scanners. Our analyses show that scanners behave differently for different attack types and there largely exist conflicting labels on detection and attack types by individual scanners as well as across scanners. We propose a machine learning based approach considering such characteristics to identify the attack label of a malicious URL from conflicting labels. Finally, we suggest that researchers first need to identify attack types of malicious URLs, depending on which appropriate thresholds should be set to collect malicious ground truth. 


\bibliographystyle{unsrt}  
\bibliography{ref}

\begin{thebibliography}{10}

\bibitem{li2019tealeave}
Vector~Guo Li, Matthew Dunn, Paul Pearce, Damon McCoy, Geoffrey~M Voelker, and
  Stefan Savage.
\newblock Reading the tea leaves: A comparative analysis of threat
  intelligence.
\newblock In {\em 28th $\{$USENIX$\}$ Security Symposium ($\{$USENIX$\}$
  Security 19)}, pages 851--867, 2019.

\bibitem{sharif2018impending}
Mahmood Sharif, Jumpei Urakawa, Nicolas Christin, Ayumu Kubota, and Akira
  Yamada.
\newblock Predicting impending exposure to malicious content from user
  behavior.
\newblock In {\em Proceedings of the 2018 ACM SIGSAC Conference on Computer and
  Communications Security}, pages 1487--1501, 2018.

\bibitem{vt1:phishing:2018}
Ke~Tian, Steve T.~K. Jan, Hang Hu, Danfeng Yao, and Gang Wang.
\newblock Needle in a haystack: Tracking down elite phishing domains in the
  wild.
\newblock In {\em Proceedings of the Internet Measurement Conference 2018}, IMC
  '18, page 429–442, New York, NY, USA, 2018. Association for Computing
  Machinery.

\bibitem{vt1:www:2018}
Najmeh Miramirkhani, Timothy Barron, Michael Ferdman, and Nick Nikiforakis.
\newblock Panning for gold.com: Understanding the dynamics of domain
  dropcatching.
\newblock In {\em Proceedings of the 2018 World Wide Web Conference}, WWW '18,
  page 257–266, Republic and Canton of Geneva, CHE, 2018. International World
  Wide Web Conferences Steering Committee.

\bibitem{vt3:oprea2018made}
Alina Oprea, Zhou Li, Robin Norris, and Kevin Bowers.
\newblock Made: Security analytics for enterprise threat detection.
\newblock In {\em Proceedings of the 34th Annual Computer Security Applications
  Conference}, pages 124--136, 2018.

\bibitem{vt1:zuo2017smartgen}
Chaoshun Zuo and Zhiqiang Lin.
\newblock Smartgen: Exposing server urls of mobile apps with selective symbolic
  execution.
\newblock In {\em Proceedings of the 26th International Conference on World
  Wide Web}, pages 867--876, 2017.

\bibitem{zhu2020:labeldynamics}
Shuofei Zhu, Jianjun Shi, Limin Yang, Boqin Qin, Ziyi Zhang, Linhai Song, and
  Gang Wang.
\newblock Measuring and modeling the label dynamics of online anti-malware
  engines.
\newblock In {\em 29th $\{$USENIX$\}$ Security Symposium ($\{$USENIX$\}$
  Security 20)}, 2020.

\bibitem{tops2021maat}
Aleieldin Salem, Sebastian Banescu, and Alexander Pretschner.
\newblock Maat: Automatically analyzing virustotal for accurate labeling and
  effective malware detection.
\newblock {\em ACM Transactions on Privacy and Security (TOPS)}, 24(4):1--35,
  2021.

\bibitem{peng2019blackbox}
Peng Peng, Limin Yang, Linhai Song, and Gang Wang.
\newblock Opening the blackbox of virustotal: Analyzing online phishing scan
  engines.
\newblock In {\em Proceedings of the Internet Measurement Conference}, pages
  478--485, 2019.

\bibitem{Kantchelian:2015:VTMalClassifier}
Alex Kantchelian, Michael~Carl Tschantz, Sadia Afroz, Brad Miller, Vaishaal
  Shankar, Rekha Bachwani, Anthony~D. Joseph, and J.~D. Tygar.
\newblock Better malware ground truth: Techniques for weighting anti-virus
  vendor labels.
\newblock In {\em Proceedings of the 8th ACM Workshop on Artificial
  Intelligence and Security}, AISec '15, pages 45--56, 2015.

\bibitem{salem2021maat}
Aleieldin Salem.
\newblock Towards accurate labeling of android apps for reliable malware
  detection.
\newblock In {\em Proceedings of the Eleventh ACM Conference on Data and
  Application Security and Privacy}, pages 269--280, 2021.

\bibitem{bouwman2022helping}
Xander Bouwman, Victor Le~Pochat, Pawel Foremski, Tom Van~Goethem, Carlos~H
  Ga{\~n}{\'a}n, Giovane Moura, Samaneh Tajalizadehkhoob, Wouter Joosen, and
  Michel van Eeten.
\newblock Helping hands: Measuring the impact of a large threat intelligence
  sharing community.
\newblock In {\em Proceedings of the 31st USENIX Security Symposium}. USENIX
  Association, 2022.

\bibitem{www:vt_level_malicious}
Onur Catakoglu, Marco Balduzzi, and Davide Balzarotti.
\newblock Automatic extraction of indicators of compromise for web
  applications.
\newblock In {\em Proceedings of the 25th international conference on world
  wide web}, pages 333--343, 2016.

\bibitem{vt1:imc:2018}
Armin Sarabi and Mingyan Liu.
\newblock Characterizing the internet host population using deep learning: A
  universal and lightweight numerical embedding.
\newblock In {\em Proceedings of the Internet Measurement Conference 2018}, IMC
  '18, page 133–146, New York, NY, USA, 2018. Association for Computing
  Machinery.

\bibitem{malwarevt:ccs18}
Binlin Cheng, Jiang Ming, Jianmin Fu, Guojun Peng, Ting Chen, Xiaosong Zhang,
  and Jean-Yves Marion.
\newblock Towards paving the way for large-scale windows malware analysis:
  Generic binary unpacking with orders-of-magnitude performance boost.
\newblock In {\em Proceedings of the 2018 ACM SIGSAC Conference on Computer and
  Communications Security}, CCS '18, page 395–411, New York, NY, USA, 2018.
  Association for Computing Machinery.

\bibitem{choi2011detecting}
Hyunsang Choi, Bin~B Zhu, and Heejo Lee.
\newblock Detecting malicious web links and identifying their attack types.
\newblock In {\em 2nd USENIX Conference on Web Application Development (WebApps
  11)}, 2011.

\bibitem{compromised:usenix:2021}
Ravindu~De Silva, Mohamed Nabeel, Charitha Elvitigala, Issa Khalil, Ting Yu,
  and Chamath Keppitiyagama.
\newblock Compromised or attacker-owned: A large scale classification and study
  of hosting domains of malicious urls.
\newblock In {\em 30th $\{$USENIX$\}$ Security Symposium ($\{$USENIX$\}$
  Security 21)}, Vancouver, B.C., 2021. $\{$USENIX$\}$ Association.

\bibitem{feal2021blocklist}
{\'A}lvaro Feal, Pelayo Vallina, Julien Gamba, Sergio Pastrana, Antonio Nappa,
  Oliver Hohlfeld, Narseo Vallina-Rodriguez, and Juan Tapiador.
\newblock Blocklist babel: On the transparency and dynamics of open source
  blocklisting.
\newblock {\em IEEE Transactions on Network and Service Management}, 2021.

\bibitem{phishtank}
Phishtank.
\newblock Phishtank.
\newblock \url{https://www.phishtank.com/}, 2021.
\newblock Accessed July 2021.

\bibitem{openphish}
OpenPhish.
\newblock Openphish.
\newblock \url{https://www.openphish.com/}.

\bibitem{urlhaus}
URLhaus.
\newblock Urlhaus.
\newblock \url{https://urlhaus.abuse.ch}, 2021.
\newblock Accessed July 2021.

\bibitem{maldomlist}
Malware~Domain List.
\newblock Malware domain list.
\newblock \url{http://www.malwaredomainlist.com/mdl.php}, 2021.
\newblock Accessed July 2021.

\bibitem{phishingbl:2020}
Simon Bell and Peter Komisarczuk.
\newblock An analysis of phishing blacklists: Google safe browsing, openphish,
  and phishtank.
\newblock In {\em Proceedings of the Australasian Computer Science Week
  Multiconference}, ACSW '20, New York, NY, USA, 2020. Association for
  Computing Machinery.

\bibitem{Zhang2021CrawlPhishLA}
Penghui Zhang, Adam Oest, Haehyun Cho, Zhibo Sun, RC~Johnson, Brad Wardman,
  Shaown Sarker, Alexandros Kapravelos, Tiffany Bao, Ruoyu Wang, Yan
  Shoshitaishvili, Adam Doup{\'e}, and Gail-Joon Ahn.
\newblock Crawlphish: Large-scale analysis of client-side cloaking techniques
  in phishing.
\newblock {\em 2021 IEEE Symposium on Security and Privacy (SP)}, pages
  1109--1124, 2021.

\bibitem{virustotal}
{VirusTotal, Subsidiary of Google}.
\newblock {Free Online Virus, Malware and URL Scanner}.
\newblock \url{https://www.virustotal.com/}.
\newblock Accessed: 04-02-2021.

\bibitem{choo2019devicewatch}
Euijin Choo, Mohamed Nabeel, Mashael Alsabah, Issa Khalil, Ting Yu, and Wei
  Wang.
\newblock Devicewatch: Identifying compromised mobile devices through network
  traffic analysis and graph inference.
\newblock {\em arXiv preprint arXiv:1911.12080}, 2019.

\bibitem{reg:2017:RAID}
Thomas Vissers, Jan Spooren, Pieter Agten, Dirk Jumpertz, Peter Janssen, Marc
  Van~Wesemael, Frank Piessens, Wouter Joosen, and Lieven Desmet.
\newblock Exploring the ecosystem of malicious domain registrations in the .eu
  tld.
\newblock In {\em Research in Attacks, Intrusions, and Defenses}, pages
  472--493. Springer International Publishing, 2017.

\bibitem{kim2020anatomy}
Sungjin Kim.
\newblock Anatomy on malware distribution networks.
\newblock {\em IEEE Access}, 8:73919--73930, 2020.

\bibitem{oest2020phishtime}
Adam Oest, Yeganeh Safaei, Penghui Zhang, Brad Wardman, Kevin Tyers, Yan
  Shoshitaishvili, and Adam Doup{\'e}.
\newblock Phishtime: Continuous longitudinal measurement of the effectiveness
  of anti-phishing blacklists.
\newblock In {\em 29th $\{$USENIX$\}$ Security Symposium ($\{$USENIX$\}$
  Security 20)}, pages 379--396, 2020.

\bibitem{oest2020sunrise}
Adam Oest, Penghui Zhang, Brad Wardman, Eric Nunes, Jakub Burgis, Ali Zand,
  Kurt Thomas, Adam Doup{\'e}, and Gail-Joon Ahn.
\newblock Sunrise to sunset: Analyzing the end-to-end life cycle and
  effectiveness of phishing attacks at scale.
\newblock In {\em 29th $\{$USENIX$\}$ Security Symposium ($\{$USENIX$\}$
  Security 20)}, pages 361--377, 2020.

\bibitem{bennett2010online}
Paul~N Bennett and Vitor~R Carvalho.
\newblock Online stratified sampling: evaluating classifiers at web-scale.
\newblock In {\em Proceedings of the 19th ACM international conference on
  Information and knowledge management}, pages 1581--1584, 2010.

\bibitem{katariya2012active}
Namit Katariya, Arun Iyer, and Sunita Sarawagi.
\newblock Active evaluation of classifiers on large datasets.
\newblock In {\em 2012 IEEE 12th International Conference on Data Mining},
  pages 329--338. IEEE, 2012.

\bibitem{moubayed2018dnssquatting}
Abdallah Moubayed, MohammadNoor Injadat, Abdallah Shami, and Hanan Lutfiyya.
\newblock Dns typo-squatting domain detection: A data analytics \& machine
  learning based approach.
\newblock In {\em 2018 IEEE Global Communications Conference (GLOBECOM)}, pages
  1--7. IEEE, 2018.

\bibitem{APWG}
APWG.
\newblock Anti-phishing working group.
\newblock \url{https://www.apwg.org/}, 2021.
\newblock Accessed July 2021.

\bibitem{siteadvisor}
McAfee.
\newblock Site advisor.

\bibitem{f1auc}
Takaya Saito and Marc Rehmsmeier.
\newblock The precision-recall plot is more informative than the roc plot when
  evaluating binary classifiers on imbalanced datasets.
\newblock {\em PLOS ONE}, 10(3):1--21, 03 2015.

\bibitem{thomas2015ad}
Kurt Thomas, Elie Bursztein, Chris Grier, Grant Ho, Nav Jagpal, Alexandros
  Kapravelos, Damon McCoy, Antonio Nappa, Vern Paxson, Paul Pearce, et~al.
\newblock Ad injection at scale: Assessing deceptive advertisement
  modifications.
\newblock In {\em 2015 IEEE Symposium on Security and Privacy}, pages 151--167.
  IEEE, 2015.

\bibitem{arp2014drebin}
Daniel Arp, Michael Spreitzenbarth, Malte Hubner, Hugo Gascon, Konrad Rieck,
  and CERT Siemens.
\newblock Drebin: Effective and explainable detection of android malware in
  your pocket.
\newblock In {\em Ndss}, volume~14, pages 23--26, 2014.

\bibitem{jaccard1912distribution}
Paul Jaccard.
\newblock The distribution of the flora in the alpine zone. 1.
\newblock {\em New phytologist}, 11(2):37--50, 1912.

\bibitem{berndt1994dtw}
Donald~J Berndt and James Clifford.
\newblock Using dynamic time warping to find patterns in time series.
\newblock In {\em KDD workshop}, volume~10, pages 359--370. Seattle, WA, USA:,
  1994.

\bibitem{horn2012matrix}
Roger~A Horn and Charles~R Johnson.
\newblock {\em Matrix analysis}.
\newblock Cambridge university press, 2012.

\bibitem{blacklist:2009}
Justin Ma, Lawrence~K. Saul, Stefan Savage, and Geoffrey~M. Voelker.
\newblock Beyond blacklists: Learning to detect malicious web sites from
  suspicious urls.
\newblock In {\em Proceedings of the 15th ACM SIGKDD International Conference
  on Knowledge Discovery and Data Mining}, KDD '09, page 1245–1254, New York,
  NY, USA, 2009. Association for Computing Machinery.

\bibitem{predictive:leet:2010}
Mark Felegyhazi, Christian Kreibich, and Vern Paxson.
\newblock On the potential of proactive domain blacklisting.
\newblock In {\em Proceedings of the 3rd USENIX Conference on Large-Scale
  Exploits and Emergent Threats: Botnets, Spyware, Worms, and More}, LEET'10,
  page~6, USA, 2010. USENIX Association.

\bibitem{vt1:thomas2016investigating}
Kurt Thomas, Juan A~Elices Crespo, Ryan Rasti, Jean-Michel Picod, Cait
  Phillips, Marc-Andr{\'e} Decoste, Chris Sharp, Fabio Tirelo, Ali Tofigh,
  Marc-Antoine Courteau, et~al.
\newblock Investigating commercial pay-per-install and the distribution of
  unwanted software.
\newblock In {\em 25th $\{$USENIX$\}$ Security Symposium ($\{$USENIX$\}$
  Security 16)}, pages 721--739, 2016.

\bibitem{vt1:wong2018tackling}
Michelle~Y Wong and David Lie.
\newblock Tackling runtime-based obfuscation in android with $\{$TIRO$\}$.
\newblock In {\em 27th $\{$USENIX$\}$ Security Symposium ($\{$USENIX$\}$
  Security 18)}, pages 1247--1262, 2018.

\bibitem{korczynski2017capturing}
David Korczynski and Heng Yin.
\newblock Capturing malware propagations with code injections and code-reuse
  attacks.
\newblock In {\em Proceedings of the 2017 ACM SIGSAC Conference on Computer and
  Communications Security}, pages 1691--1708, 2017.

\bibitem{duan2018things}
Yue Duan, Mu~Zhang, Abhishek~Vasisht Bhaskar, Heng Yin, Xiaorui Pan, Tongxin
  Li, Xueqiang Wang, and XiaoFeng Wang.
\newblock Things you may not know about android (un) packers: A systematic
  study based on whole-system emulation.
\newblock In {\em NDSS}, 2018.

\bibitem{yang2017malware}
Wei Yang, Deguang Kong, Tao Xie, and Carl~A Gunter.
\newblock Malware detection in adversarial settings: Exploiting feature
  evolutions and confusions in android apps.
\newblock In {\em Proceedings of the 33rd Annual Computer Security Applications
  Conference}, pages 288--302, 2017.

\bibitem{kwon2015dropper}
Bum~Jun Kwon, Jayanta Mondal, Jiyong Jang, Leyla Bilge, and Tudor
  Dumitra{\c{s}}.
\newblock The dropper effect: Insights into malware distribution with
  downloader graph analytics.
\newblock In {\em Proceedings of the 22nd ACM SIGSAC Conference on Computer and
  Communications Security}, pages 1118--1129, 2015.

\bibitem{vtuse2:imc:2018}
Haoyu Wang, Zhe Liu, Jingyue Liang, Narseo Vallina-Rodriguez, Yao Guo, Li~Li,
  Juan Tapiador, Jingcun Cao, and Guoai Xu.
\newblock Beyond google play: A large-scale comparative study of chinese
  android app markets.
\newblock In {\em Proceedings of the Internet Measurement Conference 2018}, IMC
  '18, page 293–307, New York, NY, USA, 2018. Association for Computing
  Machinery.

\bibitem{qiang:2018quality}
Li~Qiang, Jiang Zhengwei, Yang Zeming, Liu Baoxu, Wang Xin, and Zhang Yunan.
\newblock A quality evaluation method of cyber threat intelligence in user
  perspective.
\newblock In {\em 2018 17th IEEE International Conference On Trust, Security
  And Privacy In Computing And Communications/12th IEEE International
  Conference On Big Data Science And Engineering (TrustCom/BigDataSE)}, pages
  269--276. IEEE, 2018.

\bibitem{du2018statistical}
Pang Du, Zheyuan Sun, Huashan Chen, Jin-Hee Cho, and Shouhuai Xu.
\newblock Statistical estimation of malware detection metrics in the absence of
  ground truth.
\newblock {\em IEEE Transactions on Information Forensics and Security},
  13(12):2965--2980, 2018.

\bibitem{charlton2018measuring}
John Charlton, Pang Du, Jin-Hee Cho, and Shouhuai Xu.
\newblock Measuring relative accuracy of malware detectors in the absence of
  ground truth.
\newblock In {\em MILCOM 2018-2018 IEEE Military Communications Conference
  (MILCOM)}, pages 450--455. IEEE, 2018.

\bibitem{Kuhrer:2014:blacklists}
Marc K{\"u}hrer, Christian Rossow, and Thorsten Holz.
\newblock Paint it black: Evaluating the effectiveness of malware blacklists.
\newblock In Angelos Stavrou, Herbert Bos, and Georgios Portokalidis, editors,
  {\em Research in Attacks, Intrusions and Defenses}, pages 1--21, Cham, 2014.
  Springer International Publishing.

\bibitem{sakib2020maximizing}
Muhammad~N Sakib, Chin-Tser Huang, and Ying-Dar Lin.
\newblock Maximizing accuracy in multi-scanner malware detection systems.
\newblock {\em Computer Networks}, 169:107027, 2020.

\bibitem{Vasek:2013:malwareurls}
Marie Vasek and Tyler Moore.
\newblock {Empirical Analysis of Factors Affecting Malware URL Detection}.
\newblock In {\em Proceedings of the eCrime Researchers Summit}, pages 1--9,
  2013.

\bibitem{hurier2017euphony}
M{\'e}d{\'e}ric Hurier, Guillermo Suarez-Tangil, Santanu~Kumar Dash,
  Tegawend{\'e}~F Bissyand{\'e}, Yves Le~Traon, Jacques Klein, and Lorenzo
  Cavallaro.
\newblock Euphony: Harmonious unification of cacophonous anti-virus vendor
  labels for android malware.
\newblock In {\em 2017 IEEE/ACM 14th International Conference on Mining
  Software Repositories (MSR)}, pages 425--435. IEEE, 2017.

\bibitem{sebastian2016avclass}
Marcos Sebasti{\'a}n, Richard Rivera, Platon Kotzias, and Juan Caballero.
\newblock Avclass: A tool for massive malware labeling.
\newblock In {\em International symposium on research in attacks, intrusions,
  and defenses}, pages 230--253. Springer, 2016.

\bibitem{mohaisen2014avmeter}
Aziz Mohaisen and Omar Alrawi.
\newblock Av-meter: An evaluation of antivirus scans and labels.
\newblock In {\em International conference on detection of intrusions and
  malware, and vulnerability assessment}, pages 112--131. Springer, 2014.

\bibitem{ramanathan2020blag}
Sivaramakrishnan Ramanathan, Jelena Mirkovic, and Minlan Yu.
\newblock Blag: Improving the accuracy of blacklists.
\newblock In {\em NDSS}, 2020.

\bibitem{siraj2022}
S.~Thirumuruganatha, M.~Nabeel, E.~Choo, I.~Khalil, and T.~Yu.
\newblock Siraj: A unified framework for aggregation of malicious entity
  detectors.
\newblock In {\em 2022 IEEE Symposium on Security and Privacy}, 2022.

\bibitem{OTX}
AT\&T.
\newblock Open threat exchange.
\newblock \url{https://cybersecurity.att.com/open-threat-exchange}, 2021.
\newblock Accessed July 2021.

\end{thebibliography}

\begin{appendices}

\eat{
\section*{Appendix I - The distribution of time difference between PT first submitted and PT verified}

Figure~\ref{fig:newfeed_verified_timediff} shows that the distribution of time difference between PT first submitted and PT verified. As shown in the figure, 96.8\% of URLs are verified within 4 days. 
\begin{figure*}[tb]
\begin{center}
\parbox{1.0\textwidth}{
\centering
    \epsfig{file=figure/newfeed30_verified_distribution.pdf,width=0.5\textwidth, height=0.17\textheight}
\caption{The distribution of time difference between PT first submitted and PT verified}\label{fig:newfeed_verified_timediff}}
\end{center}
\end{figure*}
}
\section*{Appendix I - Ethics}
This work does not have any potential ethical issues.

\section*{Appendix II - List of Scanners in Dataset}
Abusix, ADMINUSLabs, AICC (MONITORAPP), Alexa, AlienVault, alphaMountain.ai, Antiy-AVL, Armis, AutoShun, Avira, BADWARE.INFO, Baidu-International, BenkowCC, BforeAi, BitDefender, Blueliv, Certego, CINS, CMC Threat Intelligence, CRDF, C-SIRT, CLEAN MX, Comodo Valkyrie Verdict, Cyan Digital Security, CyberCrime, CyRadar, desenmascara.me, DNS8, Dr.Web, EmergingThreats, Emsisoft, ESET, ESTsecurity, Forcepoint ThreatSeeker, Feodo Tracker, FraudSense, Fortinet, G-Data, Google Safebrowsing, GreenSnow, IPSum, Hoplite Industries, Lumu, K7AntiVirus, Lionic, Kaspersky, MalBeacon, Malekal, Malsilo, Malware Domain Blocklist, Malware Domain List, MalwarePatrol, Malwarebytes hpHosts, Malwared, Malwares.com, Netcraft, NotMining, OpenPhish, Palevo Tracker, Phishlabs, Phishtank, Prebytes, Quickheal, Quttera, Rising, Sangfor, SafeToOpen, Scantitan, SCUMWARE.org, SecureBrain, Sophos, Spam404, SpyEye Tracker, Spamhaus, StopBadware, Sucuri SiteCheck, ThreatHive, Trend Micro Site Safety Center, Trustwave, urlQuery, Virusdie External Site Scan, VX Vault, Web Security Guard, Wepawet, Yandex Safebrowsing, Zeus Tracker, Zvelo, Botvrij.eu, Artists Against 419, Nucleon, Ransomware Tracker, URLhaus, Webroot, ZeroCERT, securolytics

\pagebreak
\section*{Appendix III - Manual GT URL collection and Manual Labeling Process}

To collect the set of URLs for Manual GT, we employ a stratification sampling approach proposed in ~\cite{bennett2010online} and ~\cite{katariya2012active}. In doing so, we consider multiple dimensions of strata including the popularity (the number of VT rescan query made in VT feed) and VT positive count. The URLs are then manually labeled by 5 domain experts. Specifically, experts individually visit the set of URLs using multiple browsers including Chrome, Opera, Firefox, and Safari, and manually classify the attack types. To achieve a better confidence on labeling, all URLs are labeled by two experts and exclude URLs with conflicting labels. If the URL is NX, the URL is filtered from the list of URLs to analyze. If the URL is not NX, experts classify the type of URLs with the rules including the following.
\begin{itemize}
\item Check the URL address, forms, brand logos, redirections to identify phishing URLs.
\item Check for associated files hosted in the URL to identify malware URLs. Download the file and check if the file is malware or not. In doing so, we perform the similar process to ~\cite{zhu2020:labeldynamics} and we also check the file against multiple Anti-virus engines including Sophos and McAfee desktop engine.  
\item Check if popular brand names or their variants being present in the URL address. 
\item Check the screenshots saved in the historical databases such as Internet Wayback Machine and urlscan.io.
\item Check the detailed threat report by OTX~\cite{OTX} and McAfee WebAdvisor~\cite{siteadvisor}.
\item If none of the above malicious indicators of compromise are present for a URL and the URL has been operational for at least 3 months, we mark the URL as benign.
\item  if the landing page is legitimate (e.g., known popular URLs such as \url{https://outlook.live.com/owa/} and \url{https://abc7news.com/weather/)}, we mark the URL as benign.
\end{itemize}

\section*{Appendix IV - Scanner Label Certainty Score}

\begin{figure}[htbp]
\parbox{0.9\textwidth}{
\centering
\psfig{file=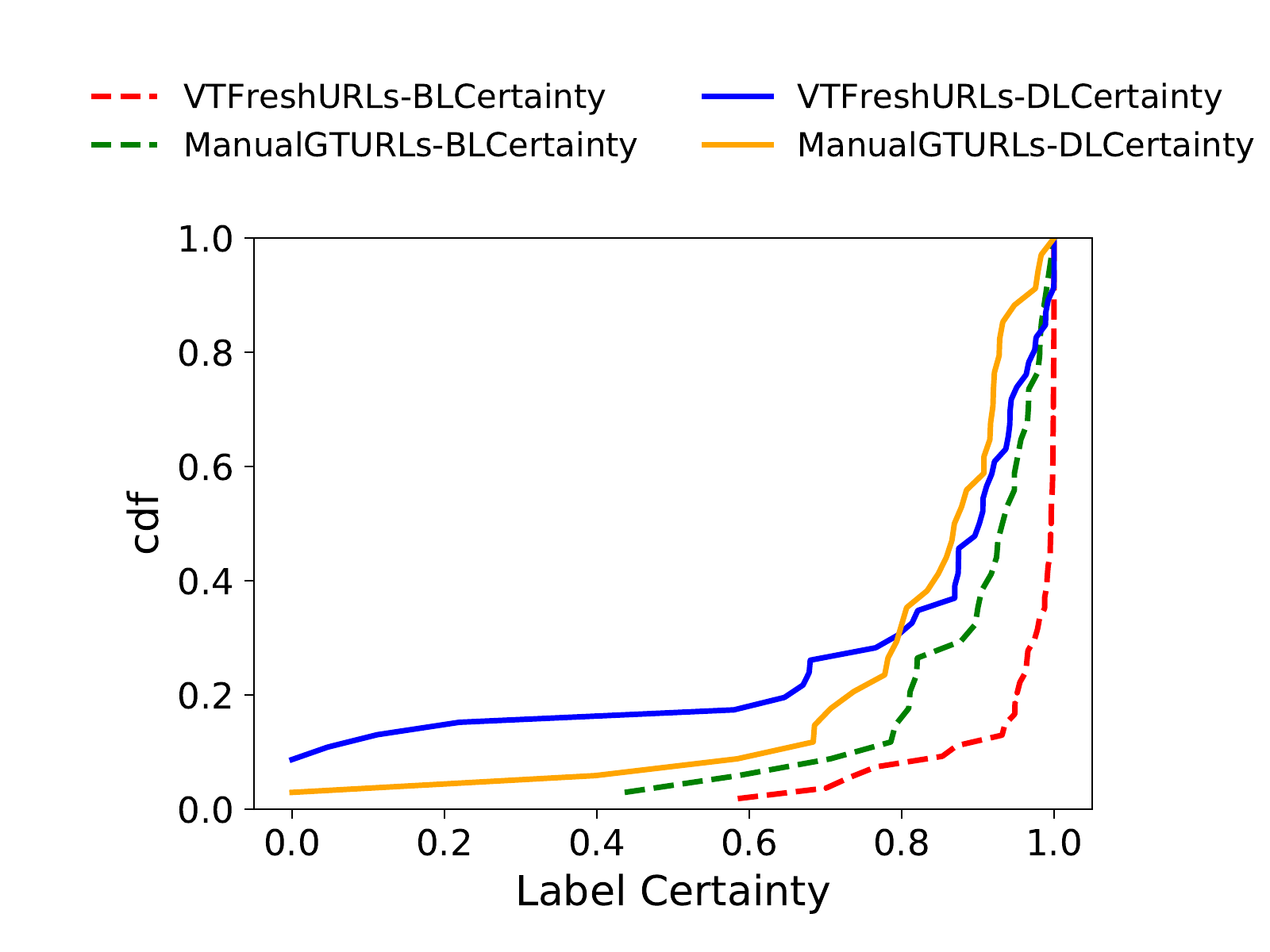,width=0.48\textwidth}\label{general_certainty}
\captionsetup{width=.8\linewidth}
\caption{Scanner label certainty score for VT Fresh and Manual GT URLs
}\label{fig:certainty_appendix}}
\end{figure}

\pagebreak
\onecolumn
\section*{Appendix V - Heatmaps for scanners' pairwise Jaccard Similarity of binary and detailed labels and scanner clustering}

\begin{figure*}[htbp]
\begin{center}
\parbox{1.0\textwidth}{
\centering
\subfigure[Phishing - Binary Label Similarity]{
\psfig{file=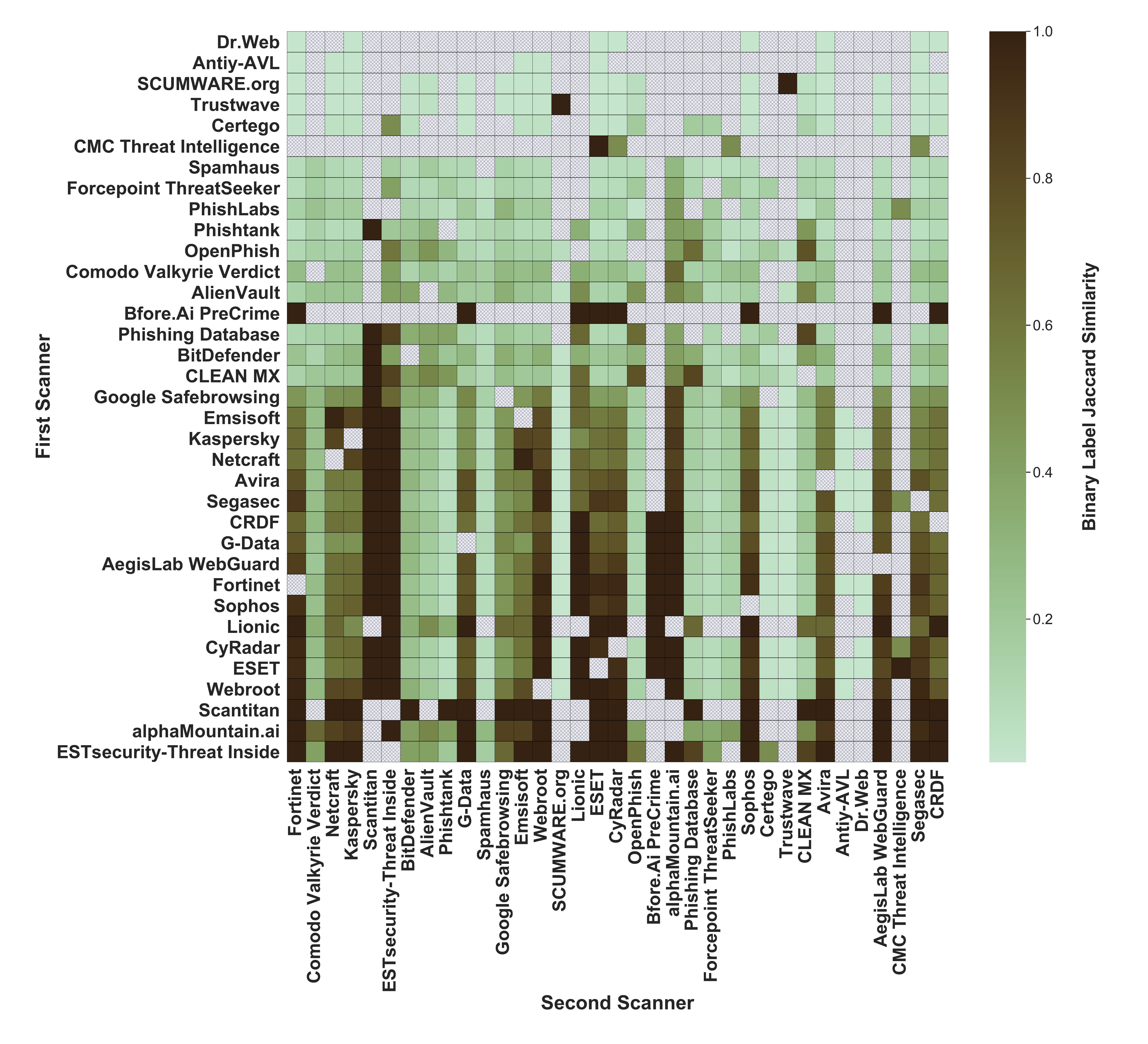,width=0.4\textwidth}\label{verified_label_heatmap}}
\subfigure[Phishing - Detail Label Similarity]{
\psfig{file=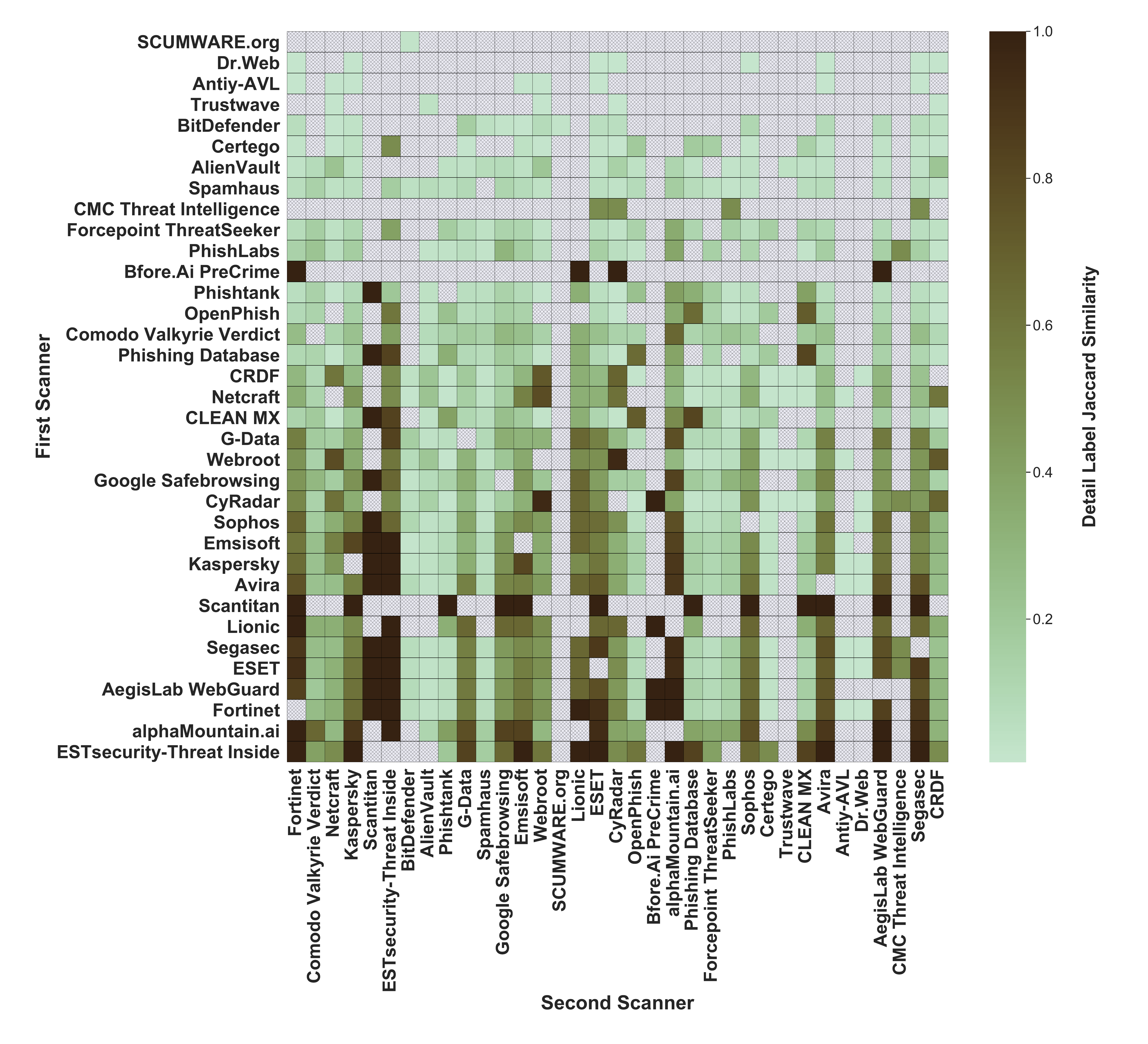,width=0.4\textwidth}\label{verified_exact_label_heatmap}}
}
\caption{Phishing - Jaccard similarity of scanners's binary and detail labels for all periods}\label{fig:label_jaccard_similarity_heatmap}
\end{center}
\end{figure*}

\begin{figure*}[htbp]
\begin{center}
\parbox{1.0\textwidth}{
\centering
\subfigure[Malware - Binary Label Similarity]{
\psfig{file=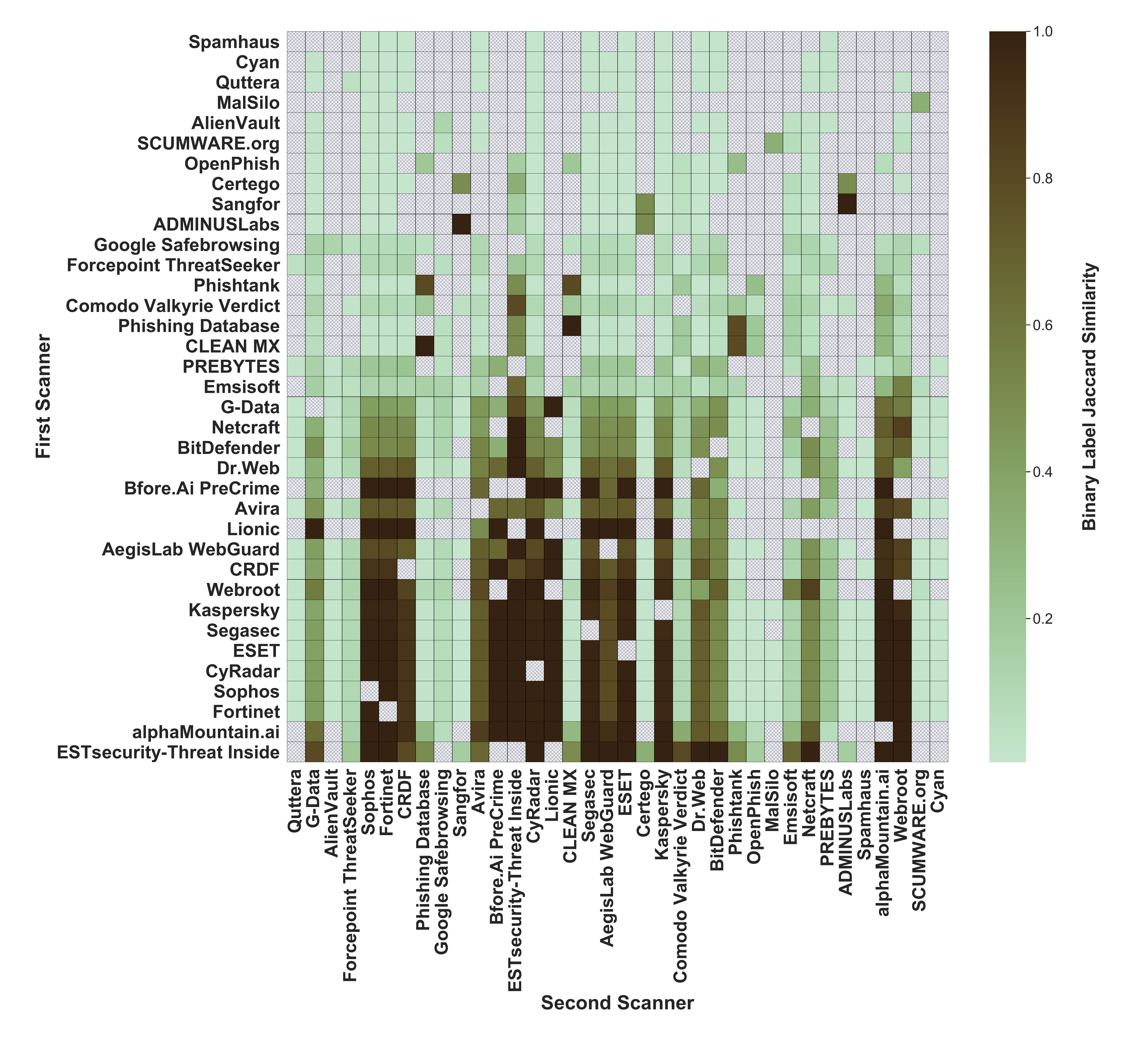,width=0.4\textwidth}\label{openphish_label_heatmap}}
\subfigure[Malware - Detail Label Similarity]{
\psfig{file=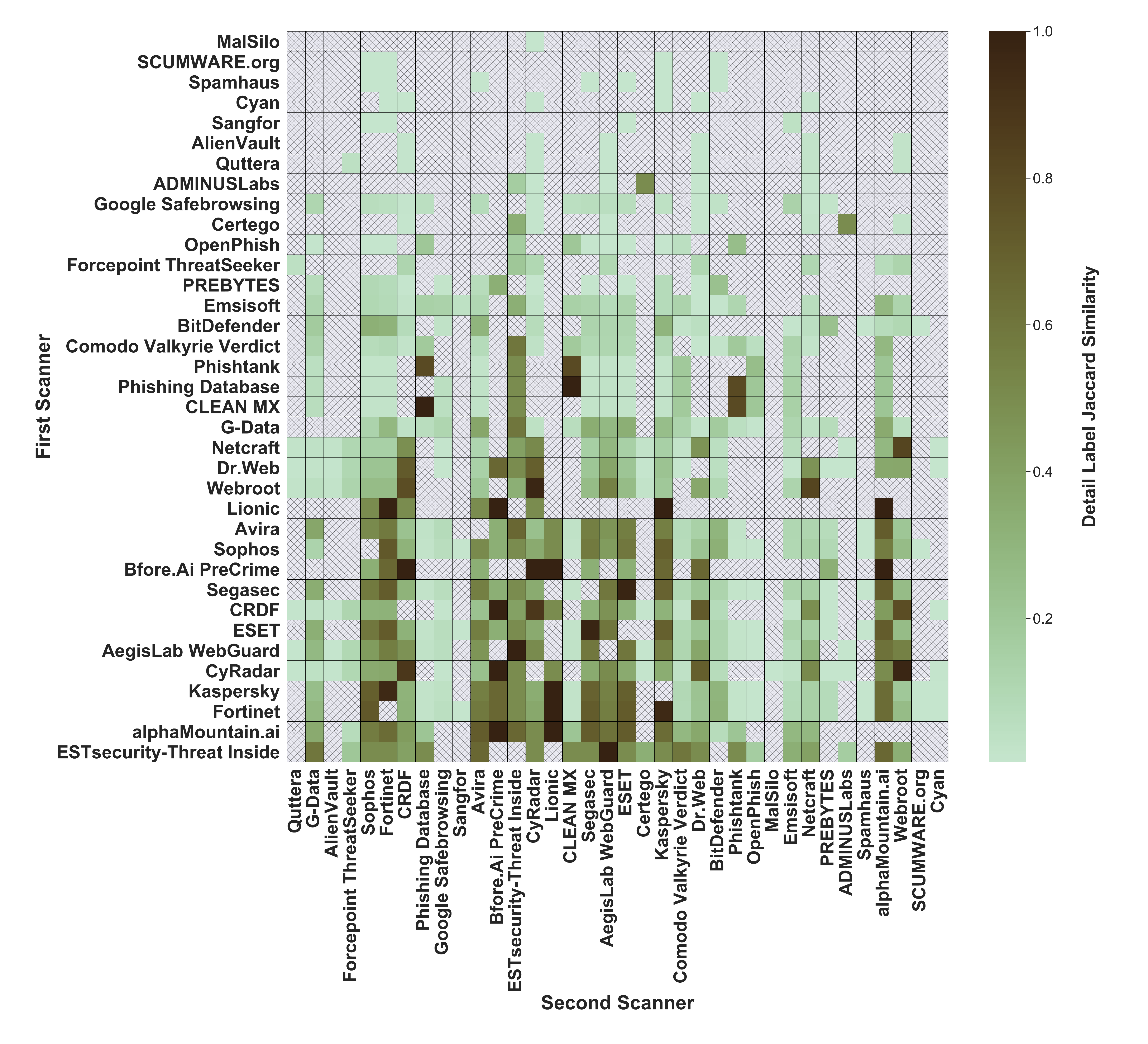,width=0.4\textwidth}\label{openphish_exact_label_heatmap}}
}
\caption{Malware- Jaccard similarity of scanners's binary and detail labels for all periods}\label{fig:label_jaccard_similarity_heatmap}
\end{center}
\end{figure*}

\begin{figure*}[htbp]
\begin{center}
\parbox{1.0\textwidth}{
\centering
\subfigure[Manual GT Malicious - Binary Label Similarity]{
\psfig{file=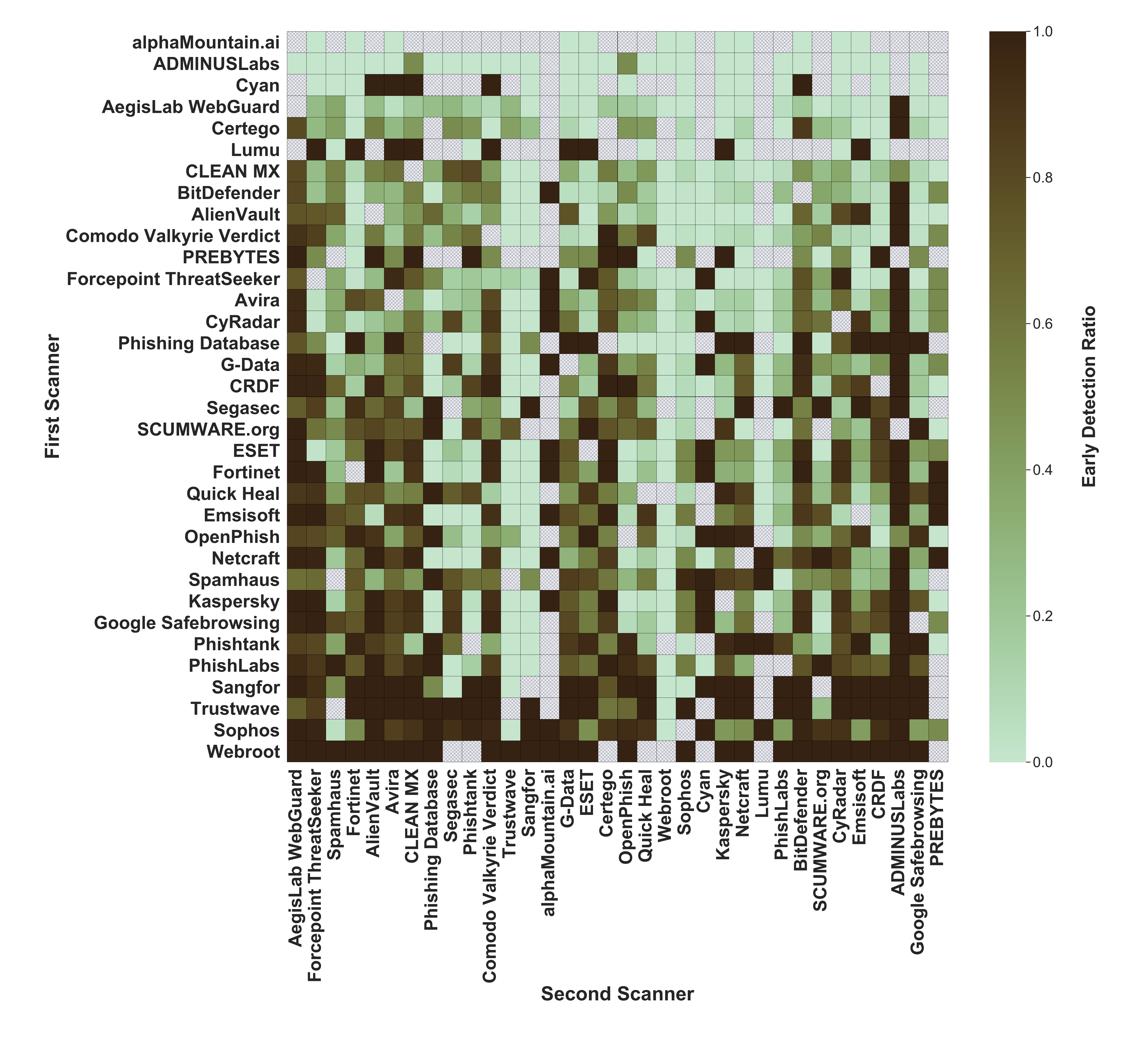,width=0.4\textwidth}\label{manualgt_heatmap}}
\subfigure[Manual GT Malicious - Detail Label Similarity]{
\psfig{file=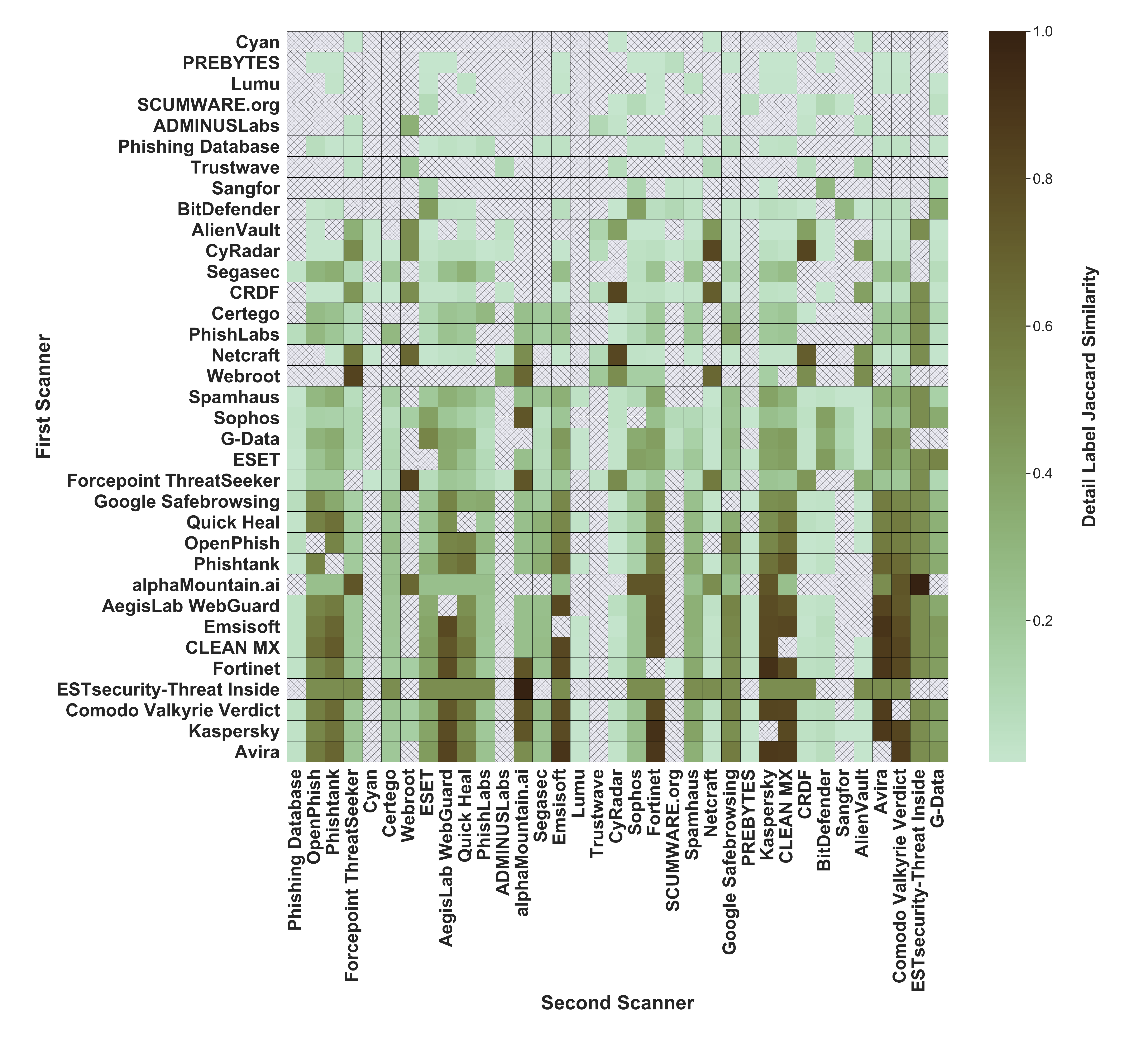,width=0.4\textwidth}\label{manualgt_exact_label_heatmap}}
}
\caption{Manual GT Malicious - Jaccard similarity of scanners's binary and detail labels for all periods}\label{fig:label_jaccard_similarity_heatmap}
\end{center}
\end{figure*}

\begin{figure*}[htbp]
\begin{center}
\parbox{1.0\textwidth}{
\centering
\subfigure[VT fresh URLs - Binary Label Similarity]{
\psfig{file=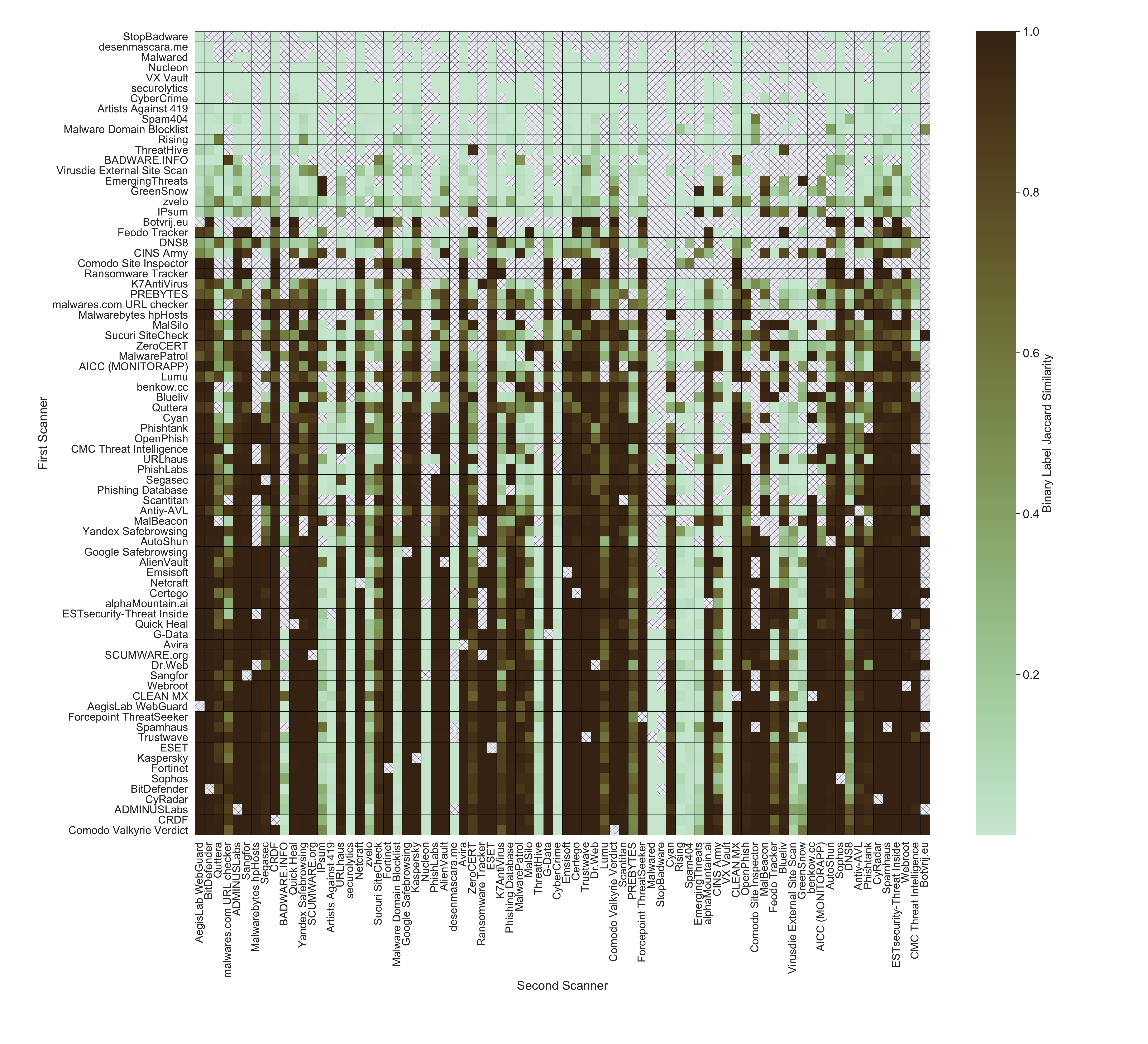,width=0.4\textwidth}\label{allengine_label_heatmap}}
\subfigure[VT fresh URLs - Detail Label Similarity]{
\psfig{file=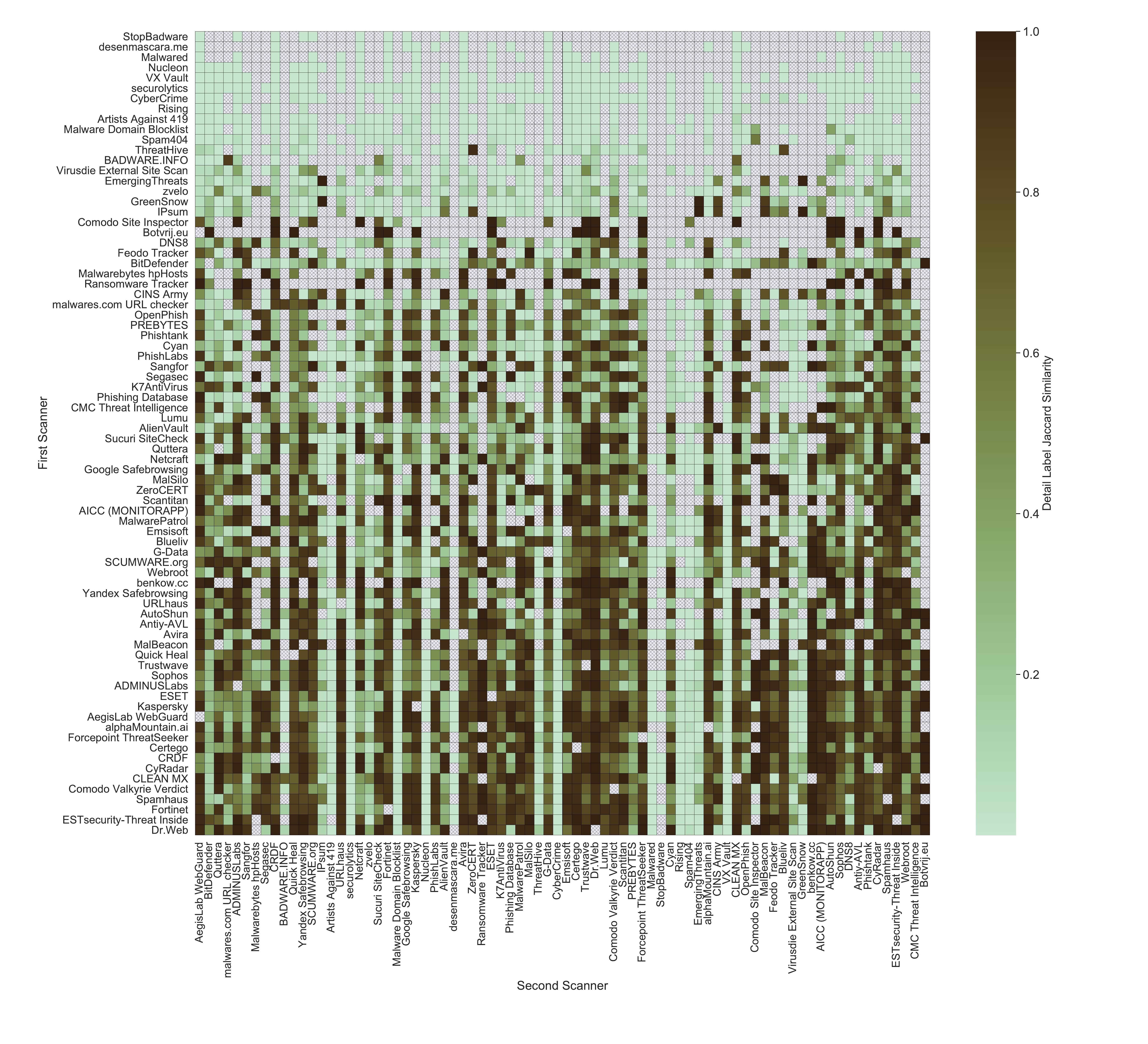,width=0.4\textwidth}\label{allengine_exact_label_heatmap}}
}
\caption{VT fresh URLs - Jaccard similarity of scanners's binary and detail labels for all periods}\label{fig:label_jaccard_similarity_heatmap}
\end{center}
\end{figure*}

\begin{figure}[htbp]
\begin{center}
\parbox{1.0\textwidth}{
\centering
\subfigure[VT Fresh]{
\psfig{file=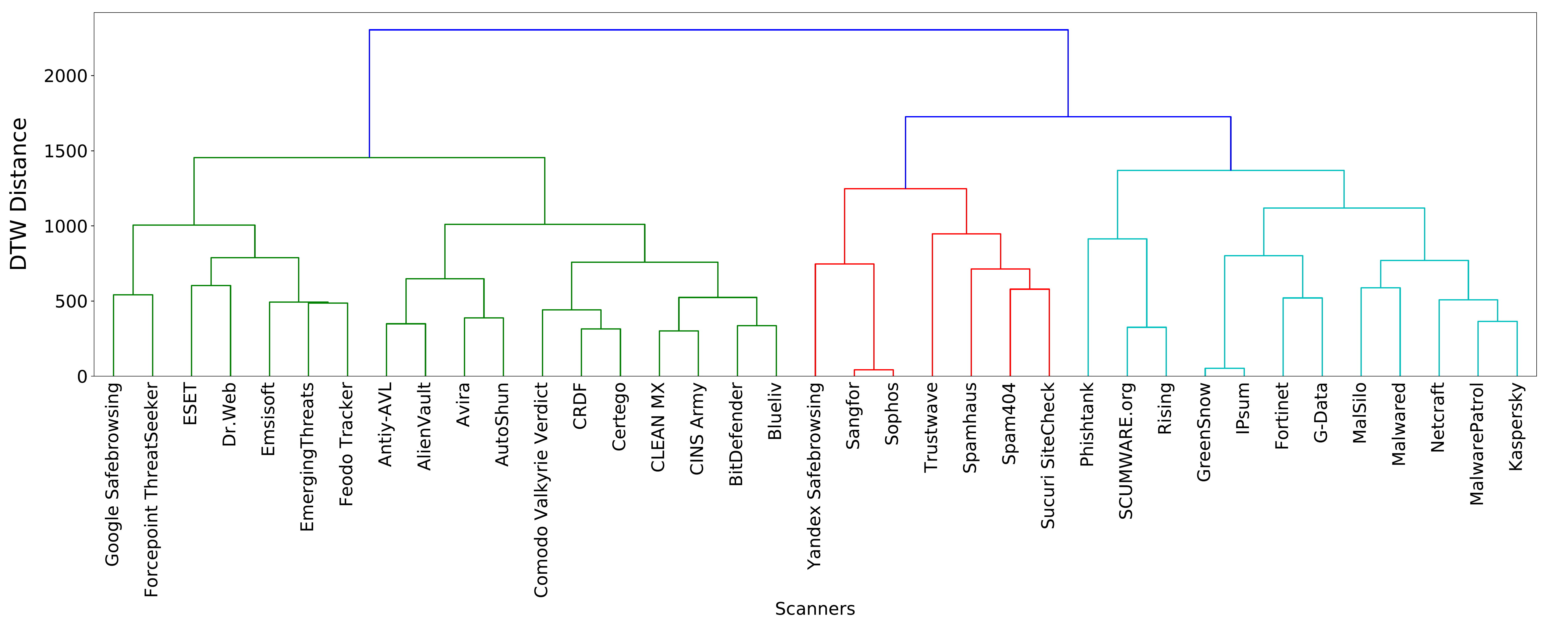,width=0.8\textwidth, height=0.18\textheight}\label{allengine_daily_clustering_dendro}}
\subfigure[Manual GT Malicious]{
\psfig{file=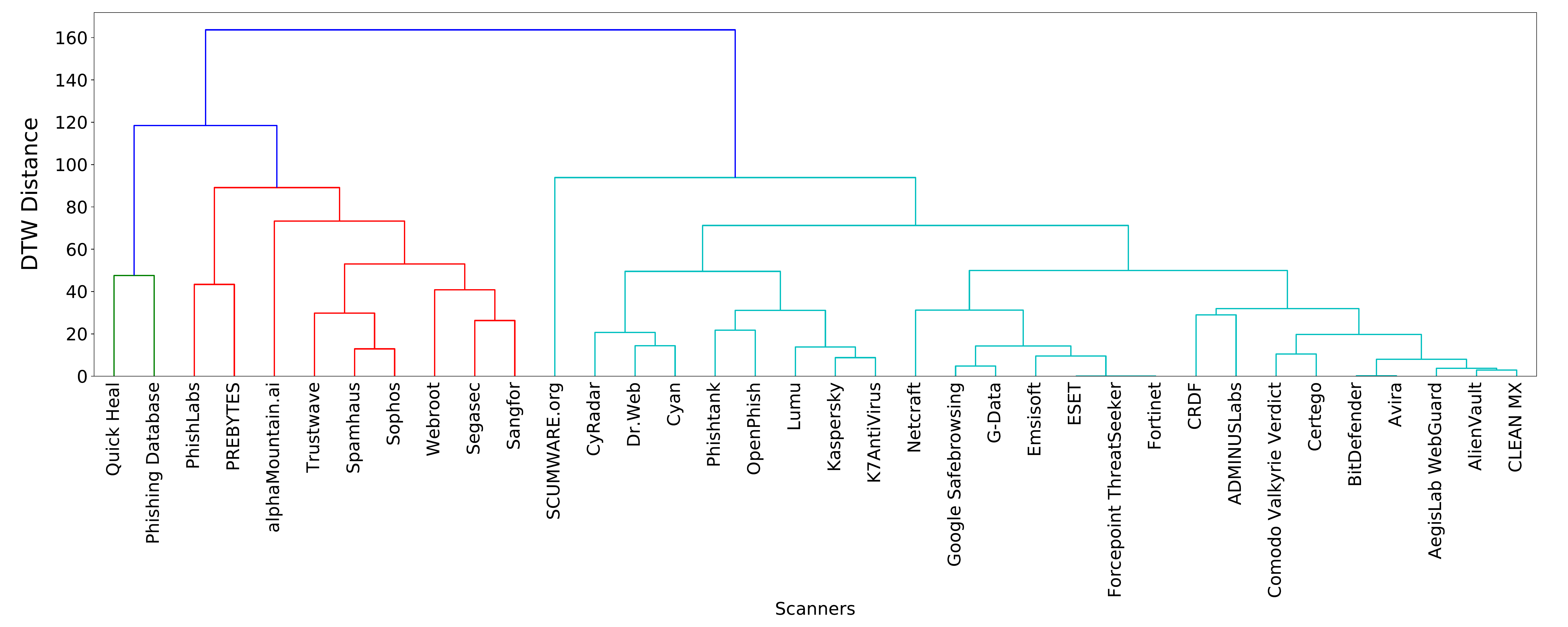,width=0.8\textwidth, height=0.18\textheight}\label{manual_daily_clustering_dendro}}
\subfigure[Phishing]{
\psfig{file=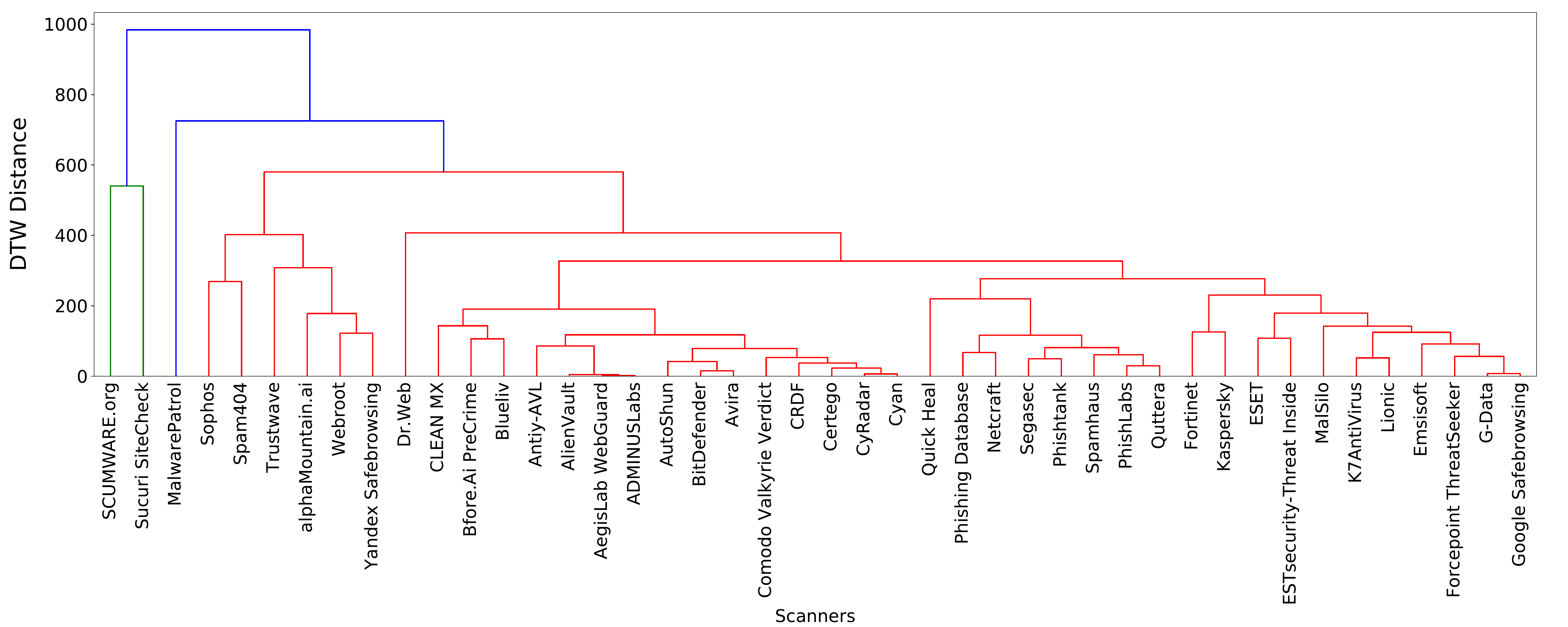,width=0.8\textwidth, height=0.18\textheight}\label{fig:phishing_clustering_dendro}}
\subfigure[Malware]{
\psfig{file=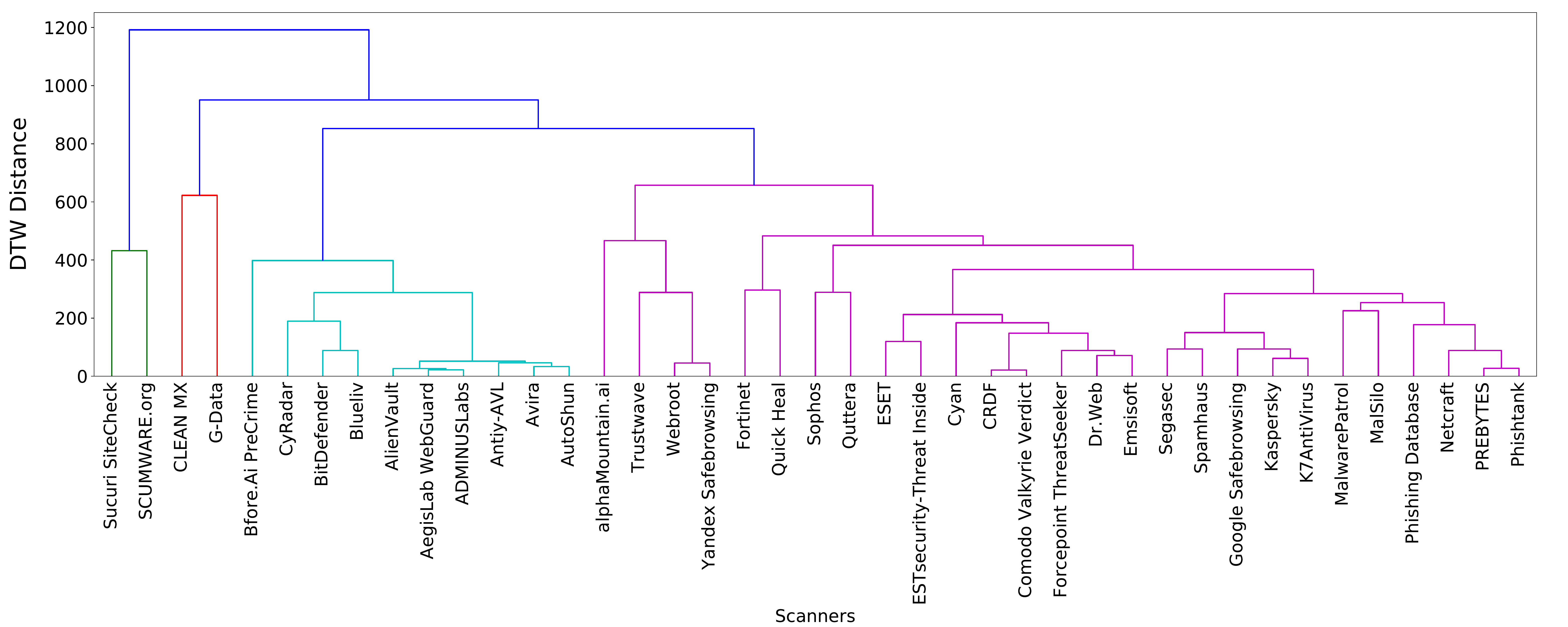,width=0.8\textwidth, height=0.18\textheight}\label{fig:malware_clustering_dendro}}
}
\caption{Scanner clustering using dynamic time warping distance}\label{fig:engine_clustering_dendro}
\end{center}
\end{figure}

\clearpage
\section*{Appendix VI - Heatmaps for scanners' early detection ratio}

\begin{figure}[htbp]
\begin{center}
\parbox{1.0\textwidth}{
\centering
\subfigure[VT fresh URLs]{
\psfig{file=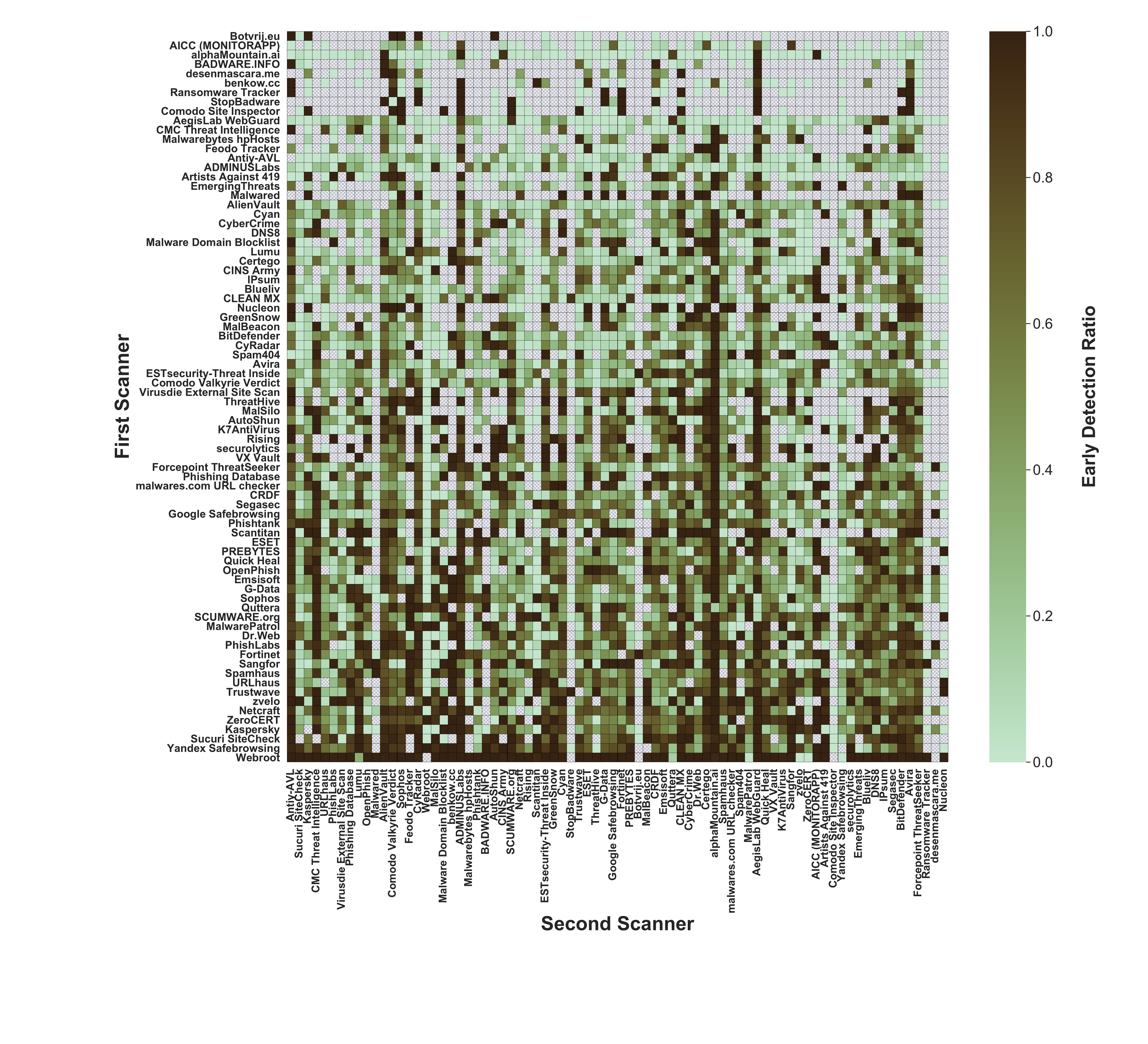,width=0.48\textwidth}\label{allengine_heatmap}}
\subfigure[Manual GT Malicious]{
\psfig{file=figure/manual_early_heatmap_sort.pdf,width=0.48\textwidth}\label{manualgt_heatmap}}
}
\caption{Early detection ratio of first scanners being earlier than the second scanner}\label{fig:early_detectratio_appendix}
\end{center}
\end{figure}
\end{appendices}



\end{document}